\documentclass[]{svjour3}

\usepackage{pgfplots,tikz-3dplot}
\usepackage{amsmath}
\usepackage{bm}
\usepackage{multirow}
\usepackage{hyperref}
\usepackage{ar}

\newcommand{\aver}[1]{ \! \left\langle {#1} \right \rangle \!}

\begin{document}

\title{The turbulent flow over the BARC rectangular cylinder: \\ a DNS study}
\titlerunning{DNS of turbulent flow over the BARC}

\author{Alessandro Chiarini \and Maurizio Quadrio}
\institute{A. Chiarini \at
              Politecnico di Milano, Dept. Aerospace Science and Technologies \\
              \email{alessandro.chiarini@polimi.it}
           \and
           M. Quadrio (corresponding author), \at
              Politecnico di Milano, Dept. Aerospace Science and Technologies \\
              \email{maurizio.quadrio@polimi.it}
}

\date{}

\maketitle

\begin{abstract}
A Direct Numerical Simulation (DNS) of the incompressible flow around a rectangular cylinder with chord-to-thickness ratio 5:1 (also known as the BARC benchmark) is presented. The work replicates the first DNS of this kind recently presented by \cite{cimarelli-leonforte-angeli-2018b}, and intends to contribute to a solid numerical benchmark, albeit at a relatively low value of the Reynolds number. The study differentiates from previous work by using an in-house finite-differences solver instead of the finite-volumes toolbox OpenFOAM, and by employing finer spatial discretization and longer temporal average. 

The main features of the flow are described, and quantitative differences with the existing results are highlighted. The complete set of terms appearing in the budget equation for the components of the Reynolds stress tensor is provided for the first time. The different regions of the flow where production, redistribution and dissipation of each component take place are identified, and the anisotropic and inhomogeneous nature of the flow is discussed. Such information is valuable for the verification and fine-tuning of turbulence models in this complex separating and reattaching flow.
\end{abstract}

\section{Introduction}

The flow around bluff bodies with sharp corners is interesting for both fundamental research and industrial applications, particularly in the field of vortex-induced oscillations \cite{williamson-govardhan-2008}. In fact, several types of structures (buildings, bridges, pylons etc) often present such cross-sectional shapes \cite{tamura-miyagi-kitagishi-1998}. 

The simplest prototype of such bodies is the rectangular cylinder. It sports a simple geometry, yet the flow around it is rich with features that are found in flows around bodies of more complex shape: a corner-induced separation, a shear layer which at a certain point becomes unstable, a detached boundary layer that may reattach, several recirculating regions and a large wake. By varying the aspect ratio $\AR$, the rectangular cylinder spans the entire set of blunt bodies, from a zero-thickness flat plate normal to the flow for $\AR=0$, to a square cylinder for $\AR=1$, and finally to a flat plate parallel to the flow for $\AR \rightarrow \infty$. 

Already at low values of the Reynolds number $Re$, the main flow features depend on $\AR$ . For small aspect ratios, say $\AR \le 2$, the body is not long enough for the flow to reattach after the leading-edge separation. For $2 \le \AR \le 3$ the flow does reattach along the top and bottom sides, but reattachment is intermittent, and vortex shedding still occurs from the leading-edge corners only. For even larger $\AR$, reattachment becomes permanent, creating a large recirculating region sometimes referred to as primary vortex, and vortex shedding occurs from both the leading- and trailing-edge corners, while the main flow features keep changing with the aspect ratio. For large enough $Re$, transition to turbulence complicates the matter further: the large scales associated with the flow instabilities coexist and non-linearly interact with the smaller scales associated to the turbulent motions. 

This variability is accompanied by a significant scatter of experimental and numerical data. This is the main reason why a benchmark on the aerodynamics of a rectangular cylinder with $\AR=5$ (Benchmark on the Aerodynamics of a Rectangular 5:1 Cylinder, or BARC) has been established \cite{bartoli-etal-2008}. Goal of the BARC benchmark is to set the standards for both numerical simulations and experiments, and to arrive at a quantitatively accurate description of the main patterns of the flow, e.g. the size of the primary vortex, the shedding frequency, the root-mean-square value of fluctuations of the vertical force. Unfortunately, as discussed by Bruno et al. in \cite{bruno-salvetti-ricciardelli-2014}, even with $\AR$ fixed, a large variability remains, which still affects the prediction of the main features of the mean flow. Besides the effect of the Reynolds number, this is due to the strong sensitivity of the BARC flow to several aspects of the experimental as well as the numerical studies. For experiments, these range from test conditions and body shape to measurement inaccuracies; for numerical simulations, they include RANS and LES turbulence modelling, discretisation choices, the numerical method itself and the computational procedure. The dispersion of BARC experimental data was considered in Ref. \cite{mariotti-etal-2016}, where uncertainties on the angle of incidence, the free-stream longitudinal turbulence intensity and the free-stream turbulence lengthscale were studied via a probabilistic method and two-dimensional URANS simulations; such uncertainties were found to not fully explain the scatter of results, with the turbulence model playing a major role. Similarly Ref. \cite{mariotti-siconolfi-salvetti-2017} presented a stochastic analysis of the sensitivity of LES results to grid resolution and details of the model filter, finding that a finer grid translates into a significantly smaller primary vortex. More recently, Ref. \cite{rocchio-etal-2020} performed a sensitivity analysis of LES simulations of the BARC flow to the rounding of the leading-edge corners, to the aim of explaining the discrepancy between well-resolved numerical simulations and experiments. They observed that introducing a very small radius of curvature is sufficient to enhance the agreement between numerical and experimental data.

As an example, we summarise below the information available for some quantities of interest, namely the frequency (expressed by the non-dimensional Strouhal number $St$) of the vortex shedding and the length $L_1$ (expressed in terms of the body height $D$) of the main recirculating region created by the leading-edge shear layer. Further information can be found in Ref.\cite{bruno-salvetti-ricciardelli-2014}. In \cite{matsumoto-etal-2003} Matsumoto and coworkers studied the BARC flow experimentally at $Re \approx 10^5$, with both smooth and turbulent incoming flows. The dominant shedding frequency was $St=0.132$ for the smooth inflow and $St=0.197$ for the turbulent inflow, with the primary vortex length varying from $L_1=1.875$ to $L_1=4.375$. Bruno et al. in \cite{bruno-etal-2010} performed a LES simulation at $Re = 4 \cdot 10^4$ finding $St \approx 0.11$ and $L_1 \approx 4.68$. Mannini et al. \cite{mannini-soda-schewe-2010} performed URANS simulations at $Re= 1 \cdot 10^5$ using a slightly modified Spalart-Allmaras turbulence model and a Linearised Explicit Algebraic model coupled with the standard $k-\omega$ model. Using the first model they found $St \approx 0.098 $ with no flow reattachment; with the second model, instead, they found $St \approx 0.105$ and $L_1 \approx 4.65$. Mannini et al. \cite{mannini-soda-schewe-2011} performed a DES study at $Re=26400$ focusing on the effect of the spanwise dimension of the domain on the main properties of the flow; they report $St \approx 0.1$ and $L_1 \approx 4.75$. Bruno et al. \cite{bruno-coste-fransos-2012} studied the combined effect of spanwise discretisation and spanwise domain size with LES. For an increasing spanwise resolution $L_1$ decreases whereas $St$ remains practically constant; increasing the spanwise domain size, instead, affects neither $St$ nor $L_1$. Patruno et al. \cite{patruno-etal-2016} studied with LES and URANS the effect of small angles of incidence at $Re \approx 10^4$. At zero incidence they found a non-symmetric mean flow with LES, with $L_1=4.01$ and $L_1=4.1$ in the upper and lower cylinder sides and $St=0.132$. Their URANS simulation, instead, using the $k-\omega-SST$ turbulence model yielded a symmetric mean flow with $L_1=4.26$ and $St=0.121$. Ricci et al. \cite{ricci-etal-2017} performed LES simulations at $Re \approx 5.5 \cdot 10^4$ to study the turbulent inflow conditions. They found that a higher level of incoming turbulence corresponds to a higher curvature of the shear layer and therefore to a shorter primary vortex, with a related upstream shift of the secondary vortex. A similar study was also carried out by Mannini et al. \cite{mannini-etal-2017} experimentally, by varying the inflow conditions and the angle of incidence in the range  $10^4 < Re < 10^5$. For zero incidence they found $St \approx 0.115$ which weakly increases with $Re$ and with the intensity of the free stream turbulence. Moore et al. \cite{moore-etal-2019} experimentally found for $Re=1.34 \cdot 10^4$ $St \approx 0.1114$ and $L_1 \approx 4.4$; moreover, they found that $St$ does not change with the Reynolds number in the range $ 1.34 \cdot 10^4 \le Re \le 1.18 \cdot 10^5$ unlike $L_1$ which has a decreasing trend. Lastly, Cimarelli et al. \cite{cimarelli-franciolini-crivellini-2020} investigated via high-order implicit LES the influence of geometrical characteristics of the body, as the sharp leading-edge corners, the presence of separation at the trailing edge and the coupling between the two sides of the plate. At $Re=3000$ they report $St=0.14$ and $L_1 \approx 4.05$.

In such scenario of highly scattered experimental and numerical information, a remarkable achievement was the recent first Direct Numerical Simulation (DNS) of the BARC flow in the turbulent regime. It was presented by Cimarelli, Leonforte and Angeli \cite{cimarelli-leonforte-angeli-2018b}, who employed the finite-volumes OpenFOAM toolbox \cite{weller-etal-1998}, and was then used in the derivative works \cite{cimarelli-leonforte-angeli-2018,cimarelli-etal-2019-negative}. Albeit at a relatively low value of $Re$, these data represent a key step towards building a robust information set for the BARC flow, since DNS has the ability to remove from the picture the uncertainties related to turbulence modeling. It remains desirable, however, to assess the robustness of their results with respect to discretization, numerical method and computational procedures. This is one of the goals of the present work, which replicates this DNS study but employs a different solver (an in-house finite-differences code), and uses different discretization choices. As a second objective of the present contribution, we also intend to advance the statistical characterization and understanding of the BARC flow, by presenting for the first time a detailed discussion of the complete set of terms involved in the budget equations for the components of the tensor of the Reynolds stresses. The availability of the complete budget for the Reynolds stresses is essential for improving LES and RANS closure models.

\section{Numerical method and computational procedures}
\label{sec:methods}
\begin{figure}
\centering
\tdplotsetmaincoords{70}{150}	
\begin{tikzpicture}[scale=0.30,tdplot_main_coords]
	
	\tikzset{myptr/.style={decoration={markings,mark=at position 1 with {\arrow[scale=3,>=stealth]{>}}},postaction={decorate}}}

        \coordinate (Aup) at ( -1, -8,  0.5);
        \coordinate (Bup) at ( 4, -8,  0.5);
        \coordinate (Cup) at ( 4,  8,  0.5);
        \coordinate (Dup) at ( -1,  8,  0.5);
        \coordinate (Alo) at ( -1, -8, -0.5);
        \coordinate (Blo) at ( 4, -8, -0.5);
        \coordinate (Clo) at ( 4,  8, -0.5);
        \coordinate (Dlo) at ( -1,  8, -0.5);

        \coordinate (A1) at ( 4,  9.5, -0.5);
        \coordinate (A2) at (-1,  9.5, -0.5);
        \coordinate (A3) at ( 4,    8, -0.5);
        \coordinate (A4) at (-1,    8, -0.5);
        \draw[<->] (A1)--(A2);
        \draw[dashed] (A1)--(A3);
        \draw[dashed] (A2)--(A4);
        \node at (1.5,11,-0.5) {$L=5D$};
        
        \coordinate (B1) at ( -2.5, 8, 0.5);
        \coordinate (B2) at ( -2.5, 8, -0.5);
        \coordinate (B3) at ( -1,   8, 0.5);
        \coordinate (B4) at ( -1,   8, -0.5);
        \draw[<->] (B1)--(B2);
        \draw[dashed] (B1)--(B3);
        \draw[dashed] (B2)--(B4);
        \node at (-3.7,8,0) {$D$};

        \draw[very thick,black] (Aup) -- (Bup);
        \draw[very thick,black] (Bup) -- (Cup);
        \draw[very thick,black] (Cup) -- (Dup);
        \draw[very thick,black] (Dup) -- (Aup);

        \draw[very thick,black,dashed] (Alo) -- (Blo);
        \draw[very thick,black] (Blo) -- (Clo);
        \draw[very thick,black] (Clo) -- (Dlo);
        \draw[very thick,black,dashed] (Dlo) -- (Alo);

        \draw[very thick,black,dashed] (Aup) -- (Alo);
        \draw[very thick,black] (Bup) -- (Blo);
        \draw[very thick,black] (Cup) -- (Clo);
        \draw[very thick,black] (Dup) -- (Dlo);

        \coordinate (AAup) at (-19, -8,  0.5);
        \coordinate (BBup) at ( 12, -8,  0.5);
        \coordinate (CCup) at ( 12,  8,  0.5);
        \coordinate (DDup) at (-19,  8,  0.5);
        \coordinate (AAlo) at (-19, -8, -0.5);
        \coordinate (BBlo) at ( 12, -8, -0.5);
        \coordinate (CClo) at ( 12,  8, -0.5);
        \coordinate (DDlo) at (-19,  8, -0.5);

        \coordinate (AAupp) at (  -1, -8,  6);
        \coordinate (BBupp) at (  4, -8,  6);
        \coordinate (CCupp) at (  4,  8,  6);
        \coordinate (DDupp) at (  -1,  8,  6);
        \coordinate (AAloo) at (  4, -8, -6);
        \coordinate (BBloo) at (  4, -8, -6);
        \coordinate (CCloo) at (  4,  8, -6);
        \coordinate (DDloo) at (  -1,  8, -6);

        \draw[dashed] (AAlo) -- (AAup);
        \draw[] (BBlo) -- (BBup);
        \draw[] (CClo) -- (CCup);
        \draw[] (DDlo) -- (DDup);

        \draw[] (AAupp) -- (BBupp);
        \draw[] (CCupp) -- (DDupp);

        \draw[dashed] (AAloo) -- (BBloo);
        \draw[] (CCloo) -- (DDloo);

        \coordinate (OAup) at (-19, -8,  6);
        \coordinate (OBup) at ( 12, -8,  6);
        \coordinate (OCup) at ( 12,  8,  6);
        \coordinate (ODup) at (-19,  8,  6);
        \coordinate (OAlo) at (-19, -8, -6);
        \coordinate (OBlo) at ( 12, -8, -6);
        \coordinate (OClo) at ( 12,  8, -6);
        \coordinate (ODlo) at (-19,  8, -6);

        \draw[] (OAup) -- (AAupp);
        \draw[] (DDupp) -- (ODup);
        \draw[] (ODup)  -- (OAup);

        \draw[dashed] (OAlo) -- (AAloo);
        \draw[] (DDloo) -- (ODlo);
        \draw[dashed] (ODlo)  -- (OAlo);

        \draw[] (BBupp) -- (OBup);
        \draw[] (OBup)  -- (OCup);
        \draw[] (OCup)  -- (CCupp);

        \draw[dashed] (OAup) -- (AAup);
        \draw[] (OCup) -- (CCup);
        \draw[] (ODup) -- (DDup);
        \draw[] (OBup) -- (BBup);

        \draw[] (OBlo) -- (OClo);
        \draw[] (BBlo) -- (OBlo);
        \draw[] (CClo) -- (OClo);

        \draw[dashed] (OBlo) -- (BBloo);
        \draw[] (OClo) -- (CCloo);
        \draw[dashed] (OAlo) -- (AAlo);
        \draw[] (ODlo) -- (DDlo);	

        \coordinate (R1) at ( -20.5, 8, 6);
        \coordinate (R2) at ( -20.5, 8, -6);
        \coordinate (R3) at ( -19, 8, 6);
        \coordinate (R4) at ( -19, 8, -6);
        \draw[<->] (R1)--(R2);
        \draw[dashed] (R1)--(R3);
        \draw[dashed] (R2)--(R4);
        \node at (-21.7,8,0) {$42D$};
        
        \coordinate (P1) at ( 1.5, 9.5, -6);
        \coordinate (P2) at (-19, 9.5, -6);
        \coordinate (P3) at ( 1.5, 8, -6);
        \coordinate (P4) at ( -19, 8, -6);
        \draw[<->] (P1)--(P2);
        \draw[dashed] (P1)--(P3);
        \draw[dashed] (P2)--(P4);
        \node at (-5.75,11.7,-6) {$40D$};
        
        \coordinate (Q1) at ( 12, 9.5, -6);
        \coordinate (Q2) at (  1.5, 9.5, -6);
        \coordinate (Q3) at ( 12, 8, -6);
        \coordinate (Q4) at ( 1.5, 8, -6);
        \draw[<->] (Q1)--(Q2);
        \draw[dashed] (Q1)--(Q3);
        \draw[dashed] (Q2)--(Q4);
        \node at (5.25,11.7,-6) {$22.5D$};
        
        \coordinate (S1) at ( 13.5,-8, -6);
        \coordinate (S2) at ( 13.5, 8, -6);
        \coordinate (S3) at ( 12,-8, -6);
        \coordinate (S4) at ( 12, 8, -6);
        \draw[<->] (S1)--(S2);
        \draw[dashed] (S1)--(S3);
        \draw[dashed] (S2)--(S4);
        \node at (14.7,0,-6) {$5D$};
        
        \coordinate (F1) at (18,0,0);
        \coordinate (F2) at (12,0,0);
        \draw [->,>=stealth] (F1)--(F2);
        \node at (15,0,1) {$U_{\infty}$};

        \coordinate (O0) at (21,0,-6);
        \coordinate (O1) at (18,0,-6);
        \coordinate (O2) at (21,0,-4);
        \coordinate (O3) at (21,2,-6);

        \node at (18,0,-6.5) {$x$};
        \node at (20.5,0,-4) {$y$};
        \node at (20.5,2,-6) {$z$};
        \draw [->] (O0)--(O1);
        \draw [->] (O0)--(O2);
        \draw [->] (O0)--(O3);
\end{tikzpicture}	
\caption{Sketch of the BARC geometry, with the reference system and the computational domain employed in the present work.}
\label{fig:sketch}
\end{figure}
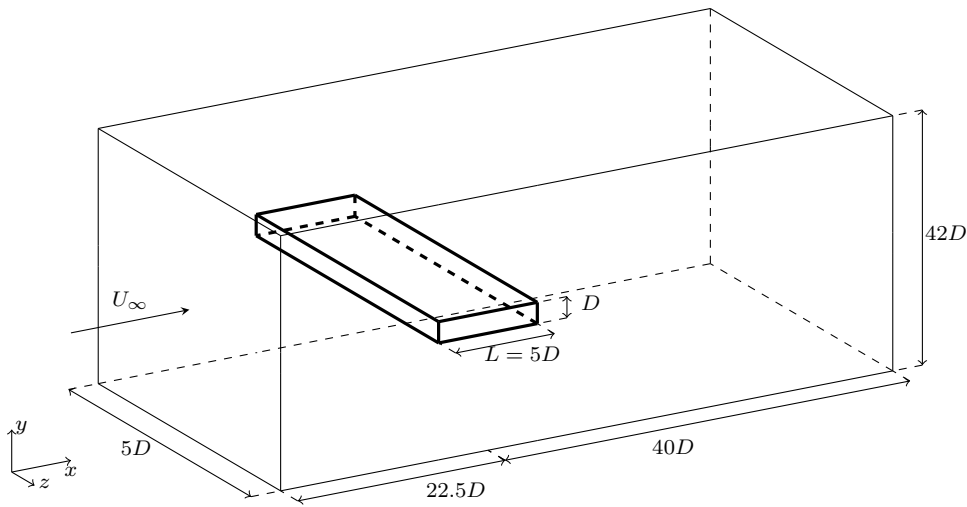
The BARC test case is a two-dimensional rectangular cylinder with streamwise length $L$ and cross-stream dimension $D$, with aspect ratio $\AR = L/D$ set at $\AR=5$. Figure \ref{fig:sketch} sketches the geometry and the reference frame employed in the present work. A Cartesian coordinate system is placed at the center of the cylinder, with the $x$ axis aligned with the flow direction, the $y$ axis denoting the cross-stream direction and the $z$ axis the spanwise direction. The incoming velocity is uniform upstream, aligned with the $x$ axis and set at $U_\infty$. The Reynolds number $Re = U_\infty D/\nu$ is based on $U_\infty$, $D$ and the kinematic viscosity $\nu$ of the fluid. As in the reference DNS work \cite{cimarelli-leonforte-angeli-2018b}, the value of the Reynolds number considered here is $Re=3000$. The incompressible Navier--Stokes equations are:
\begin{equation}
  \begin{cases}
  \displaystyle
  \frac{\partial \bm{u}}{\partial t} + \bm{u} \cdot \bm{\nabla} \bm{u} = -\bm{\nabla} p + \frac{1}{Re} \bm{\nabla}^2 \bm{u} \\
  \bm{\nabla} \cdot \bm{u} = 0
  \end{cases}
\end{equation}
where $u$, $v$ and $w$ denote the streamwise, cross-stream and spanwise components of the velocity, whereas $p$ is the pressure. The mean field is indicated with capital letters, i.e. $\bm{U}=(U,V,W)$, and the fluctuations with an apex, i.e. $\bm{u}=(u', v', w')$. The boundary conditions of the problem are the no-slip and no-penetration conditions on the surface of the cylinder together with an unperturbed velocity $(U_{\infty},0,0)$ enforced at the far field, periodic conditions at the spanwise boundaries on account of spanwise homogeneity, and a convective outlet condition for the velocity, i.e. $\partial \bm{u} / \partial t = -U_\infty \partial \bm{u} / \partial x$.

The DNS code, introduced by Luchini in \cite{luchini-2016}, solves the incompressible Navier--Stokes equations on a staggered Cartesian grid. Second-order finite differences are used in every direction. The momentum equations are advanced in time by a fractional time-stepping method that employs a third-order Runge--Kutta scheme. The Poisson equation for the pressure is solved by an iterative SOR algorithm. The presence of the cylinder is dealt with via an implicit second-order accurate immersed-boundary method, implemented in staggered variables to be continuous with respect to boundary crossing and numerically stable at all distances from the boundary \cite{luchini-2013,luchini-2016}. Hereinafter, all variables are in dimensionless form, with $D$ as length scale, $U_\infty$ as velocity scale and $D/U_\infty$ as time scale. 

The computational domain extends for $-22.5 \le x \le 40$ in the streamwise direction, $-21 \le y \le 21 $ in the vertical direction and $-2.5 \le z \le 2.5$ in the spanwise direction, with the cylinder placed at $-2.5 \le x \le 2.5$ and $-0.5 \le y \le 0.5$. The computational domain is discretised with $N_x=1776$, $N_y=942$ and $N_z=150$ points in the three directions, for a total of more than 250 millions grid points. The distribution of points is uniform in the spanwise direction, whereas geometrically varying grid spacing is employed in the streamwise and vertical directions to yield a higher resolution near the leading- and trailing-edge corners. There, the finest spacing occurs with $\Delta x = \Delta y \approx 0.0015$. The ratio between neighboring streamwise cells is 1.005 over the body from the corners towards $x=0$, and 1.02 before and after the body; the ratio between neighboring vertical cells is 1.01 for $|y| \le 1.75 $ and 1.02 otherwise. This distributions leads to $892$ points placed along the streamwise edge of the cylinder and $217$ along the cross-stream edge. 

When compared to the computational domain used in \cite{cimarelli-leonforte-angeli-2018b}, i.e. $-37.5 \le x \le 74.5 $ and $ -25 \le y \le 25 $, the present domain is slightly smaller, but discretized with a significantly finer mesh. Although known to be marginal for an accurate calculation of some high-order statistics \cite{bruno-coste-fransos-2012}, the spanwise dimension of the cylinder is identical to that of the reference study. In comparison, almost $15$ times more points are placed over the body: about 8 times more points along the streamwise edge, and twice the number of points along the cross-stream edge. 

To accumulate well-converged statistics, we have exploited both temporal averaging and ensemble averaging, by running several independent simulations. The first simulation is started from the solution corresponding to a steady potential flow, with its symmetry broken by injecting just after the leading-edge corners localized random noise of small amplitude, i.e. $ <  U_{\infty}/10$, during the first 5 time units. The simulation is then advanced for a considerably long time, approximately $400 D/U_\infty$, to allow for the flow to fully loose memory of the initial transition and reach a truly statistically-stationary state. Indeed, this flow takes a long time to develop as confirmed by \cite{cimarelli-leonforte-angeli-2018b}, where a comparably long initial transient of $250 D/U_\infty$ was discarded too. Once the flow reaches the desired statistically steady state, the main simulation enters the production stage and accumulates statistics for further $341$ time units. At the same time, from the latest stages of the preliminary run, $10$ independent flow fields are saved and later used to produce independent initial conditions for $10$ additional simulations. At the time of writing, they have accumulated useful statistics for additional $2004$ time units, leading to a total averaging time of $ 2345 D/U_\infty$. The convergence of the statistics has been assessed by testing different sample sizes.

The creation of independent initial conditions is based on the idea of removing the large-scale coherent structures populating the flow, but without excessively distorting its spatial structure, so that the required initial transient can be kept to a minimum. The process is based on Fourier transforming the velocity values along each spanwise line, followed by a random phase change, and a final inverse transform. This procedure preserves the structure of the mean flow but effectively removes the large-scale structures, which are then regenerated quickly but independently on each sample. Although by visual observation the fields become fully independent after a very small simulation time, a conservative approach is adopted, and 20 further time units are discarded before starting the accumulation of statistics. 

The temporal discretization uses a varying time step, to ensure that the Courant-Frederic-Levy CFL number remains at $CFL \le 1$; this condition produces an average value of the time step of $ \Delta t \approx 0.0013$. The simulations have been run on the GALILEO supercomputer at CINECA. Each simulation uses $32$ cores of a single computing node, and subdivides the spatial domain in smaller subdomains along the streamwise and spanwise directions. Each simulation needs approximately $14$ hours to advance the flow by one time unit.

%
%
%

\section{Instantaneous and mean flow fields}
\label{sec:meanflow}

\subsection{Instantaneous flow}

We start by a qualitative characterization of a typical snapshot of the flow, which provides the opportunity of describing its main structures. They are visualised with the $\lambda_2$ criterion \cite{jeong-hussain-1995}, based on the second largest eigenvalue $\lambda_2$ of the tensor $S_{ik} S_{kj} + \Omega_{ik}\Omega_{kj}$, where
\begin{equation}
S_{ij}=\frac{1}{2}\left(\frac{\partial u_i}{\partial x_j} + \frac{\partial u_j}{\partial x_i}\right), \ \ \ \ \ \
\Omega_{ij}=\frac{1}{2}\left( \frac{\partial u_i}{\partial x_j} - \frac{\partial u_j}{\partial x_i} \right)
\end{equation}
are the symmetric and antisymmetric part of the velocity gradient tensor $\partial u_i / \partial x_j$. The spatial orientation of the structures is additionally described by means of isosurfaces of the vorticity components $\omega_x$ and $\omega_z$.

\begin{figure}
\centering
\includegraphics[width=0.49\textwidth]{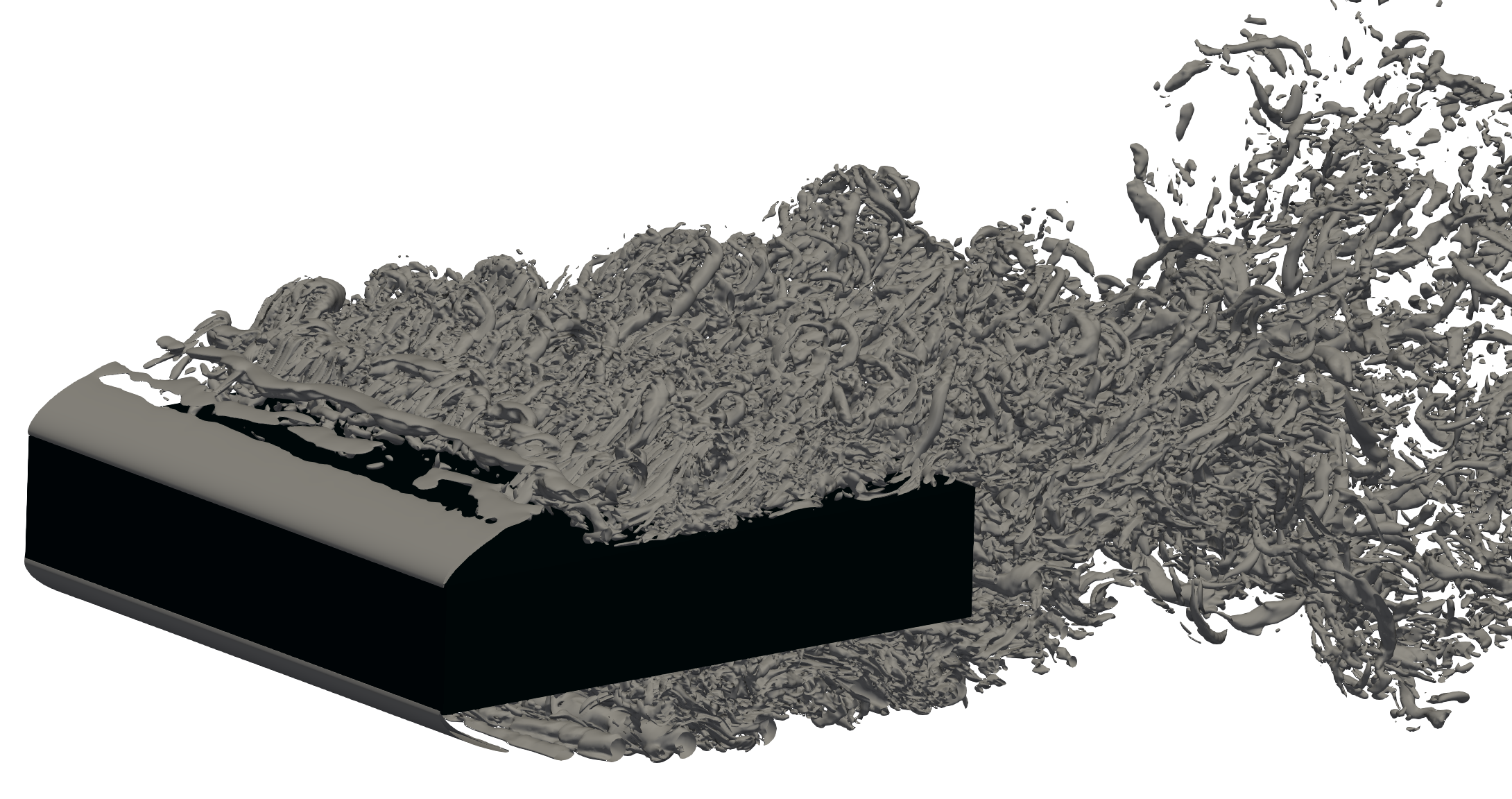}
\includegraphics[width=0.49\textwidth]{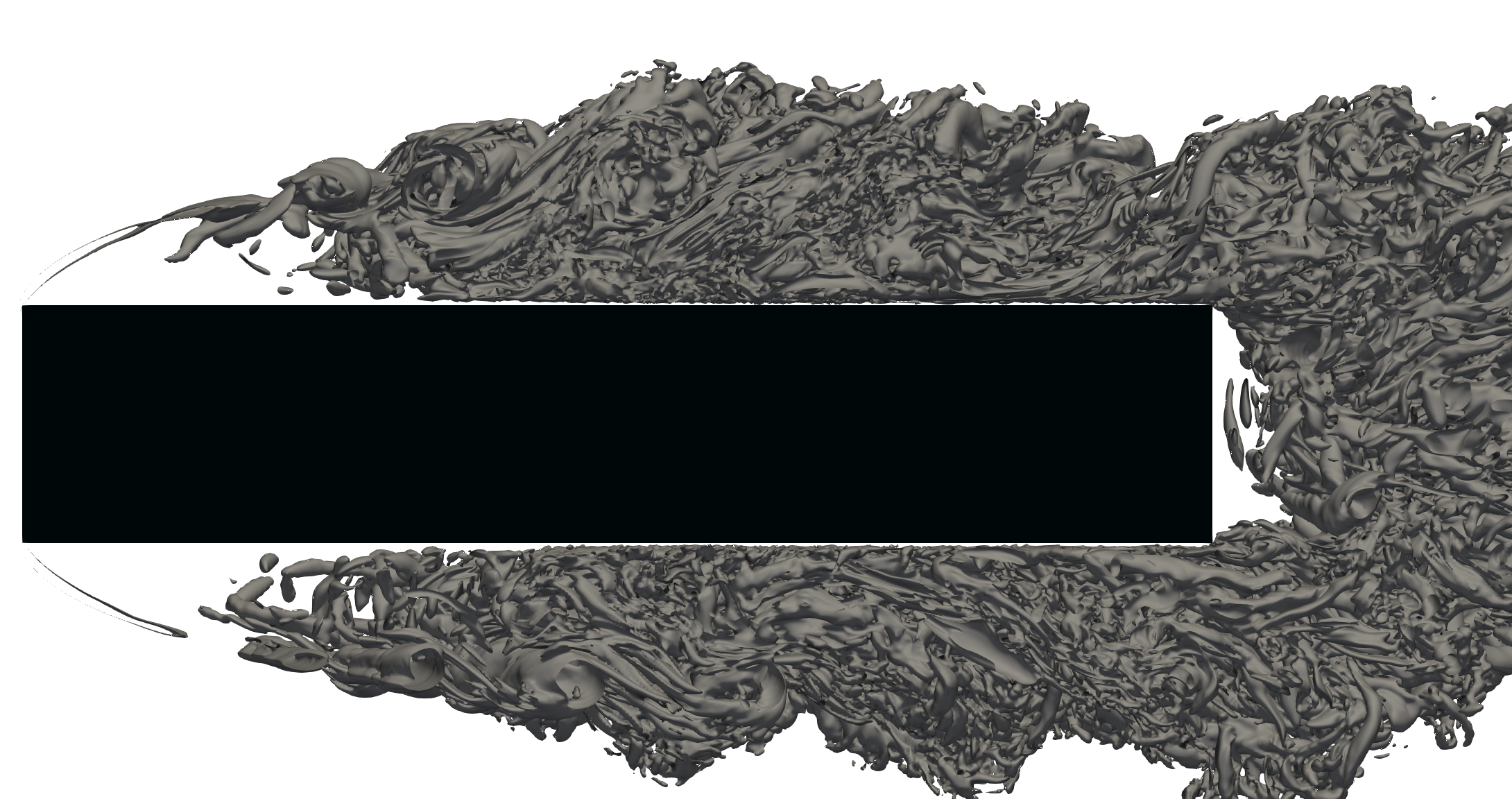}
\includegraphics[width=0.49\textwidth]{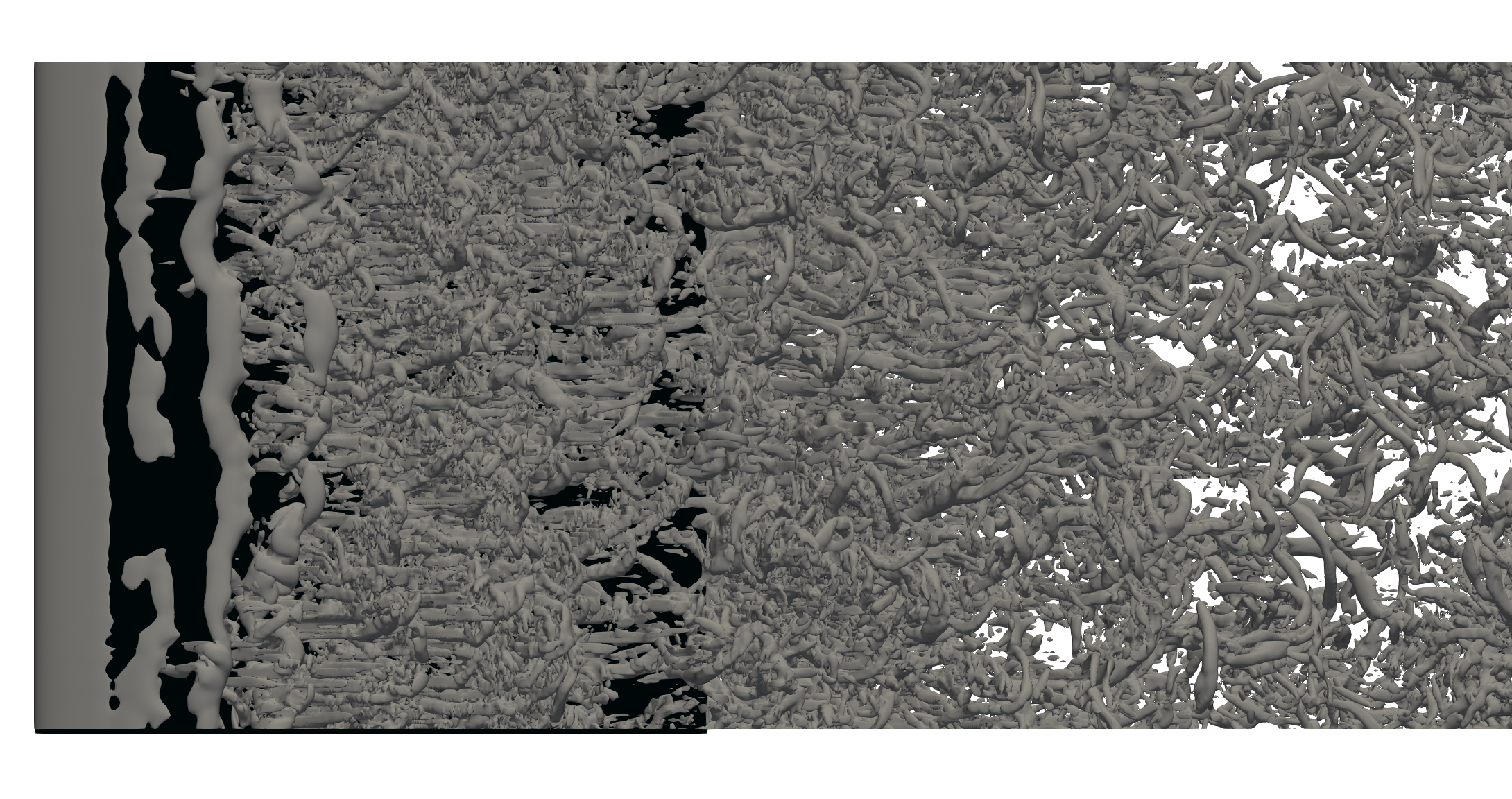}
\includegraphics[width=0.49\textwidth]{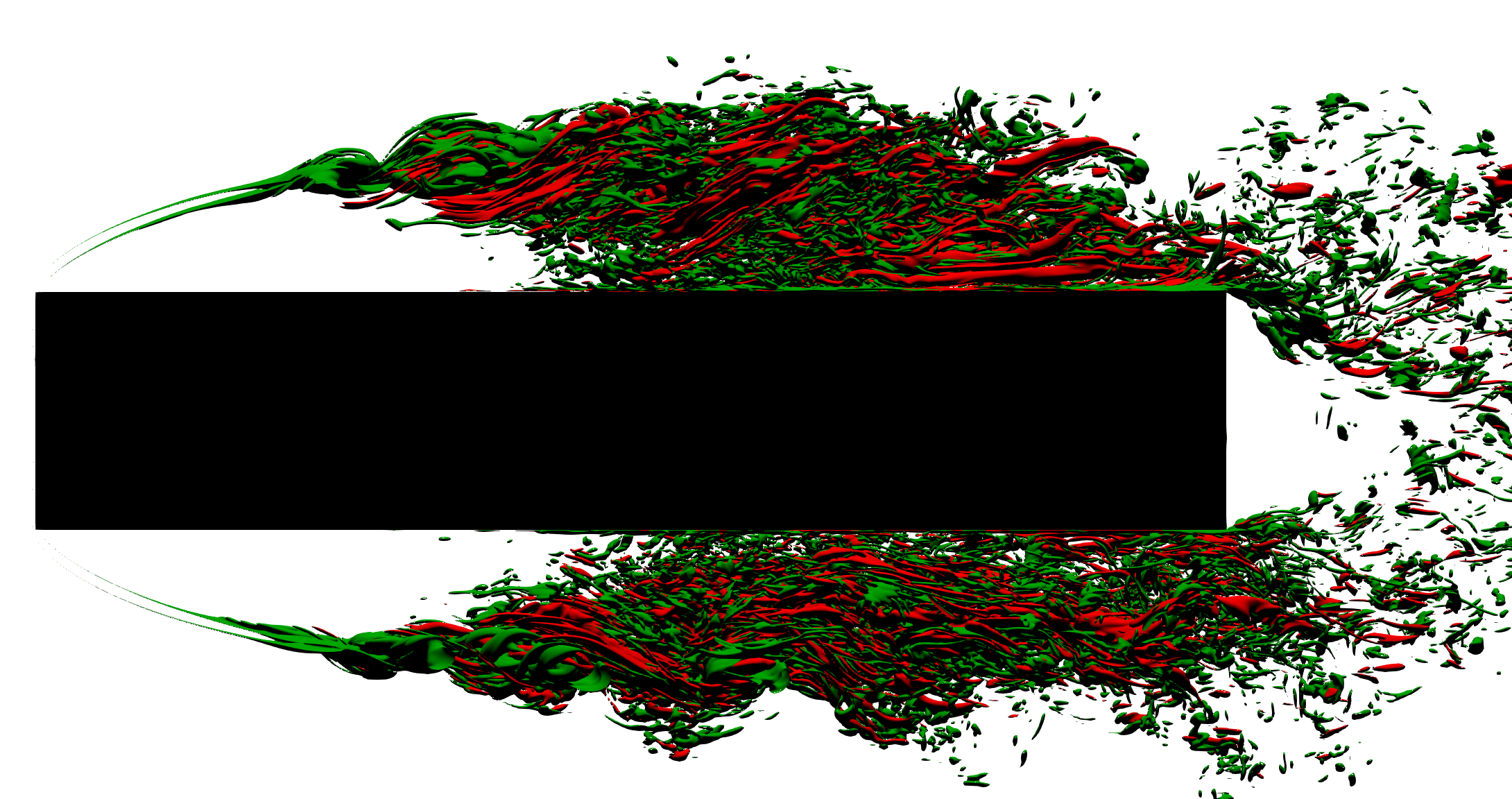}
\caption{The instantaneous flow. Top left: three-dimensional view of the isosurface $\lambda_2=-5$. Top right: side view. Bottom left: top view. Bottom right: side view for isosurfaces of $|\omega_z|=15$ (green) and $|\omega_x|=15$ (red).}
\label{fig:lambda2-vorticity}
\end{figure}
Figure \ref{fig:lambda2-vorticity} plots isosurfaces for $\lambda_2$, $\omega_x$ and $\omega_z$. The sharp leading-edge corner determines the detachment point. In the initial part for $-2.5<x<-2$ the separated shear layer remains two-dimensional and laminar, as demonstrated by comparing the contours of $\lambda_2$ with those of $\omega_z$, and then transitions to turbulence. The separatrix originating from the corner defines the spatial extent of a large recirculating region, a.k.a. the primary vortex; a secondary smaller counter-rotating vortex is formed around $x=-1.5$, where no significant three-dimensional structures are observed. As already described in previous works \cite{cimarelli-leonforte-angeli-2018,sasaki-kiya-1991,tenaud-etal-2016}, first at $x \approx -2 $ a Kelvin-Helmoltz like instability of the shear layer appears, leading to a breakdown into large-scale spanwise tubes. Further downstream, the spanwise tubes stretched by the mean flow roll up and originate hairpin-like vortices. A sudden transition to turbulence occurs at $x \approx -1.3$. Then, further downstream ($x \ge 0$), the hairpin-like vortices are stretched and break down to elongated streamwise vortices, identified by the contours of $\omega_x$, confirming previous findings \cite{cimarelli-leonforte-angeli-2018,sasaki-kiya-1991}. At these streamwise positions the flow is now fully turbulent, and the coexistence between small- and large-scales structures is clearly observed. The flow then proceeds towards the trailing edge, where again the sharp trailing-edge corner fixes separation in space, and a large turbulent wake ensues. 

\subsection{Temporal evolution of global and local quantities}

\begin{figure}
\centering
\includegraphics[width=0.49\textwidth]{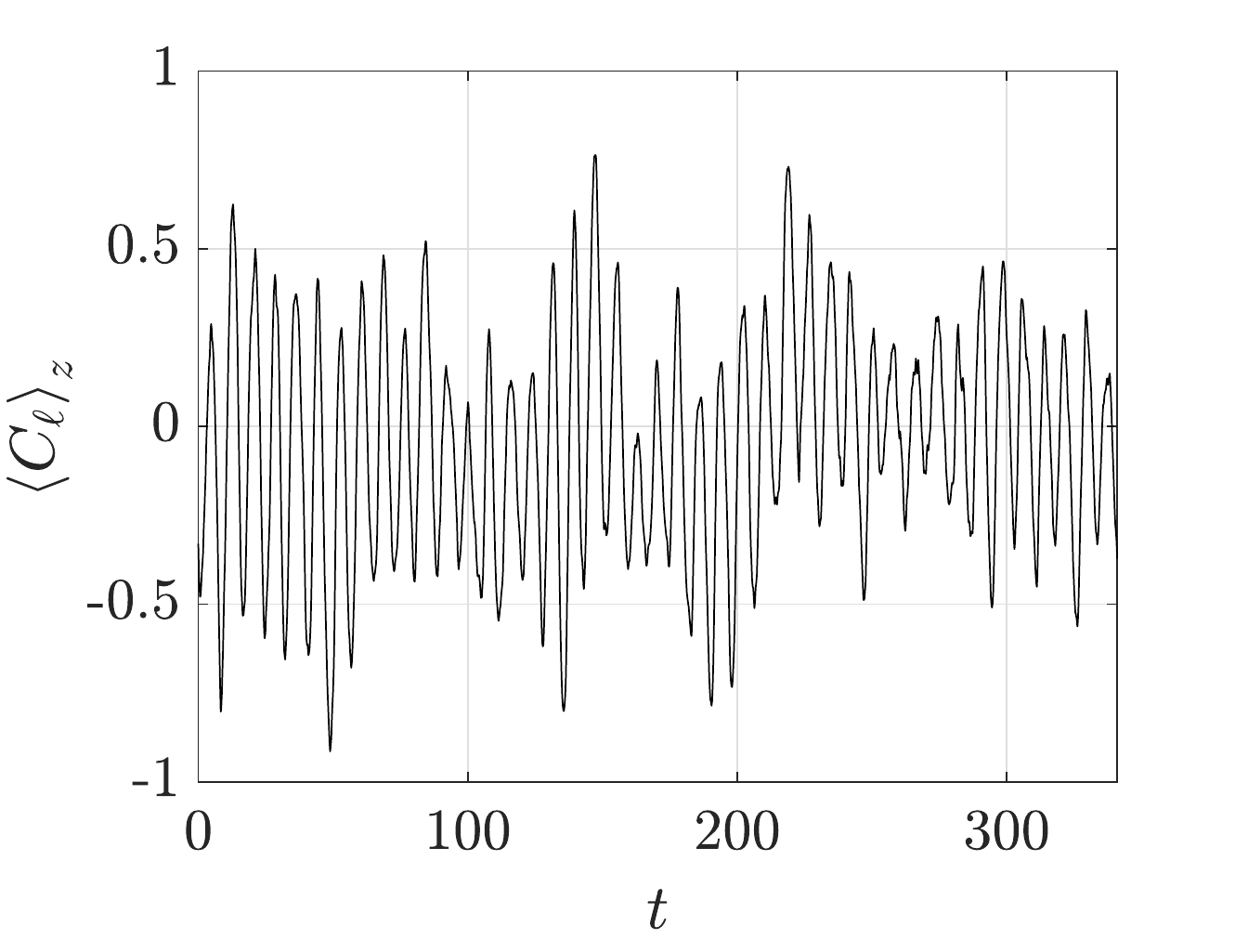}
\includegraphics[width=0.49\textwidth]{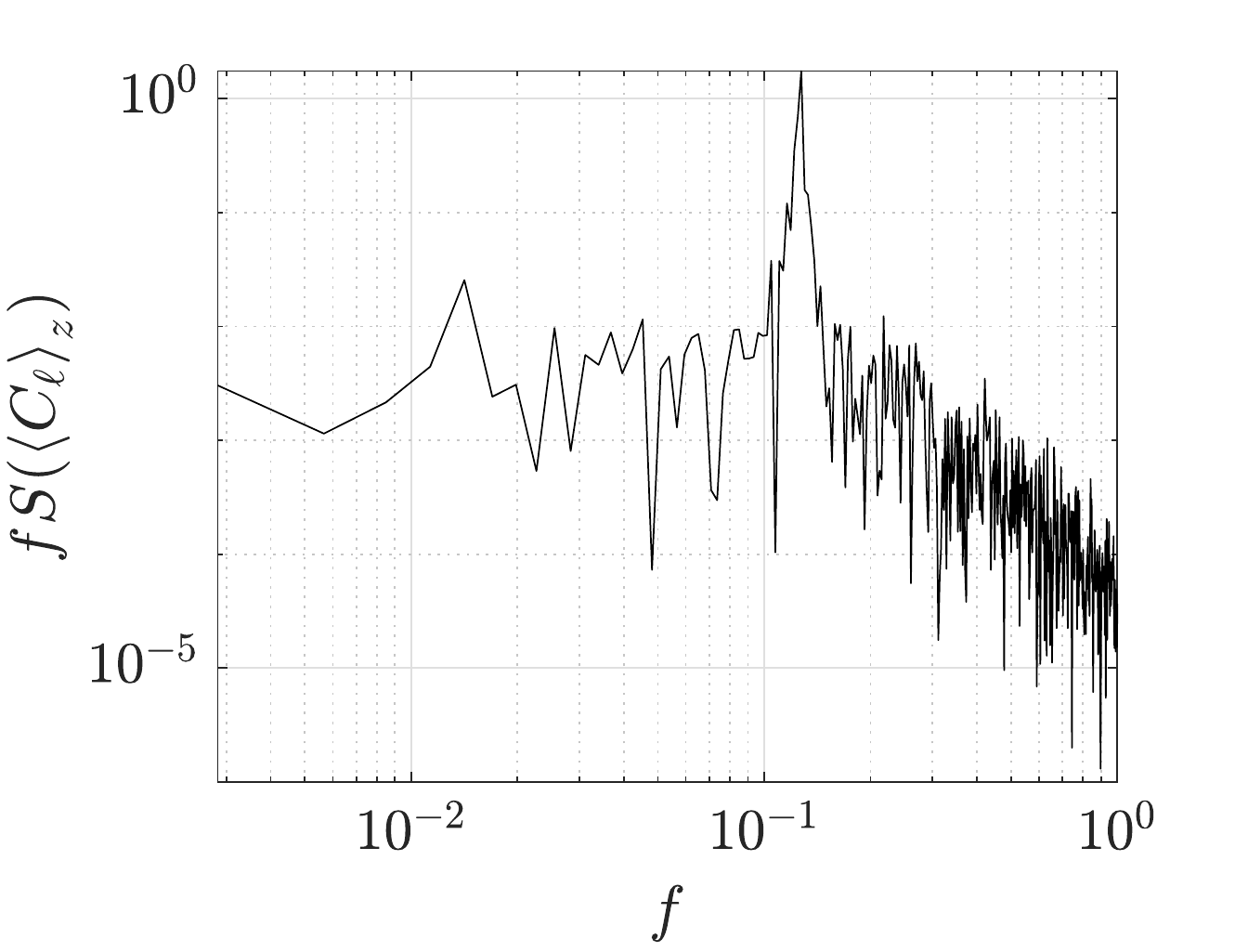}
\includegraphics[width=0.49\textwidth]{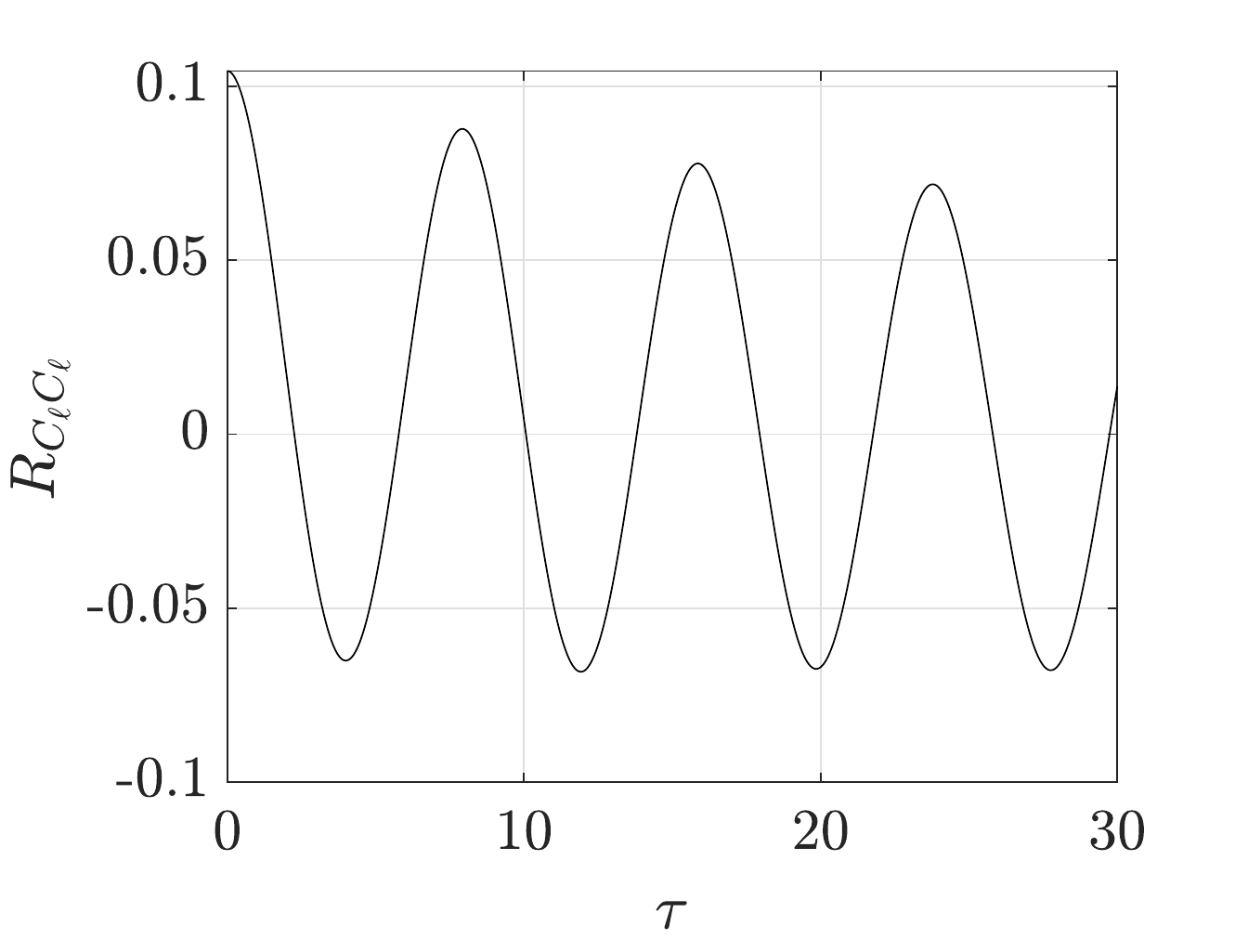}
\caption{Temporal characterization of the spanwise-averaged lift coefficient $\aver{C_\ell}_z$. Top left: time history of $\aver{C_\ell}_z$ from the primary simulation. Top right: premultiplied frequency spectrum. Bottom: temporal autocovariance.}
\label{fig:Cl-tf}
\end{figure}

A quantitative description of the flow begins with global quantities related to the mean flow field. Here and in the following, mean quantities, indicated by the operator $\aver{\cdot}$, are computed after time average and also by exploiting the homogeneity of the spanwise direction. The operators $\aver{\cdot}_z$ or $\aver{\cdot}_t$ will be used to indicate quantities averaged only along $z$, or in time respectively. For the BARC, the key quantities are the traditional dimensionless lift and drag coefficients, $C_\ell$ and $C_d$, i.e. the vertical and horizontal components of the aerodynamic force per unit width, normalised with $0.5 \rho U_\infty^2 D$. Obviously for symmetry reasons $\aver{C_\ell}=0$ for an infinitely long simulation. Ref. \cite{cimarelli-leonforte-angeli-2018b} reports $\aver{C_d}=0.96$ and does not mention the value of $\aver{C_\ell}$; our simulation provides a drag coefficient which is in agreement with theirs, at $\aver{C_d}=0.9437$, and a pretty small residual value of $\aver{C_\ell}=-0.0155$ which keeps decreasing with the integration time. Quantitative comparison of these and other quantities of interest with those of Ref. \cite{cimarelli-leonforte-angeli-2018b} are reported in Table \ref{tab:diff-time-hystory}, together with the range of values from other (experimental and numerical) studies collected and analysed in Ref. \cite{bruno-salvetti-ricciardelli-2014}. These studies are typically carried out at higher $Re$.

\begin{table}
\caption{Comparison between the present results and those from Ref.\cite{cimarelli-leonforte-angeli-2018b}. Additional columns report the data range (with their mean value in parentheses) for the experimental and numerical studies collected by Ref. \cite{bruno-salvetti-ricciardelli-2014}.}
\label{tab:diff-time-hystory}
\centering
\begin{tabular}{ccccc}
 \hline
                                   & Present     & Ref. \cite{cimarelli-leonforte-angeli-2018b} &  Other experimental data \cite{bruno-salvetti-ricciardelli-2014} & Other numerical data \cite{bruno-salvetti-ricciardelli-2014} \\
 \cline{1-5}
$f_1$   & $0.1274$  & $0.14$  & $0.105$--$0.132$ ($0.1135$) & $0.073$--$0.16$ ($0.109$) \\
$\aver{C_\ell}$ & $-0.0155$  & - & - & $-0.33$--$0.42$ ($-0.0141$) \\
$\aver{C_\ell}_{z,rms}$ & $0.2893$  &  - & $0.4$ ($0.4$) & $0.108$--$1.465$ ($0.65$) \\
$\aver{C_d}$ & $0.9437$ & $0.96$ & $1$--$1.029$ ($1.0072$) & $0.96$--$1.39$ ($1.074$) \\
 \hline
 \end{tabular}
\end{table}

Figure \ref{fig:Cl-tf} shows the temporal evolution of the spanwise-averaged $\aver{C_\ell}_z$ recorded in the primary (longest) simulation. Its excursions appear to be somewhat larger than the reference study, with the present signal reaching instantaneous values of up to $+0.762$ and $-0.913$, whereas in \cite{cimarelli-leonforte-angeli-2018b} they remain below 0.5 in absolute value. This is attributed to the much finer vertical grid spacing used in the present work, that allows to better capture the separation at the leading-edge corners; the same trend was observed also for the low-Reynolds laminar flow around a square cylinder in Ref. \cite{sohankar-norberg-davidson-1998} (see table IV in their paper), and in Ref. \cite{anzai-etal-2017} (see their figure 18). The root-mean-square value of the fluctuations for $\aver{C_\ell}_z$ is 0.2893. Reported values for this quantity, which Ref.\cite{bruno-coste-fransos-2012} finds to be significantly affected by LES spanwise discretization, are for example 0.7319 (Ref.\cite{bruno-etal-2010}, LES at $Re=4000$), 0.26 (Ref.\cite{mannini-soda-schewe-2010}, URANS at $Re=100,000$), 0.42--1.07 (Ref.\cite{mannini-soda-schewe-2011}, DES at $Re=26400$). Ref.\cite{patruno-etal-2016} mentions 0.19 for LES and 0.81 for URANS. 

The dominant time scales present in the flow are often \cite{bruno-etal-2010,mannini-soda-schewe-2011,ricci-etal-2017,mannini-etal-2017} extracted by looking at localised peaks in the frequency spectrum $S$. The premultiplied periodogram of $\aver{C_\ell}_z$, shown in the top right panel and computed with data from the primary simulation only, shows a peak for the frequency $f_1=0.1274$. It is known from previous work \cite{okajima-1982,nakamura-ohya-tsuruta-1991,ozono-etal-1992,mills-etal-1995,cimarelli-leonforte-angeli-2018b} that this peak is associated to the vortex shedding in the wake. A more precise view of the dominant frequency can be obtained from the temporal autocovariance  $R_{C_\ell C_\ell}(\tau)$ of the signal, shown in the bottom panel of figure \ref{fig:Cl-tf}. The first peak after the maximum at zero time separation is located at a time separation of $\tau = 1/f_1$. Overall, the identified frequency is quite similar to the time scale identified in \cite{cimarelli-leonforte-angeli-2018b}, who measured $f_1 \approx 0.14$. A frequency analysis of $\aver{C_d}_z$ (not shown) confirms the presence of a localised peak at a frequency of $2 f_1$ in the spectrum of the drag coefficient, induced by the alternate vortex shedding. Other studies, with different numerical approaches and at different $Re$, report for $f_1$ values of 0.098 or 0.105 \cite{mannini-soda-schewe-2010}, 0.1 \cite{mannini-soda-schewe-2011}, 0.11 \cite{bruno-etal-2010}, 0.132 \cite{patruno-etal-2016}. Table \ref{tab:diff-time-hystory} shows that often experiments and simulations tend to underestimate the shedding frequency. This is attributed to the sensitivity of the BARC flow to slight deviations from the ideal setting, or to simulation inaccuracies; the finite $Re$ should not affect the computed value, because according to Ref. \cite{nakamura-ohya-tsuruta-1991,schewe-2013,moore-etal-2019} at $Re \ge 3000$ the Strouhal number has already reached its asymptotic value.

Several authors found that this flow is also characterised by a large-scale low-frequency unsteadiness associated to a shrinkage and enlargement of the main recirculating region. Kiya and Sasaki in \cite{kiya-sasaki-1983} and \cite{kiya-sasaki-1985} report for this frequency a value of approximately $1/6$ of the shedding frequency for a blunt flat plate at $Re \approx 10^4$. Similar results are also reported by \cite{cherry-1984} for the same flow and by \cite{eaton-johnson-1982} for a backward facing step. Ref. \cite{cimarelli-leonforte-angeli-2018b}, instead, report $f \approx 0.042$, that is approximately $1/3$ of their shedding frequency. This low frequency is not detected clearly in the present data. However, the low Reynolds number and the amount of integration time preclude any firm statement. If such low frequency indeed exists at the present $Re$, it would require an extremely long integration to be reliably detected, as shown in \cite{lehmkuhl-etal-2013} for the low-$Re$ turbulent flow past a circular cylinder.

\begin{figure}
  \centering
  \includegraphics[width=0.49\textwidth]{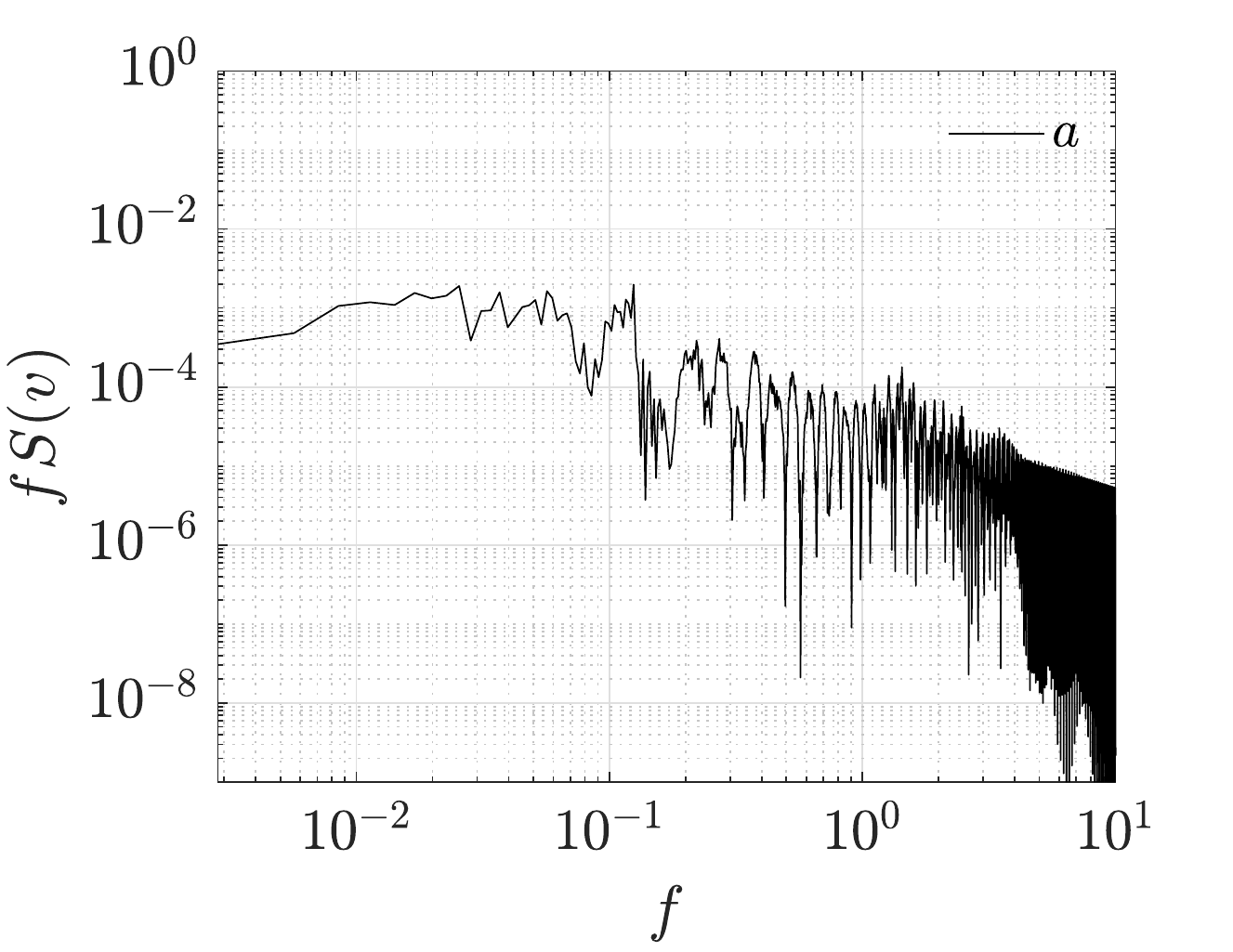}
  \includegraphics[width=0.49\textwidth]{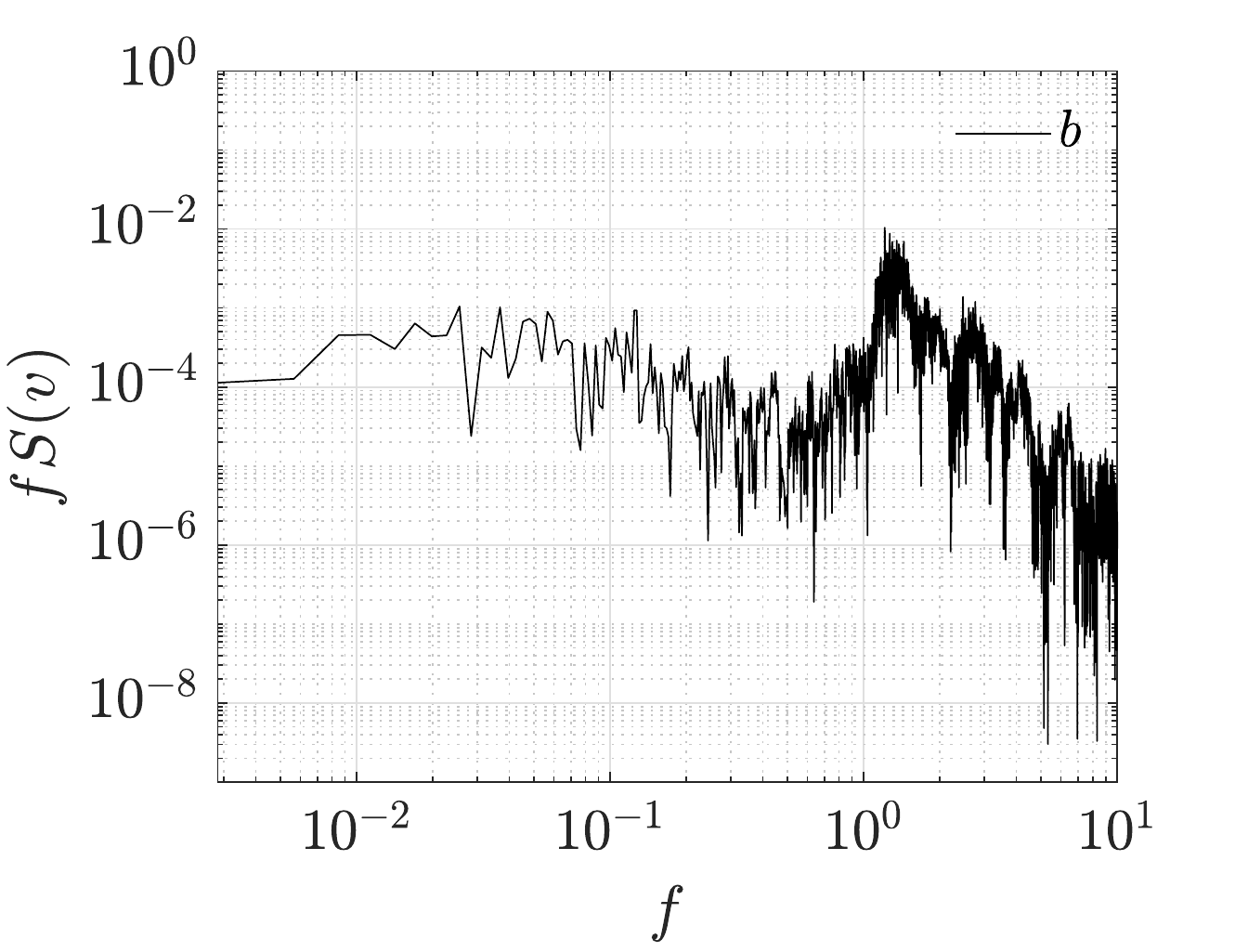}
  \includegraphics[width=0.49\textwidth]{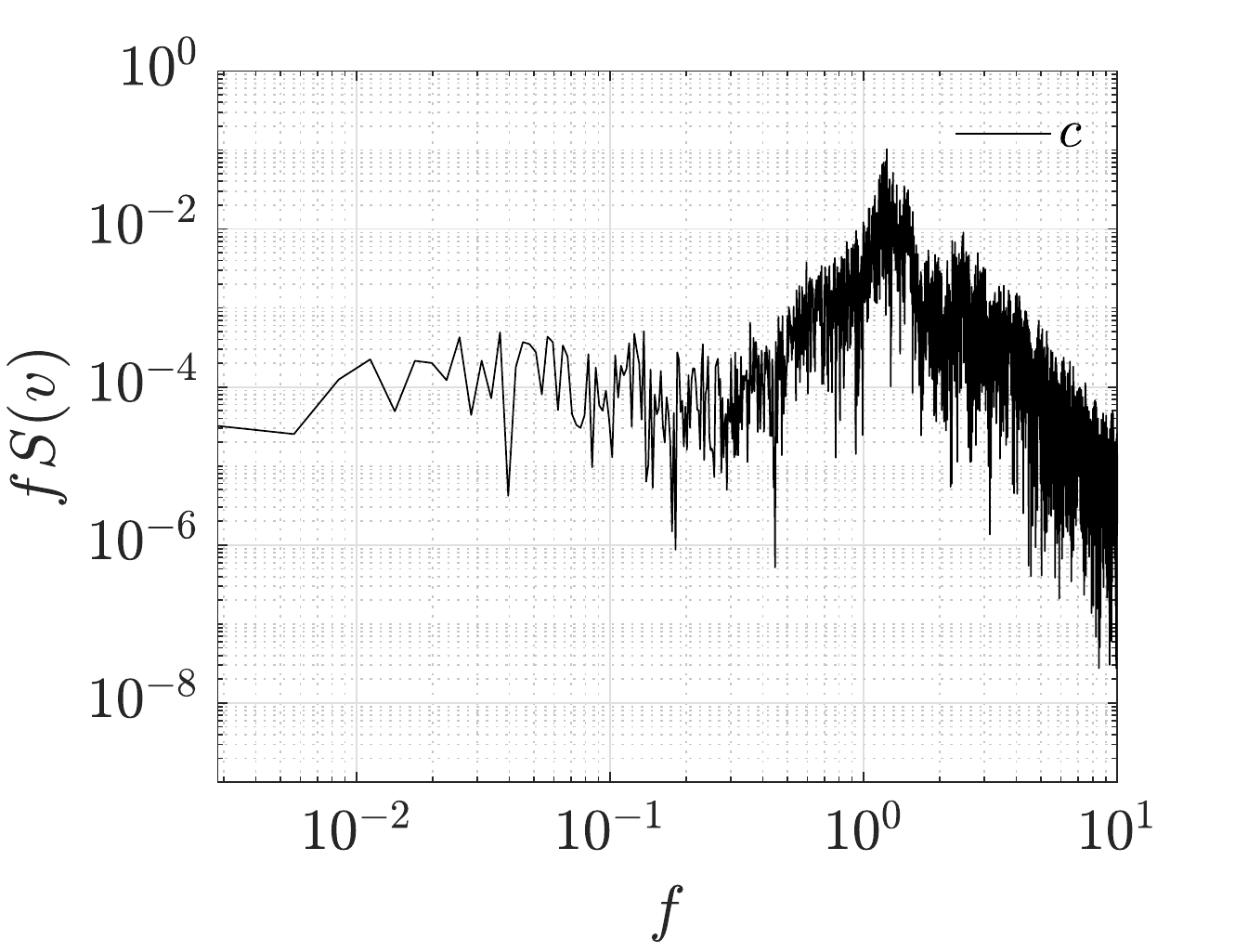}
  \includegraphics[width=0.49\textwidth]{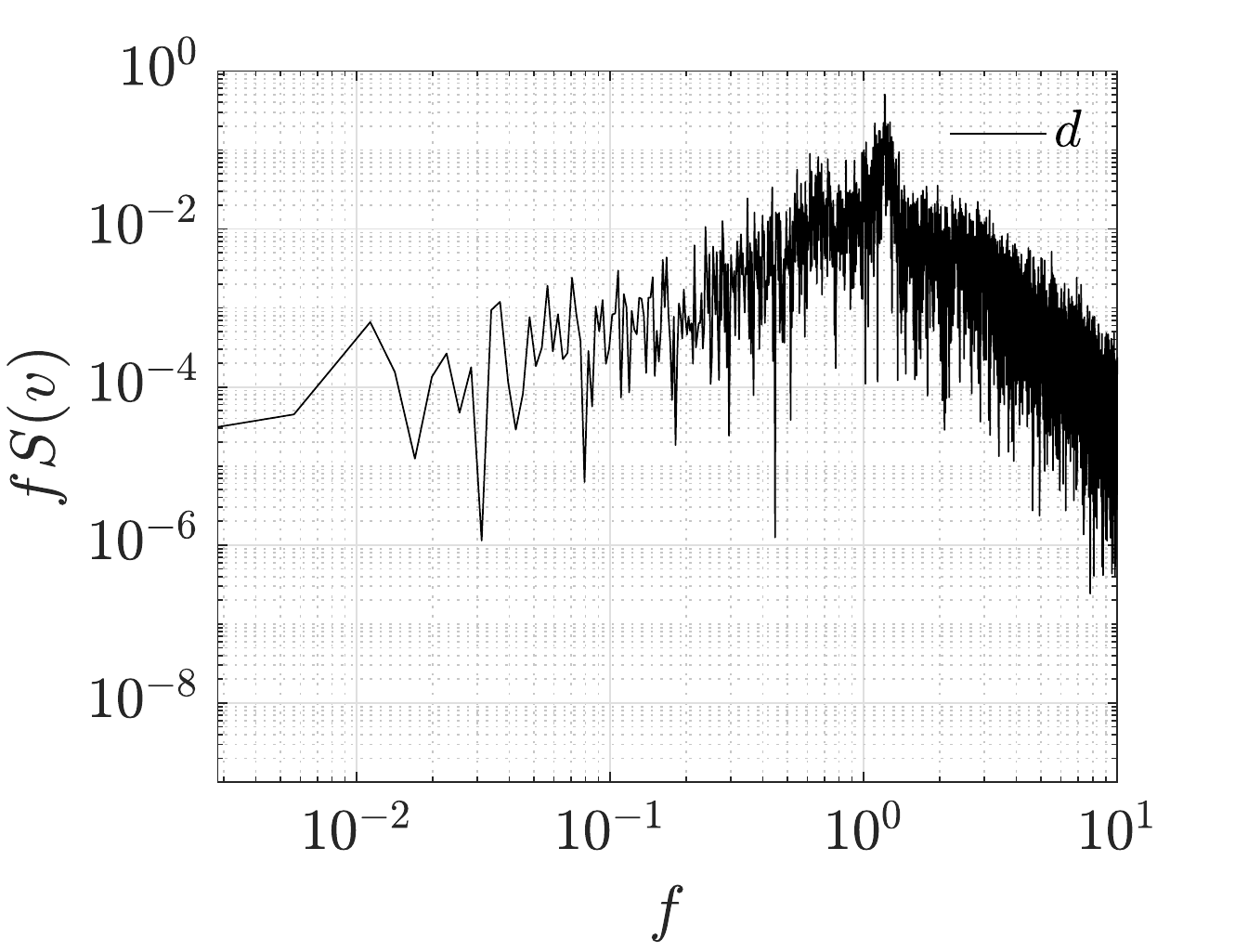}
  \includegraphics[width=0.49\textwidth]{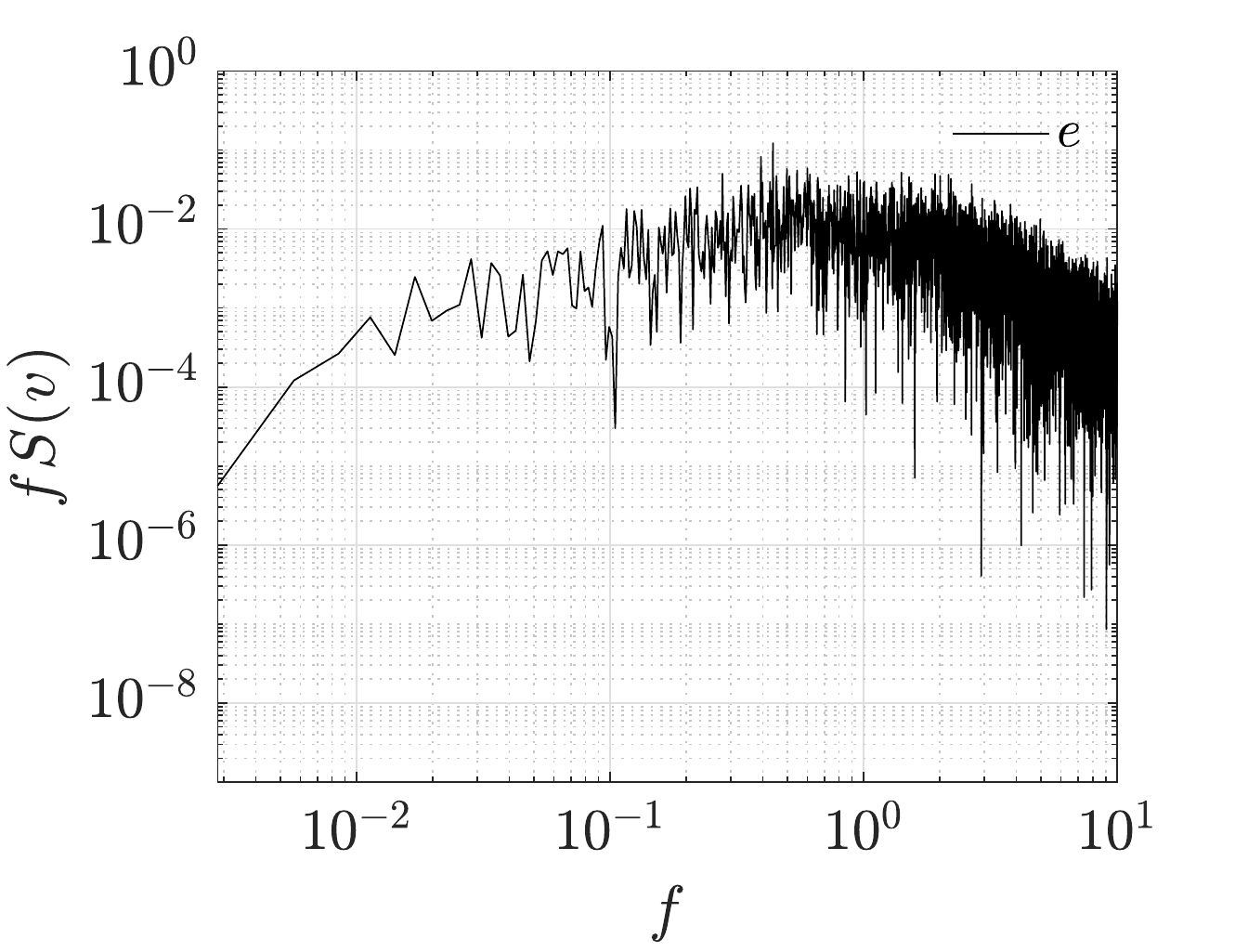}
  \includegraphics[width=0.49\textwidth]{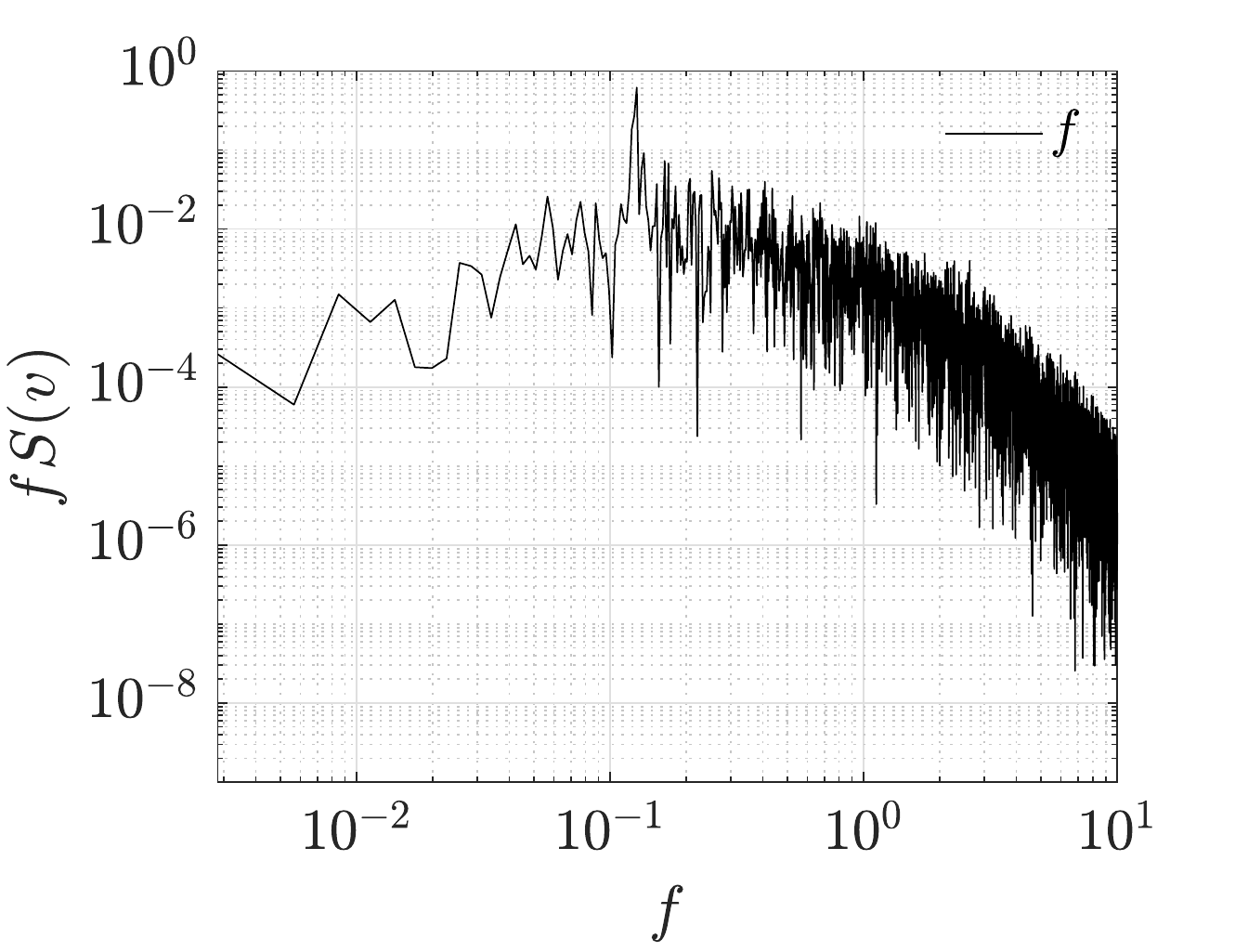}
  \caption{Premultiplied frequency spectra of the vertical velocity component $v$ evaluated in the six points highlighted in figure \ref{fig:lambda2-vorticity}.}
  \label{fig:v-StD}
\end{figure}

The complex BARC flow contains different time scales which vary across the flow domain. Mimicking a typical experiment, where a fixed-point sensor would be placed in different positions, numerical probes are placed at six points in the midplane $z=0$ to record the time history of velocity and pressure. These points are drawn in figure \ref{fig:meanVel}, and have the following coordinates: $(x_a,y_a)=(-2.05,0.805)$, $(x_b,y_b)=(-1.716,0.9)$, $(x_c,y_c)=(-1.437,0.94)$, $(x_d,y_d)=(-0.98 0.98)$, $(x_e,y_e)=(1.5,0.98)$ and $(x_f,y_f)=(3.3,0.2)$. The first four points lie on the main shear layer; point $e$ is outside the boundary layer at a position where the flow has already reattached; and point $f$ is in the wake, near the trailing edge of the recirculating region. As an example, we discuss below and show in figure \ref{fig:v-StD} the frequency spectrum of the cross-stream velocity component $v$, although an equivalent picture is obtained when the streamwise velocity component is examined (not shown).
The power spectrum at point $a$ near the leading edge lacks a clearly dominating frequency, since the flow is essentially laminar. However, a small local peak is observed at the frequency $f_1$ that also emerges in the spectrum of $\aver{C_\ell}_z$, associated with the vortex shedding at the trailing edge. This suggests that at this Reynolds number the wake still influences the leading-edge region, as it happens at much lower $Re$ \cite{hourigan-thompson-tan-2001}. The vortex-shedding frequency has been detected at these upstream sections also in Ref. \cite{moore-etal-2019} for $Re=13400$, but not by Ref. \cite{rocchio-etal-2020} at $Re=40000$; this suggests that the influence of the vortex shedding from the trailing edge on the first part of the shear layer gradually fades away as $Re$ increases. Moving along the shear layer, a higher dominant frequency $f \approx 1.3$ emerges, associated to the amplification of the velocity fluctuations in the shear layer \cite{cimarelli-leonforte-angeli-2018b} owing to the Kelvin-Helmotz instability \cite{moore-etal-2019}. The peak in the spectrum moves towards lower frequencies along the shear layer ($f \approx 1.44,1.23,1.21$ for $b$, $c$, $d$ respectively) and at the same time broadens. This frequency range is not entirely in agreement with Ref. \cite{cimarelli-leonforte-angeli-2018b}, where the range is reportedly $f \approx 0.9-1.8$. Similar results have been found also at larger Reynolds numbers; Ref. \cite{rocchio-etal-2020} report frequencies in the range of $f \approx 0.75-1.52 $ for $x<-0.5$ and $ f \approx 0.375$ for $ x \approx 0$. Point $e$ is downstream the reattachment point, and the flow there is fully turbulent. The slow time scales associated to the unsteadiness of the flow coexist with faster turbulent scales, resulting in broad spectra without evident dominant peaks. Point $f$ is placed in the wake, and experiences all the time scales; the local vortex shedding of large structures at $f=f_1$ becomes visible again.

\subsection{The mean flow}

\begin{figure}
\centering
\includegraphics[trim=290 100 290 80,clip,width=0.49\textwidth]{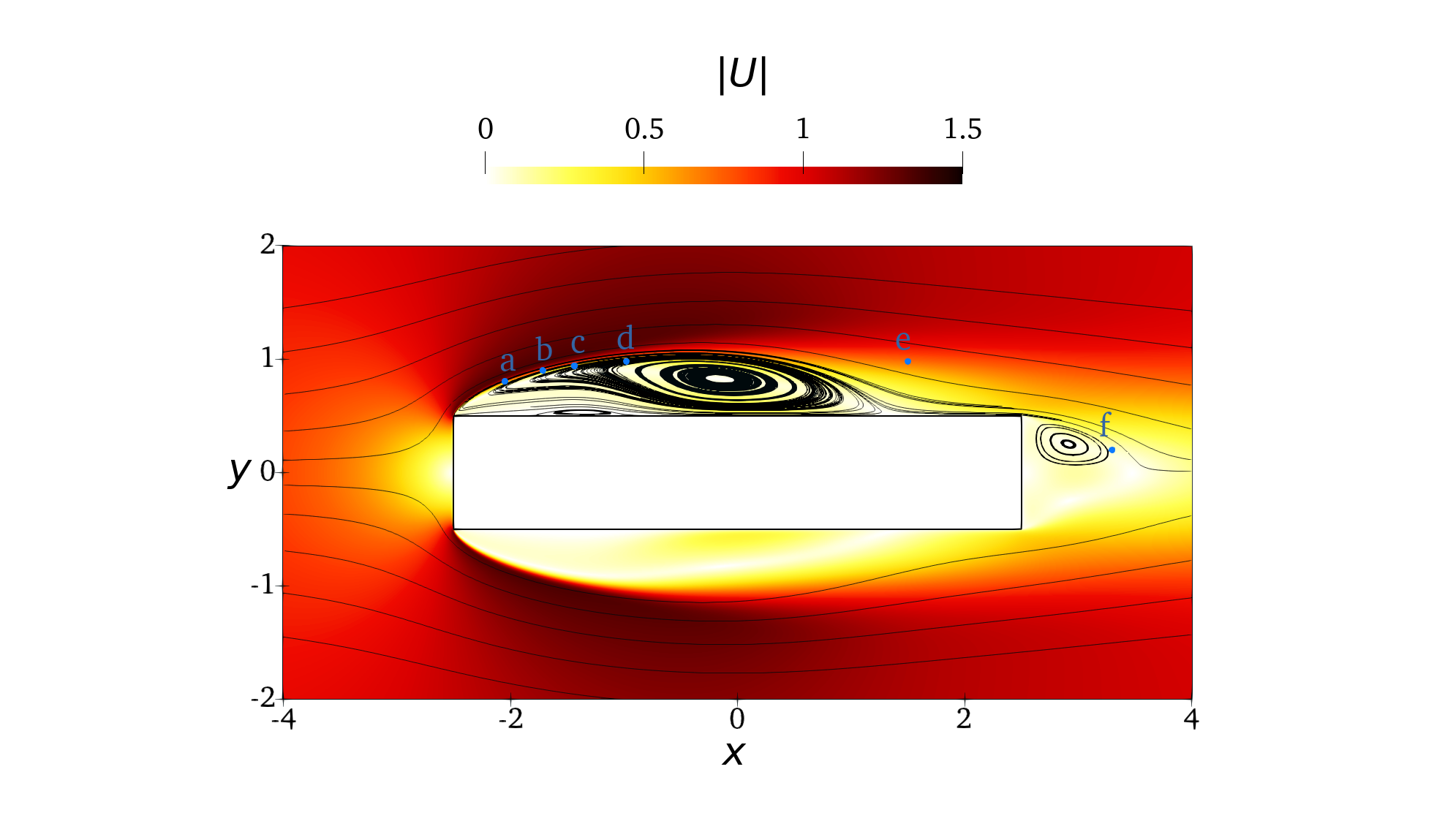}
\includegraphics[trim=290 100 290 80,clip,width=0.49\textwidth]{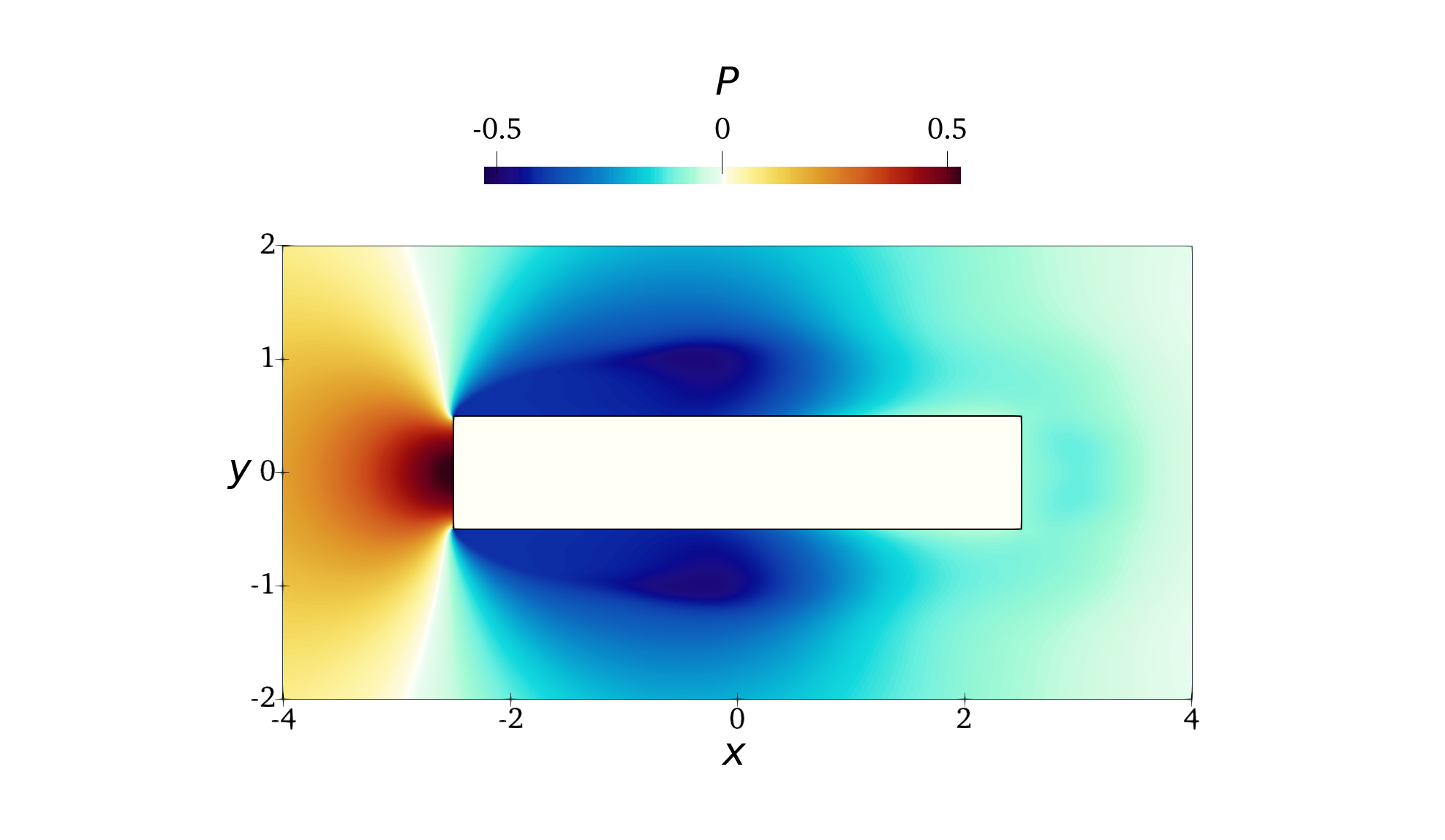}
\caption{Left: mean velocity field, with mean streamlines drawn over a colormap of the magnitude of the mean velocity. Right: colormap of the mean pressure field. Points labelled as $a,b,c,d,e,f$ are positioned in interesting positions along the flow which are discussed in the text.}
\label{fig:meanVel}
\end{figure}

Figure \ref{fig:meanVel} shows the fields of the mean velocity (left) and pressure (right), obtained after spanwise and temporal average of the entire dataset (2345 time units). The mean flow separates at the sharp leading edge and reattaches downstream, before eventually separating again at the trailing-edge corner. Three recirculating regions are identified. The first is the large region identified by the shear layer separating at the leading edge. It starts from $x_{s,1}=-2.5$ and extends down to the reattachment point located at $x_{e,1}=1.455$, with a length of $L_1 \equiv x_{s,1}-x_{e,1}=3.955$. The centre of rotation of the primary vortex, defined as the stagnation point with $U=V=0$ is found at $(x_{c,1},y_{c,1})=(-0.143,0.83)$. In Ref.\cite{cimarelli-leonforte-angeli-2018b} it is noted that this region is associated with large negative values of pressure. Indeed, a pressure minimum is found close to the bubble centre, at $(x_{p,1},y_{p,1})=(-0.3,0.99)$. Quantitative comparison of these and other quantities as measured here with those of the reference study \cite{cimarelli-leonforte-angeli-2018b}, where $L_1=3.65$, are reported in Table \ref{tab:diff-meanflow}. The larger size of the primary vortex may descend from the finer grid used here \cite{mariotti-siconolfi-salvetti-2017}. In terms of $L_1$, literature values are 4.68 \cite{bruno-etal-2010}, 4.65 \cite{mannini-soda-schewe-2010}, 4.75 \cite{mannini-soda-schewe-2011}, 4.01 or 4.26 \cite{patruno-etal-2016}, 0.81 \cite{mannini-soda-schewe-2011}. 

Within the large primary bubble, a second smaller counter-rotating recirculation bubble is observed. This secondary vortex is associated with the detachment of the reverse boundary layer caused by the adverse pressure gradient. Its characteristic length-scale is smaller: the secondary bubble extends between $x_{s,2}=-1.87$ and $x_{e,2}=-0.91$, with a shorter length of $L_2 \approx 0.96$, with the centre of rotation placed at $(x_{c,2},y_{c,2})=(-1.3,0.541)$. In Ref.\cite{cimarelli-leonforte-angeli-2018b} the length $L_2$ of this structure is $L_2=1$. Other literature values are 1.88 \cite{bruno-etal-2010}, 0.31 \cite{mannini-soda-schewe-2010}, 0.75 \cite{cimarelli-franciolini-crivellini-2020}. 

The third recirculating region is observed in the wake region, just after the trailing edge. This wake vortex extends from $x_{s,3}=2.5$ up to $x_{e,3}=3.475$, corresponding to a length of $L_3 \approx 0.975$, and its centre of rotation is placed at $(x_{c,3},y_{c,3})=(2.915,0.25)$. In Ref.\cite{cimarelli-leonforte-angeli-2018b} the length $L_3$ of this structure is $L_3=1.2$. Other literature values are 0.76 \cite{bruno-etal-2010}, 1.4 or 0.7 \cite{mannini-soda-schewe-2010}, 0.94 \cite{mannini-soda-schewe-2011}, 0.9 or 0.81 \cite{patruno-etal-2016}, 1 \cite{cimarelli-franciolini-crivellini-2020}.

\begin{table}
\caption{Characterization of the three recirculation regions: comparison between the present simulations and Ref. \cite{cimarelli-leonforte-angeli-2018b}.}
\label{tab:diff-meanflow}
\centering
\begin{tabular}{ccccc}
 \hline
                                   &             & Present         & & Ref.\cite{cimarelli-leonforte-angeli-2018b} \\
 \cline{1-5}                                  
 \multirow{4}{*}{Primary vortex}   & $x_s$        & $-2.5$          & & $-2.5$ \\
                                   & $x_e$        & $1.455$         & & $1.15$ \\
                                   & $L$          & $3.955$         & & $3.65$ \\
                                   & $(x_c,y_c)$ & $(-0.143,0.83)$  & & $(-0.46,0.85)$ \\
 \cline{1-5}                                  
 \multirow{4}{*}{Secondary vortex} & $x_s$       & $-1.87$         & & $-2.1$ \\
                                   & $x_e$       & $-0.91$         & & $-1.1$ \\
                                   & $L  $       & $0.96$           & & $ 1 $  \\
                                   & $(x_c,y_c)$ & $(-1.3,0.541)$  & & $-$    \\
 \cline{1-5}                                  
 \multirow{4}{*}{Wake vortex}      & $x_s$       & $2.5$           & & $2.5$  \\ 
                                   & $x_e$       & $3.475$          & & $3.7$  \\
                                   & $L$         & $0.975$          & & $1.2$  \\
                                   & $(x_c,y_c)$ & $(2.915,0.25)$   & & $(3,0.23$) \\
  \hline
\end{tabular}
\end{table}

\begin{figure}
\centering
\includegraphics[width=0.24\textwidth]{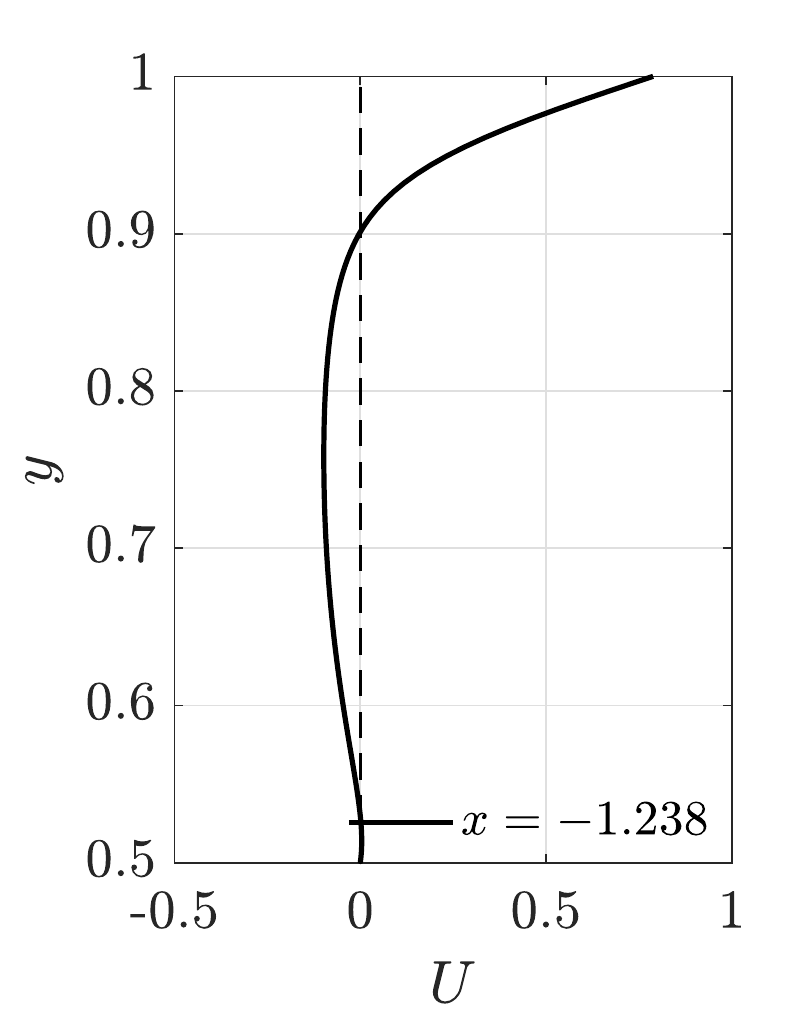}
\includegraphics[width=0.24\textwidth]{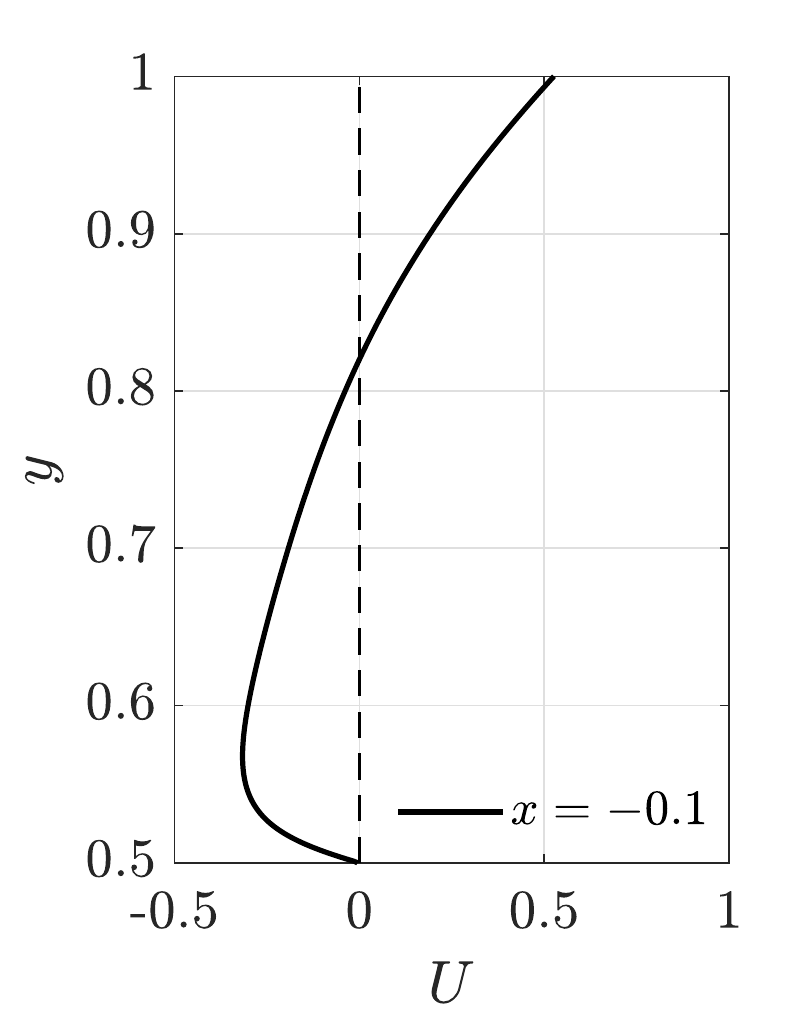}
\includegraphics[width=0.24\textwidth]{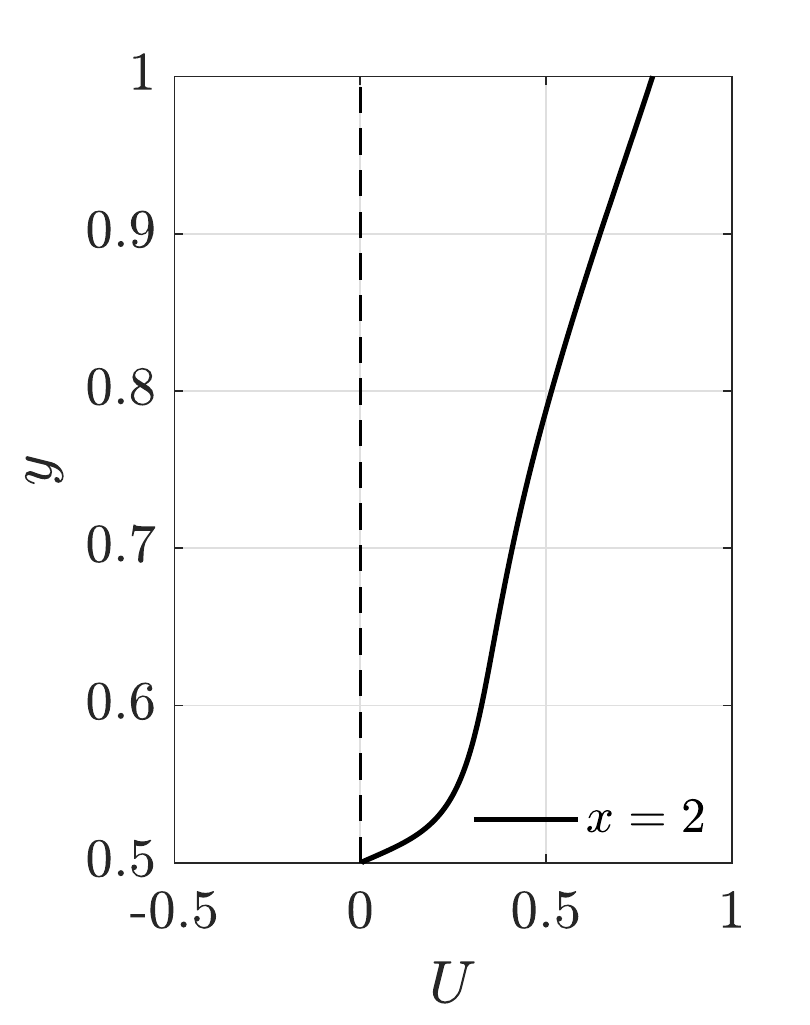}
\includegraphics[width=0.24\textwidth]{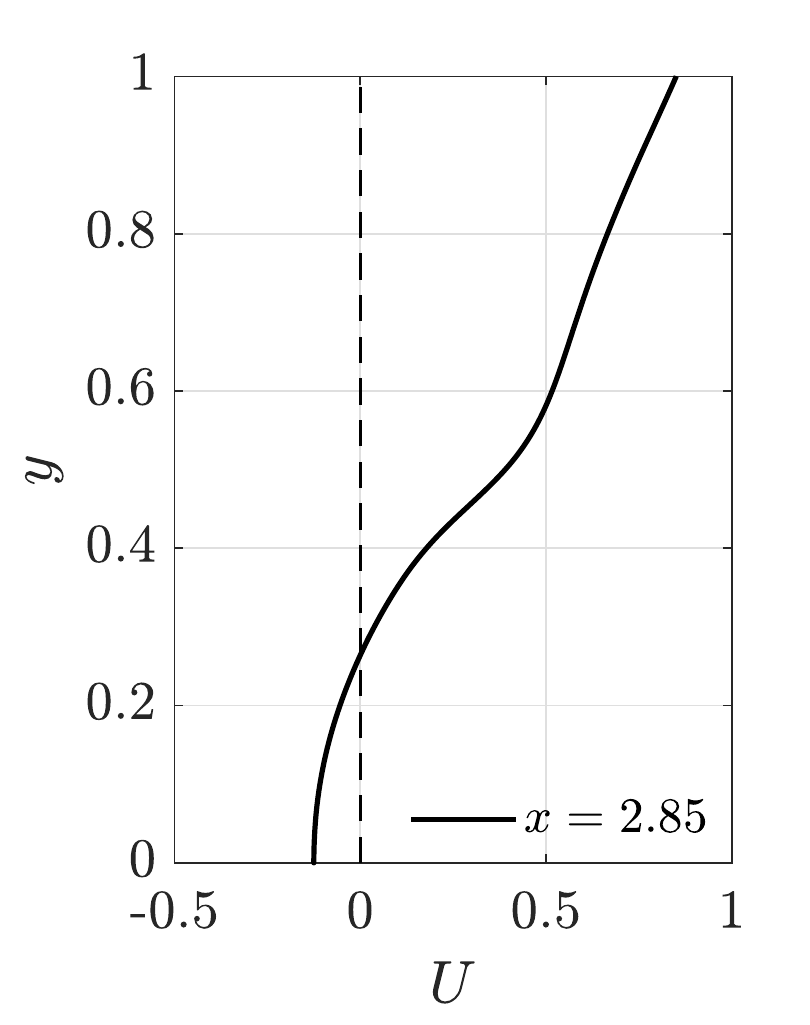}
\includegraphics[width=0.24\textwidth]{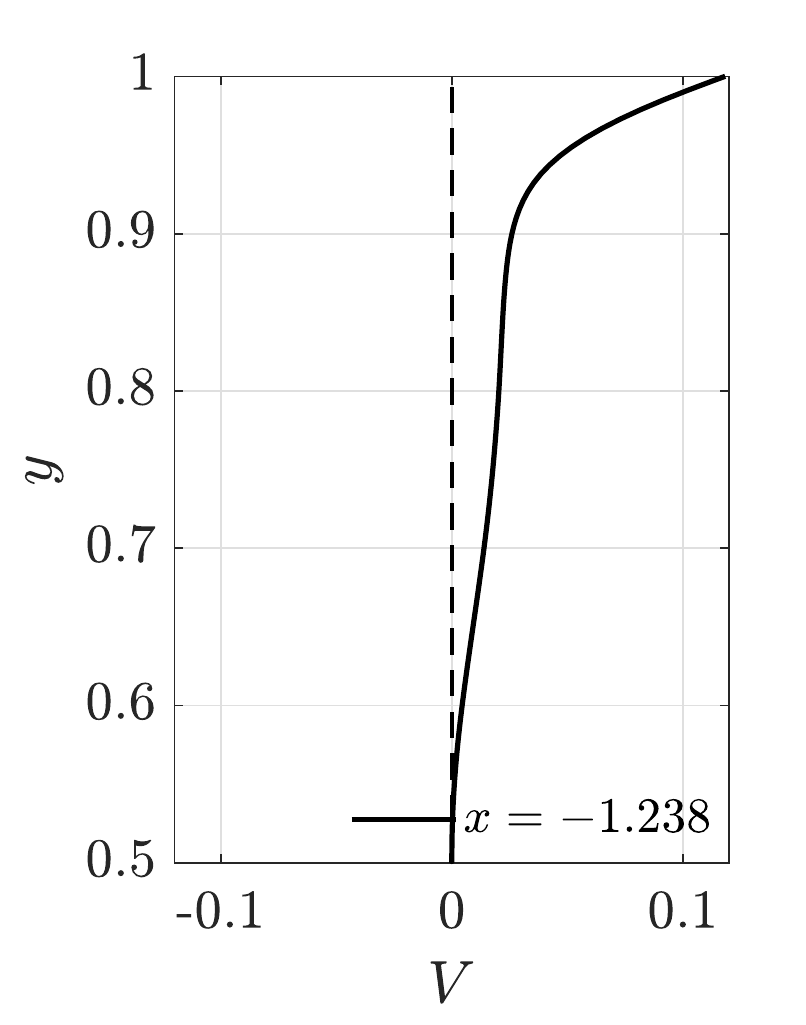}
\includegraphics[width=0.24\textwidth]{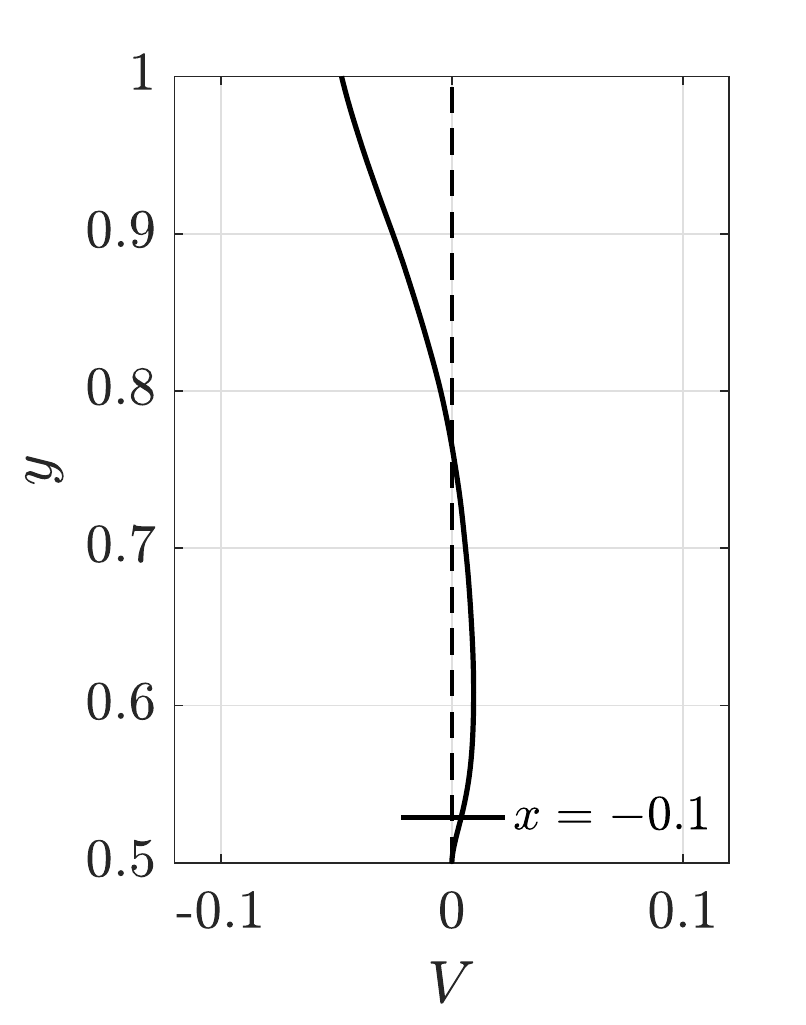}
\includegraphics[width=0.24\textwidth]{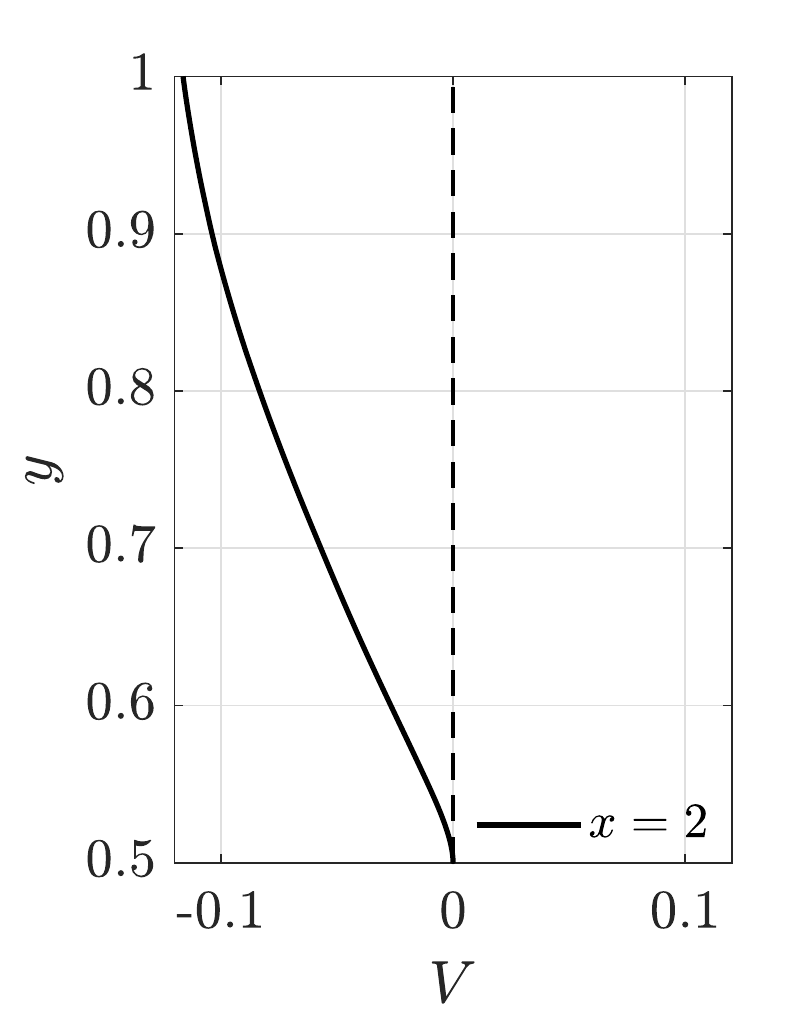}
\includegraphics[width=0.24\textwidth]{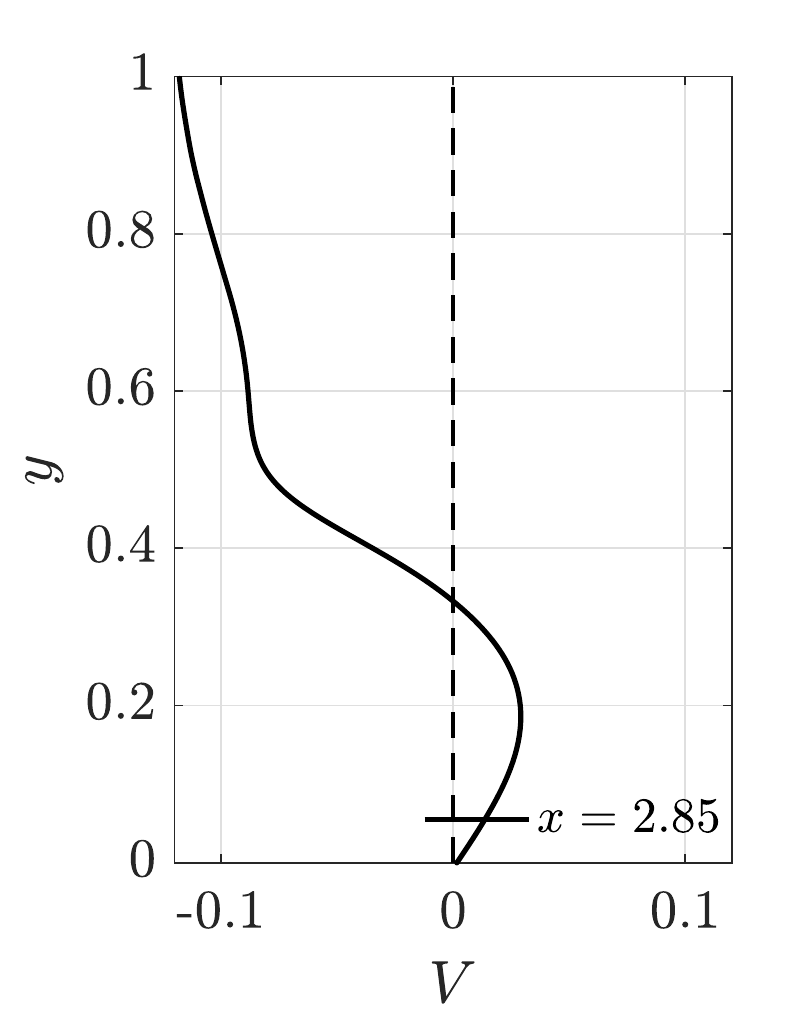}
\caption{Profiles of the longitudinal (top) and cross-stream (bottom) mean velocity components $U$ and $V$ at four $x$ coordinates along the body and in the near wake.}
\label{fig:velocity-profiles}
\end{figure}
Figure \ref{fig:velocity-profiles} plots the two components of the mean velocity vector at four streamwise stations, i.e. at $x=-1.238$ (corresponding to the small recirculation region), $x=-0.1$ (corresponding to the main recirculating region), $x=2$ (after the reattachment of the primary bubble) and $x=2.85$ (in correspondence of the recirculation region in the wake). These velocity profiles are presented to provide a reference for experimental measurements, where hot-wire traverses could be employed, as for example in Refs. \cite{mannini-etal-2017} and \cite{ricci-etal-2017}.

\begin{figure}
\centering
\includegraphics[width=0.49\textwidth]{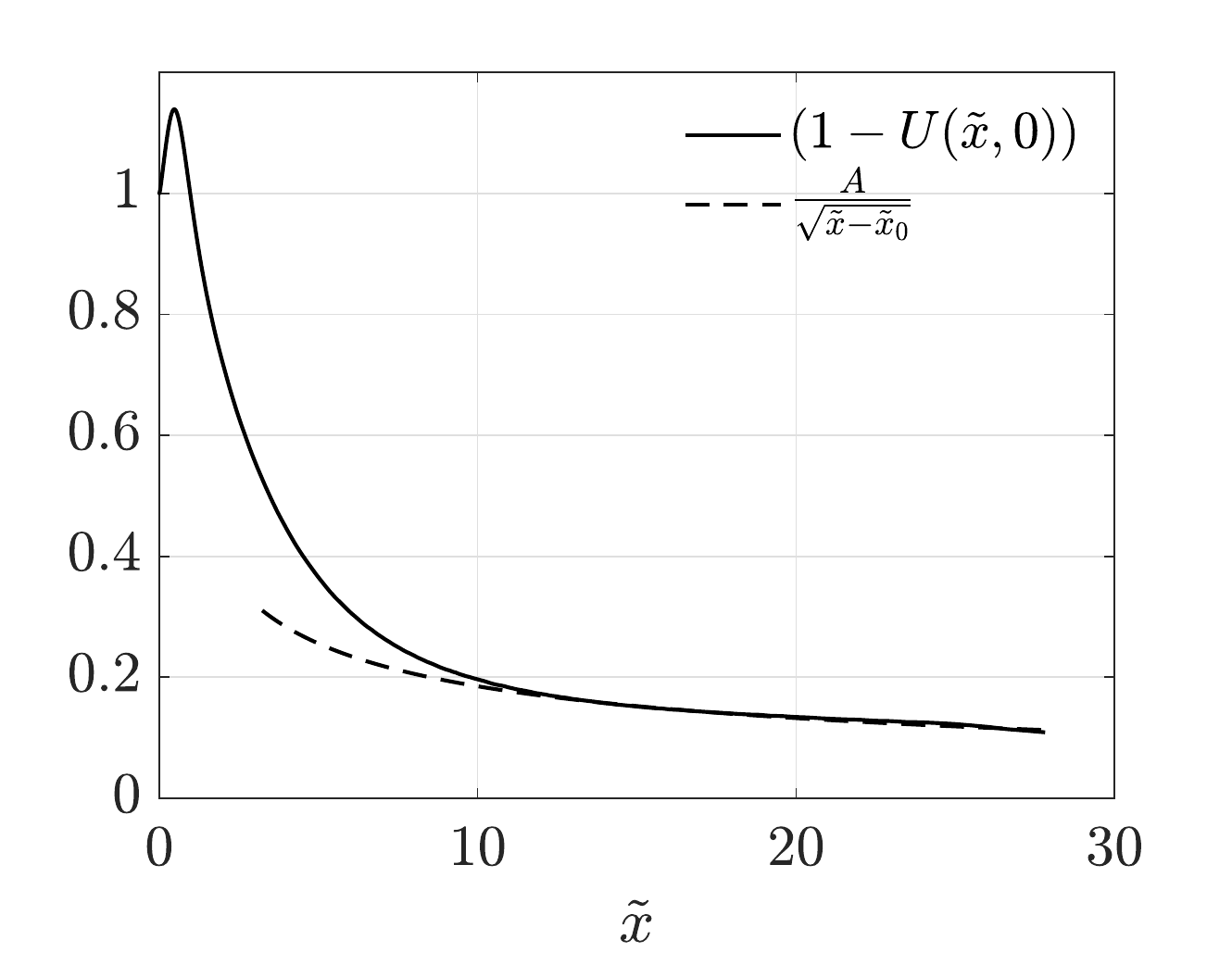}
\includegraphics[width=0.49\textwidth]{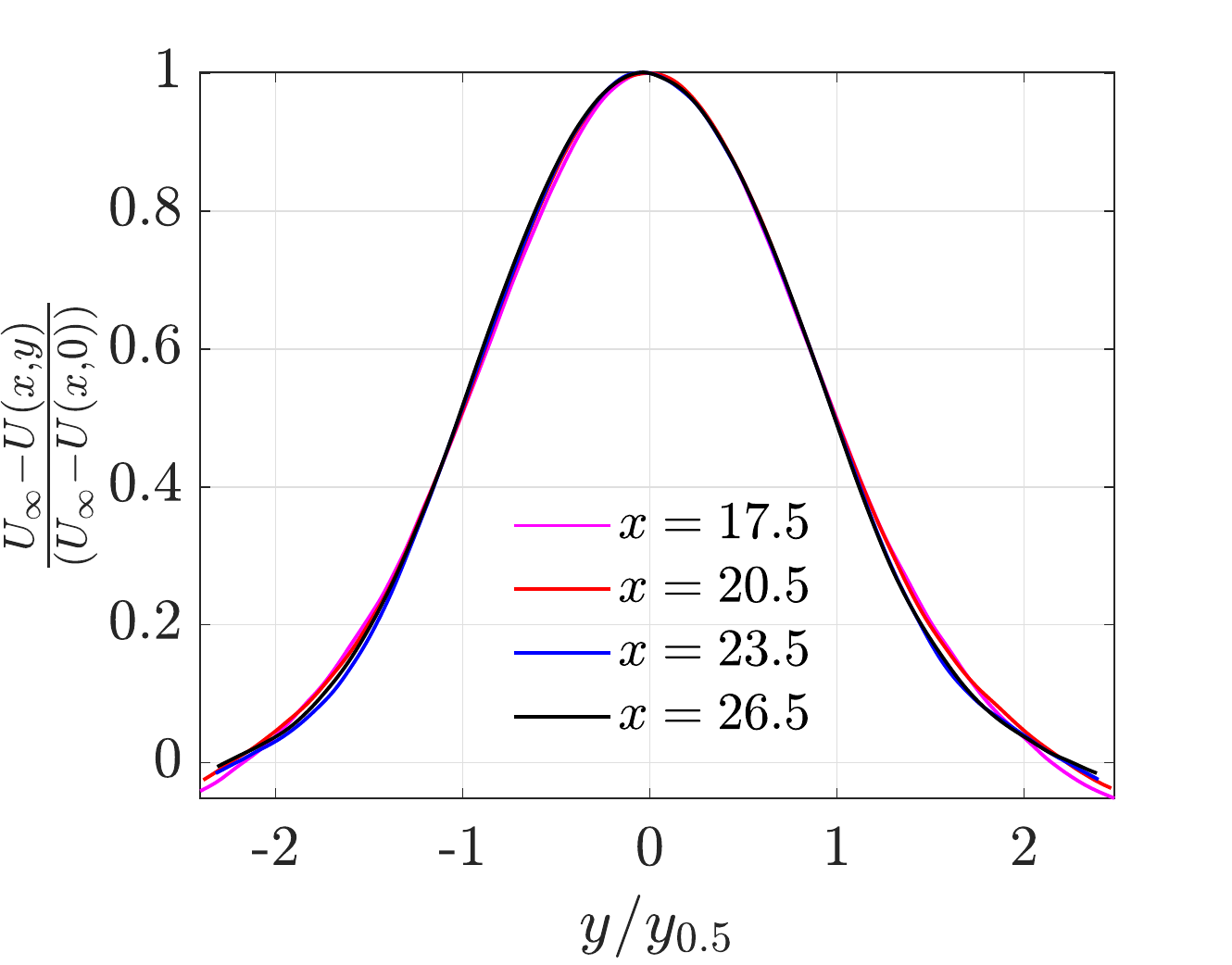}
\caption{The mean velocity in the wake. Left: velocity defect $1-U(\tilde{x},0)$ for the mean velocity at the centerline, with $\tilde{x}=x-2.5$. The self-similar power law $\frac{A}{\sqrt{\tilde{x}-\tilde{x}_0}}$ is also plotted with $A=0.6$ and $\tilde{x}=-0.5$. Right: cross-stream profiles of the normalized velocity defect $\frac{U_\infty - U(x,y)}{U_\infty - U(x,0)}$ as a function of the rescaled cross-stream coordinate $y/y_{0.5}$.}
\label{fig:wake-mean}
\end{figure}

The evolution of the turbulent wake is described in figure \ref{fig:wake-mean}, which assesses to what degree it obeys self-similarity. The left panel plots $1-U(\tilde{x},0)$ where the abscissa $\tilde{x}=x-2.5$ measures the distance from the trailing edge. The centerline velocity defect follows a self-similar decay for $\tilde{x} > 10$, decaying proportionally to $x^{-1/2}$, and follows the curve
\begin{equation}
1-U(\tilde{x},0)=\frac{A}{\sqrt{\tilde{x}-\tilde{x}_0}},
\end{equation}
where the two free parameters are fitted to $A=0.6$ and $\tilde{x}_0=-0.5$ (in Ref. \cite{cimarelli-leonforte-angeli-2018} the same fit yielded $A=0.66$ and $\tilde{x}_0=4$). The right panel shows profiles of the normalised defect, with the cross-stream coordinate  $y_{0.5}$ scaled with the characteristic wake thickness. The latter is defined such that $U(x,y_{0.5})=0.5(1+U(x,0))$. The profile
\begin{equation}
\frac{U_\infty - U(x,y)}{U_\infty - U(x,0)}
\end{equation}
is plotted at different streamwise stations downstream of $x=10$. The profiles collapse reasonably well in the core of the wake, lending further support to the self-similarity of the wake after some distance from the body. The implied power-law spreading of $y_{0.5} \sim x^{1/2}$ is that of the plane turbulent wake. It should be noted, however, that the collapse is only marginal for $|y/y_{0.5}|>1$, indicating that at this distance from the body the self-similarity may not be complete. Moreover, the finite downstream domain length and the grid resolution in the far region might also be not enough to precisely capture the evolution of the turbulent wake.

\begin{figure}
\centering
\includegraphics[width=0.49\textwidth]{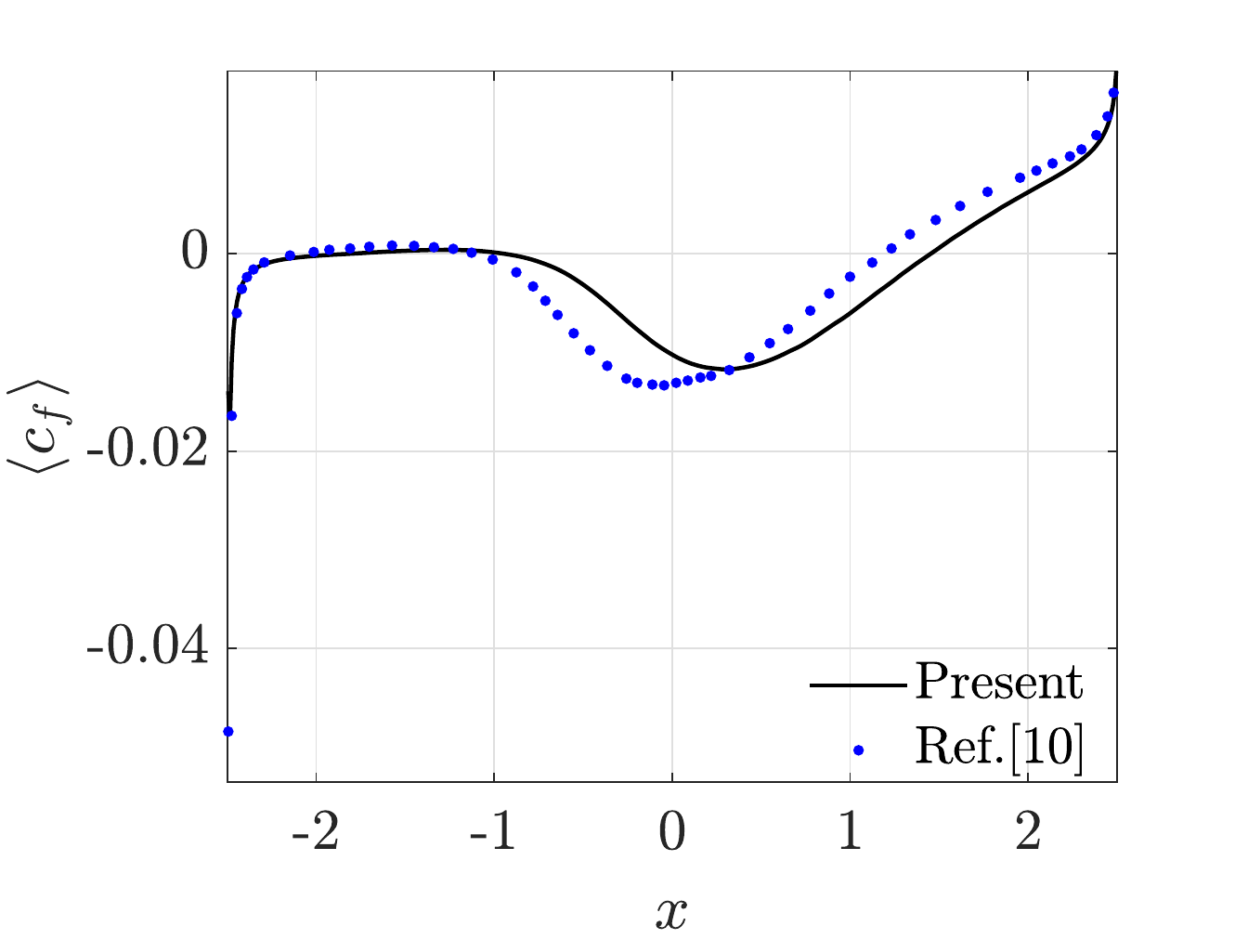}
\includegraphics[width=0.49\textwidth]{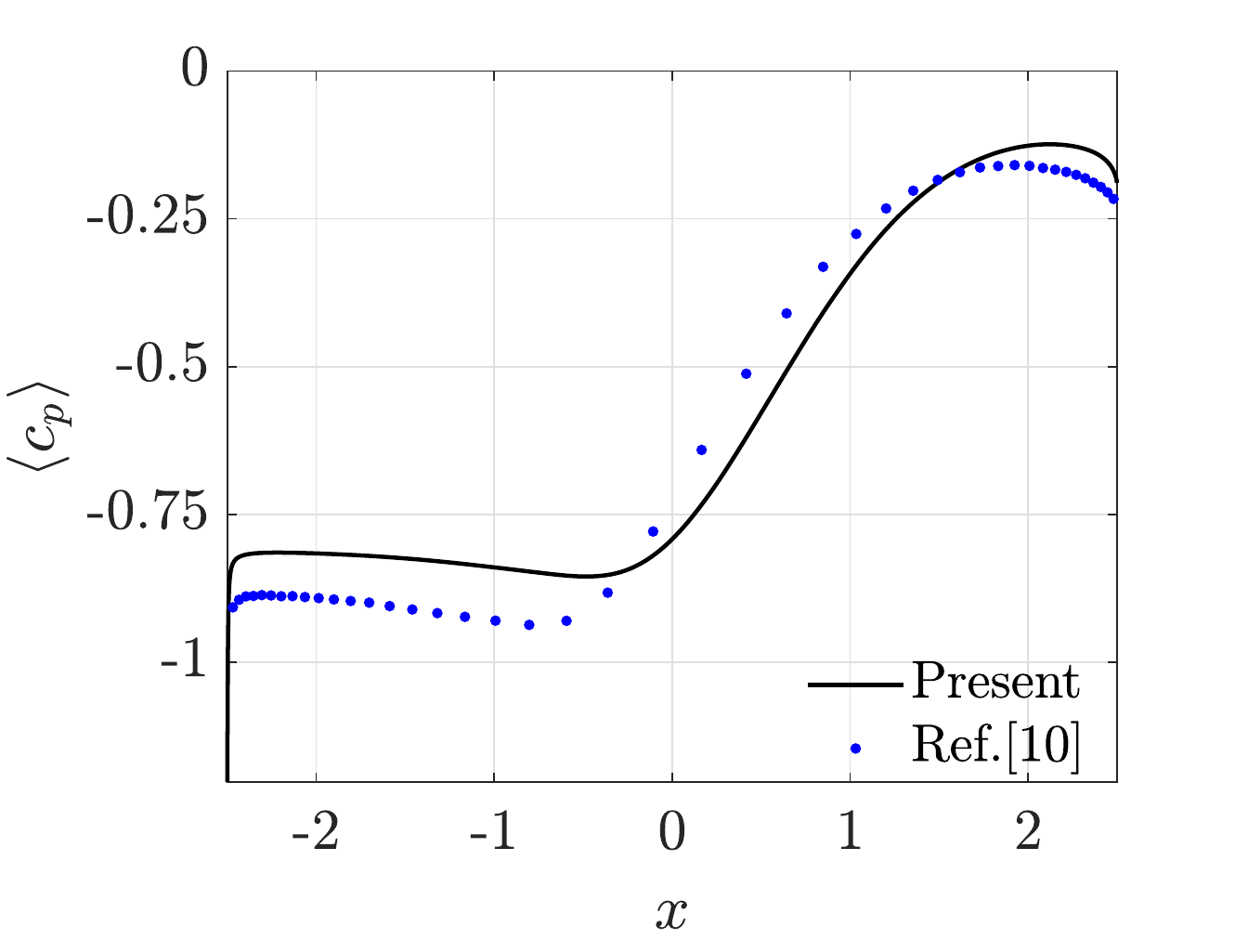}
\caption{Evolution of the friction (left) and pressure (right) coefficients along the body. Black line: present simulation; blue circles are data taken from figure 6 in Ref. \cite{cimarelli-leonforte-angeli-2018b}}
\label{fig:CfCpx}
\end{figure}
The mean flow field exerts its influence on the body, and determines the streamwise distribution of the coefficients $\aver{c_f}(x)$ and $\aver{c_p}(x)$, which express the longitudinal wall shear and the wall pressure made dimensionless with $0.5 \rho U_\infty^2$. These quantities are plotted in figure \ref{fig:CfCpx}. The friction coefficient (left panel) has its largest negative value of $-0.0166$ at the leading edge, which is less than the value reported in Ref. \cite{cimarelli-leonforte-angeli-2018b}, suggesting a different resolution of the interaction between the near-wall portion of the main shear layer and the secondary vortex. Friction then increases quickly to reach a plateau where it remains slightly positive for $-1.87 < x < -0.91 $: this region is the trace at the wall of the small-scale recirculating bubble. Further downstream, $\aver{c_f}$ becomes negative again, with a local minimum of value $-0.0118$ placed at $x=0.32$, associated with the strong reverse flow observed in the large-scale recirculating region. When the trailing edge is approached, $\aver{c_f}$ increases again to become positive for $x \ge x_{e,1}$, i.e. downstream of the primary vortex. When compared to Ref. \cite{cimarelli-leonforte-angeli-2018b}, the differences can be traced to the differences in the mean flow. The shift of the secondary vortex, for example, leads to a downstream shift of the first $\aver{c_f}>0$ region, whereas the larger extension of the primary vortex results in a downstream shift of both the local minimum, which in Ref. \cite{cimarelli-leonforte-angeli-2018b} occurs at $x \approx -0.045$ with a value of $-0.013$, and of the crossover point where $\aver{c_f}$ is zero.

The pressure coefficient $\aver{c_p}(x)$ (right panel) is negative everywhere, in agreement with available information. A minimum of $-1.205$ is found in the region of the leading edge. After a sharp and localised increase, a mild decline starts at $x \ge -2.21$ towards a local minimum of $-0.855$ at $x=-0.470$. This minimum is the footprint at the wall of the large negative pressure observed within the large-scale recirculating region. The pressure coefficient then increases again, and reaches its maximum of $-0.124$ at $x \approx 2.12$, not long before the trailing edge separation. The computed $\aver{c_p}(x)$ is in good general agreement with Ref. \cite{cimarelli-leonforte-angeli-2018b}, although some differences are evident from figure \ref{fig:CfCpx}. For example, for $x<0$ our values are smaller (in absolute value), and the local minimum occurs later downstream. For $x>0$, instead, the opposite is observed before the maximum, which takes place closer to the trailing edge and is less negative.

\section{Single-point budget of the Reynolds stresses}
\label{sec:SinglePointBudget}

The turbulent fluctuations in the flow are now described by studying the budget equation for the six independent components of the tensor for the Reynolds stresses. Terms for these budgets have been discussed e.g. in \cite{mansour-kim-moin-1988} for the channel flow, and have an obvious importance in turbulence modelling, but have not been fully documented yet for the BARC flow. As in the channel flow, two components, namely $\aver{uw}$ and $\aver{vw}$, are zero because of statistical symmetry. Note that, in this Section, primes to indicate the fluctuating velocity components will be omitted for conciseness.

\begin{figure}
\centering
\includegraphics[trim=290 100 290 80,clip,width=0.49\textwidth]{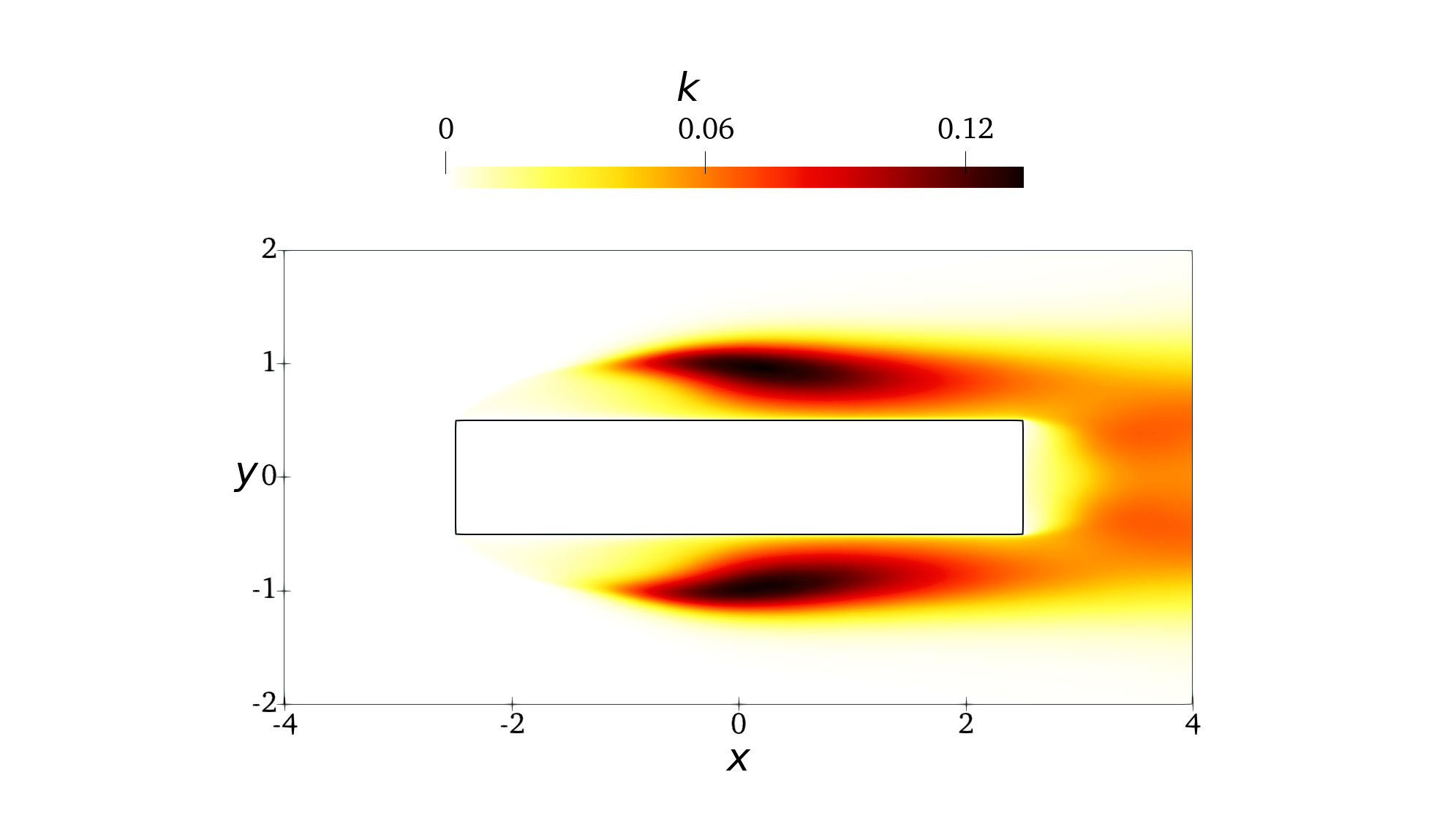}
\caption{Color map of the turbulent kinetic energy.}
\label{fig:k}
\end{figure}

Very recently, Refs. \cite{cimarelli-etal-2019-negative,moore-etal-2019,rocchio-etal-2020} presented and discussed the production term in the equation for the turbulent kinetic energy, with the latter two studies focusing on the region close to the upstream corners. Hence, before presenting the Reynolds stresses, we start with figure \ref{fig:k}, which shows the spatial map of the turbulent kinetic energy $k = \aver{u_i u_i}/2$ (a repeated index imply summation), which is proportional to the tensor trace. The map is perfectly symmetrical on the two sides of the body, again supporting the adequacy of the statistical sample. Before $x < -1.3$, $k$ is essentially zero, confirming the laminar state of the flow at the leading-edge separation and within the secondary vortex. For $x > -1.3$, however, $k$ rapidly increases, signifying a quick transition to the turbulent state: the maximum is observed at $(x,y) \approx (0.2,0.96)$. A further local maximum is observed in the wake at $(x,y) \approx (3.65,0.36)$. However, this is only half the value of the maximum in the primary vortex. Overall, the map of $k$ resembles the one reported in Ref. \cite{cimarelli-leonforte-angeli-2018}: for example their global and local maxima are at $(0.2,0.9)$ and $(3.7,0.35)$. However, the intensity of $k$ in Ref. \cite{cimarelli-leonforte-angeli-2018} is larger than here: for example their maximum in the primary vortex is $k \approx 0.40$ instead of the present maximum $k \approx 0.135$.

We now proceed to describing the budget of the full Reynolds stress tensor. In tensorial notation, a compact form of the budget equation, stemming from the manipulation of the incompressible Navier--Stokes equations, and specialised to the present spanwise-homogeneous problem, can be written as:
\begin{equation}
\frac{\partial}{\partial x} \psi_{x,ij} + \frac{\partial}{\partial y} \psi_{y,ij}  = P_{ij} + \Pi_{ij} - \epsilon_{ij} .
\label{eq:budget}
\end{equation}
In Eq.\eqref{eq:budget}, $\psi_{x,ij}$ and $\psi_{y,ij}$ are the fluxes in the $x$ and $y$ directions, defined as follows:
\begin{equation}
\psi_{k,ij} = \underbrace{\aver{u_i u_j u_k}}_{\text{turbulent transport}} + 
              \underbrace{\aver{pu_k}(\delta_{i,k}+\delta_{j,k})}_{\text{pressure transport}} +
              \underbrace{U_k \aver{u_i u_j}}_{\text{mean transport}} +
              \underbrace{\nu \frac{\partial}{\partial x_k} \aver{u_i u_j}}_{\text{viscous diffusion}} \ \ \text{with} \ k = x,y
\end{equation}
whereas $P_{ij}$, $\Pi_{ij}$ and $ \epsilon_{ij}$ denote the production, the pressure-strain and the pseudo-dissipation tensors, defined as:
\begin{equation}
P_{ij} = -\aver{u_i u_k} \frac{\partial U_j}{\partial x_k} - \aver{u_j u_k} \frac{\partial U_i}{\partial x_k} ,
\label{eq:prod}
\end{equation}
\begin{equation}
\Pi_{ij} = \aver{p \frac{\partial u_i}{\partial x_j}} + \aver{ p \frac{\partial u_j}{x_i} } ,
\end{equation}
\begin{equation}
\epsilon_{ij} = \nu \aver{\frac{\partial u_i}{\partial x_k} \frac{\partial u_j}{\partial x_k} } .
\label{eq:diss}
\end{equation}

\begin{figure}
\centering
\includegraphics[trim=290 100 290 80,clip,width=0.49\textwidth]{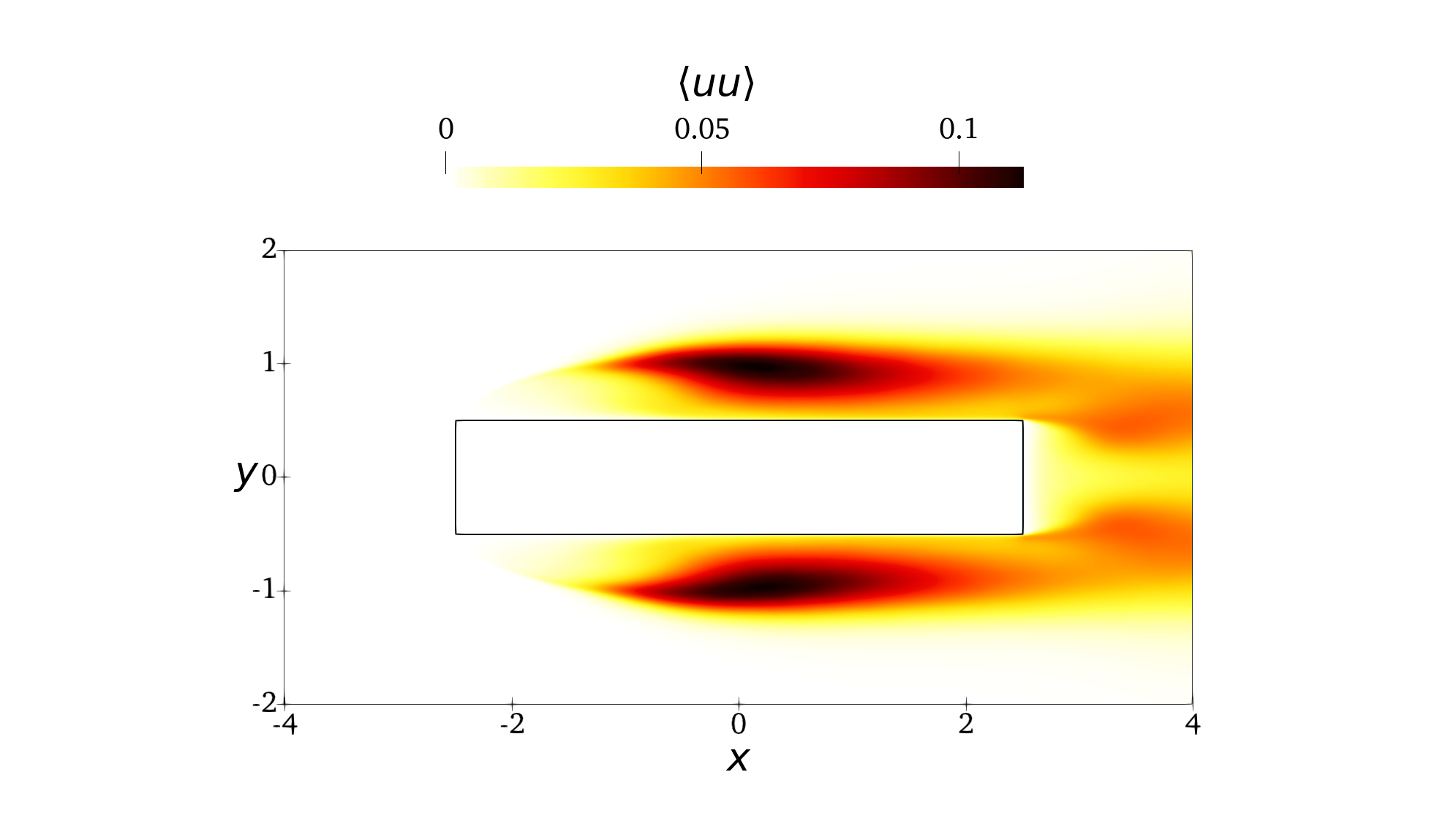}
\includegraphics[trim=290 100 290 80,clip,width=0.49\textwidth]{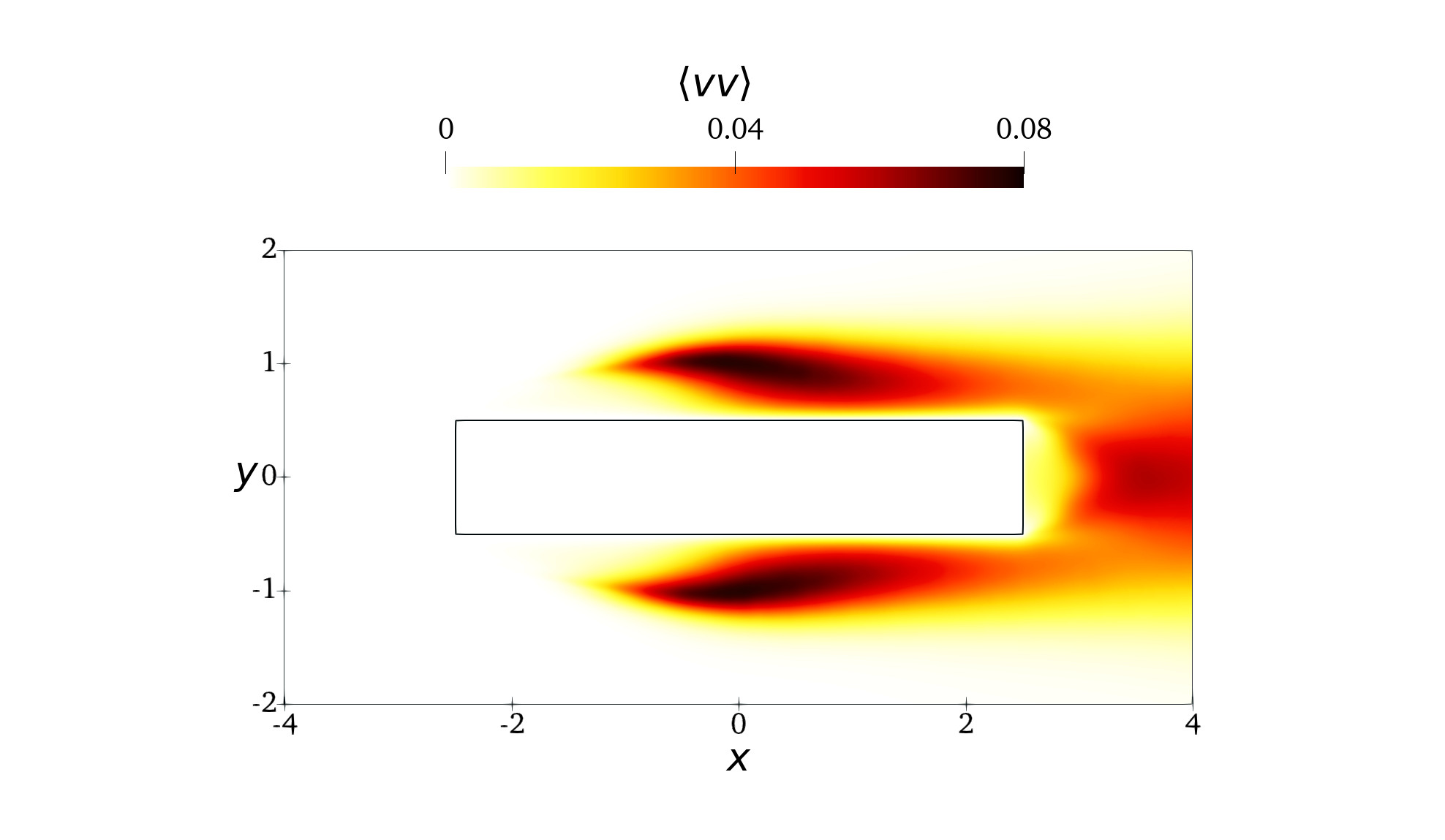}
\includegraphics[trim=290 100 290 80,clip,width=0.49\textwidth]{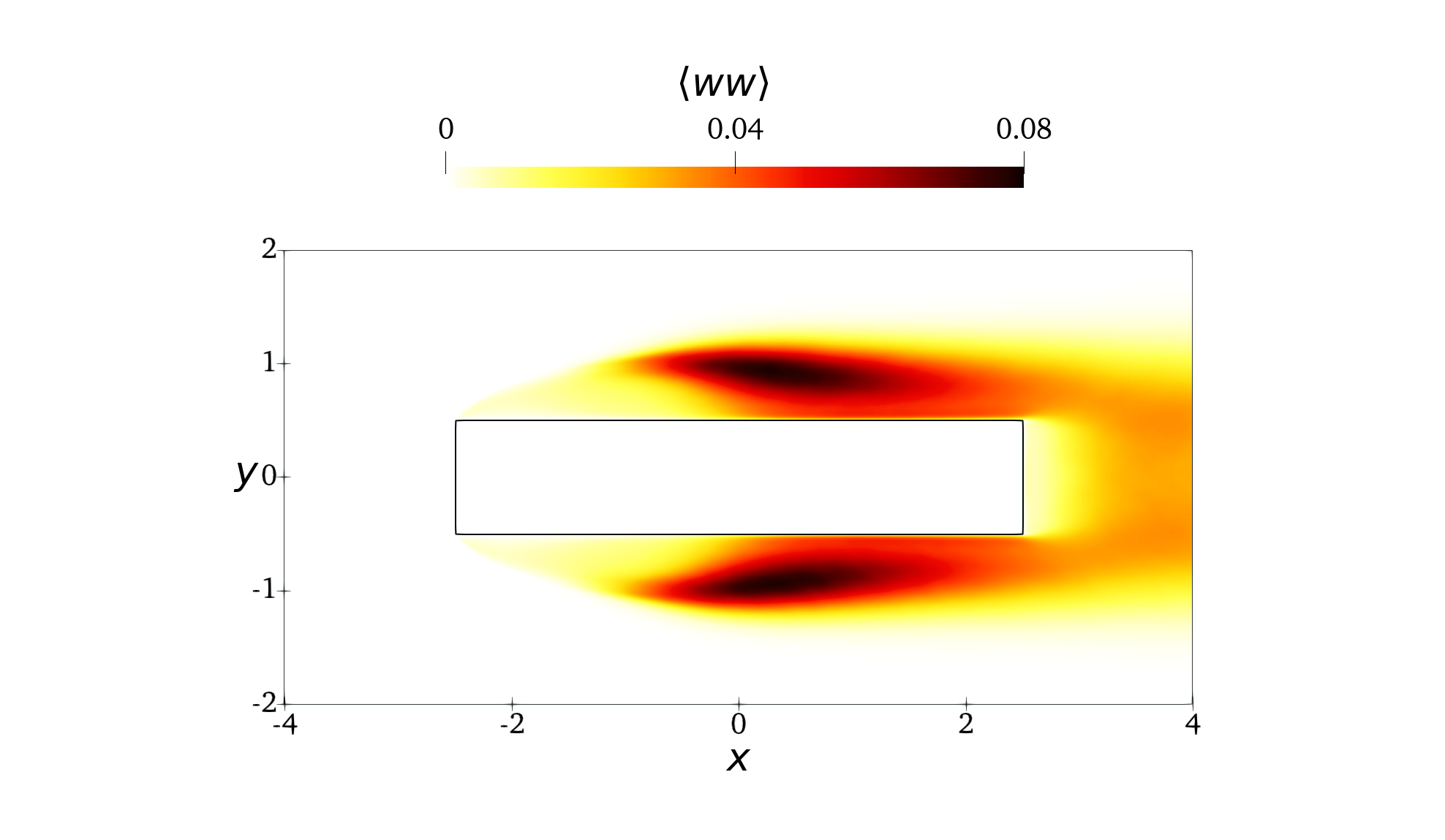}
\includegraphics[trim=290 100 290 80,clip,width=0.49\textwidth]{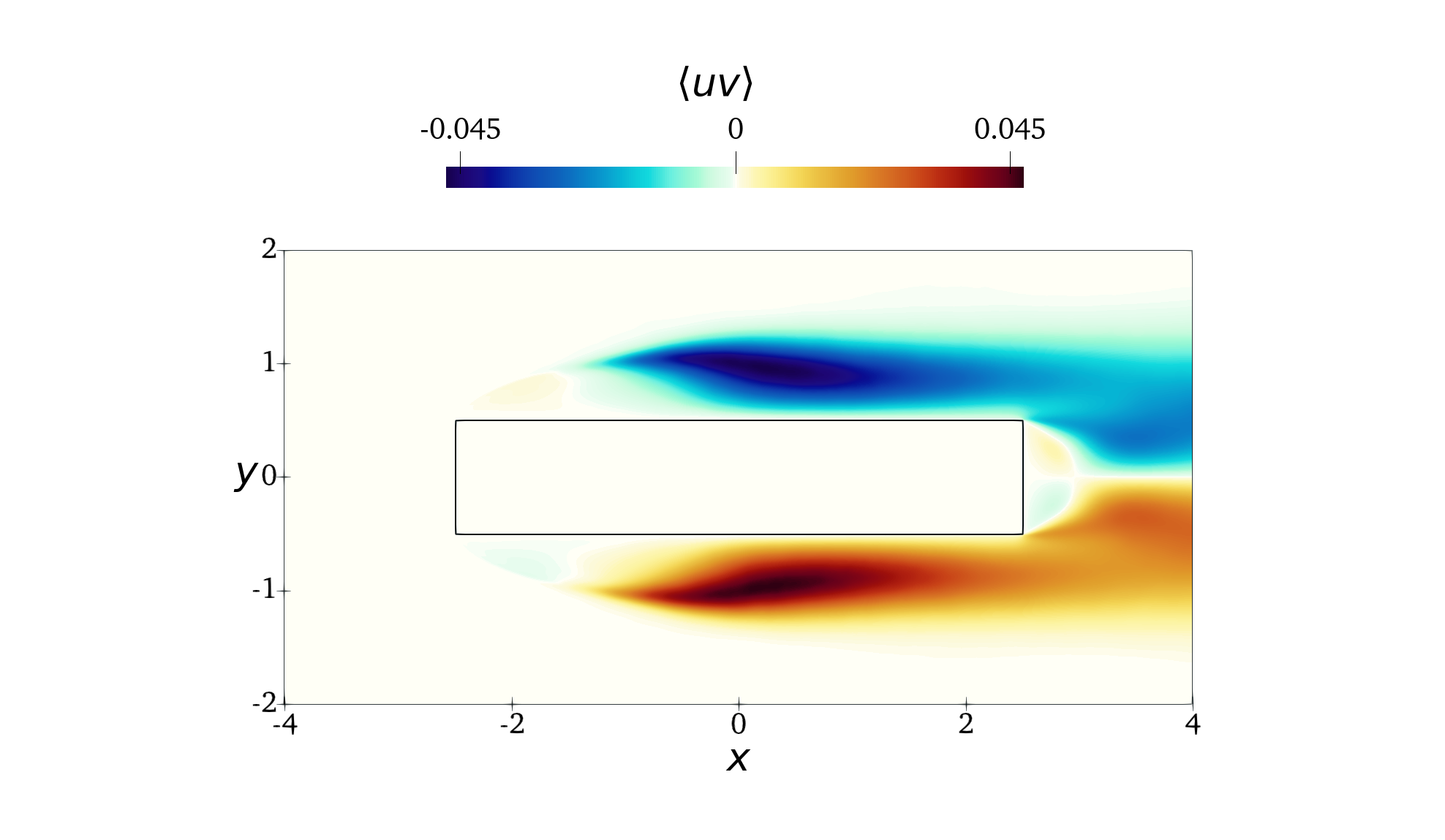}
\caption{Contours of the Reynolds stresses in the $x-y$ plane. Top left: $\aver{uu}$; top right: $\aver{vv}$; bottom left: $\aver{ww}$; bottom right: $\aver{uv}$.}
\label{fig:Rss}
\end{figure}

Figure \ref{fig:Rss} plots the four non-zero components of the Reynolds stress tensor in the $x-y$ plane. Their general features have been already described in \cite{cimarelli-leonforte-angeli-2018b}. Once the flow becomes turbulent for $x \ge -2$, the largest fluctuations are those of the streamwise component, in both the shear layer and the primary vortex, but the maxima of the three diagonal components are of comparable magnitude as also observed at larger $Re$ \cite{moore-etal-2019}. The peak of $\aver{uu}$ is at $(x,y)=(0.21,0.97)$, i.e. above the centre of the primary vortex, and shows small values close to the cylinder's wall, confirming its dynamical association to the shear layer. The maximum of $\aver{vv}$ occurs slightly upstream than that of $\aver{uu}$, i.e. at $(x,y)=(0.05,1.01)$, but it remains almost negligible near the wall, and is associated to the core of the primary vortex. Lastly, the position of the maximum of $\aver{ww}$ is almost identical to that of $\aver{uu}$, at $(x,y)=(0.24,0.95)$. Interestingly, near the wall $\aver{ww}$ is the dominant component. This strikingly differs from e.g. plane Poiseuille flow, and is in agreement with the pattern seen in figure \ref{fig:lambda2-vorticity}. In fact, large streamwise fluctuations are associated to the spanwise structures originated by the instability of the shear layer, whereas large spanwise and cross-stream fluctuations are linked to the streamwise-aligned structures populating the inner part of the large recirculating region. In the wake, $\aver{uu}$ is large in the shear layer separating from the trailing edge, with a peak at $(x,y)=(3.46,0.48)$, whereas smaller values are observed along the symmetry line $y=0$. The other components, on the other hand, are large in core of the wake, with $\aver{vv}$ peaking at $(x,y)=(3.59,0)$. The $\aver{ww}$ component has quite a broad distribution, without a distinct peak. The general appearance is consistent with known results \cite{cimarelli-leonforte-angeli-2018b}, but significant qualitative and quantitative differences emerge. Since the mean velocity field and in particular the extent of the primary vortex differ, different positions of the maxima are to be expected. Specifically, in \cite{cimarelli-leonforte-angeli-2018b} maxima at $(0.07,0.94), (0.46,0.81)$ and $(0.46,0.78)$ are found for $\aver{uu}$, $\aver{vv}$ and $\aver{ww}$ respectively. Hence the largest streamwise fluctuations occur before the other components, and the largest spanwise fluctuations occur closer to the cylinder side.

The off-diagonal component $\aver{uv}$ is also plotted in figure \ref{fig:Rss}. The colour map is antisymmetric: here and in the following, in the text we refer to the top side of the cylinder. The most negative values are observed within the primary vortex, where the streamwise fluctuations correlate significantly to the vertical ones resulting in a minimum of $\aver{uv}$ located at $(x,y) \approx (0.19,0.97)$. On the contrary, in the upstream portion for $x < -1.5$ below the separated shear layer, slightly positive $\aver{uv}$ are observed. Downstream the trailing edge $\aver{uv}$ is positive within the wake vortex, and negative elsewhere with a local minimum at $(x,y) \approx (3.5,0.37)$. The shear stress $\aver{uv}$ plays a fundamental role in the production of the turbulent kinetic energy, which has been suggested \cite{cimarelli-etal-2019-negative} to become negative in a localised region in the BARC flow. This point will be discussed later. 

\begin{figure}
\centering
\includegraphics[trim=290 100 290 80,clip,width=0.49\textwidth]{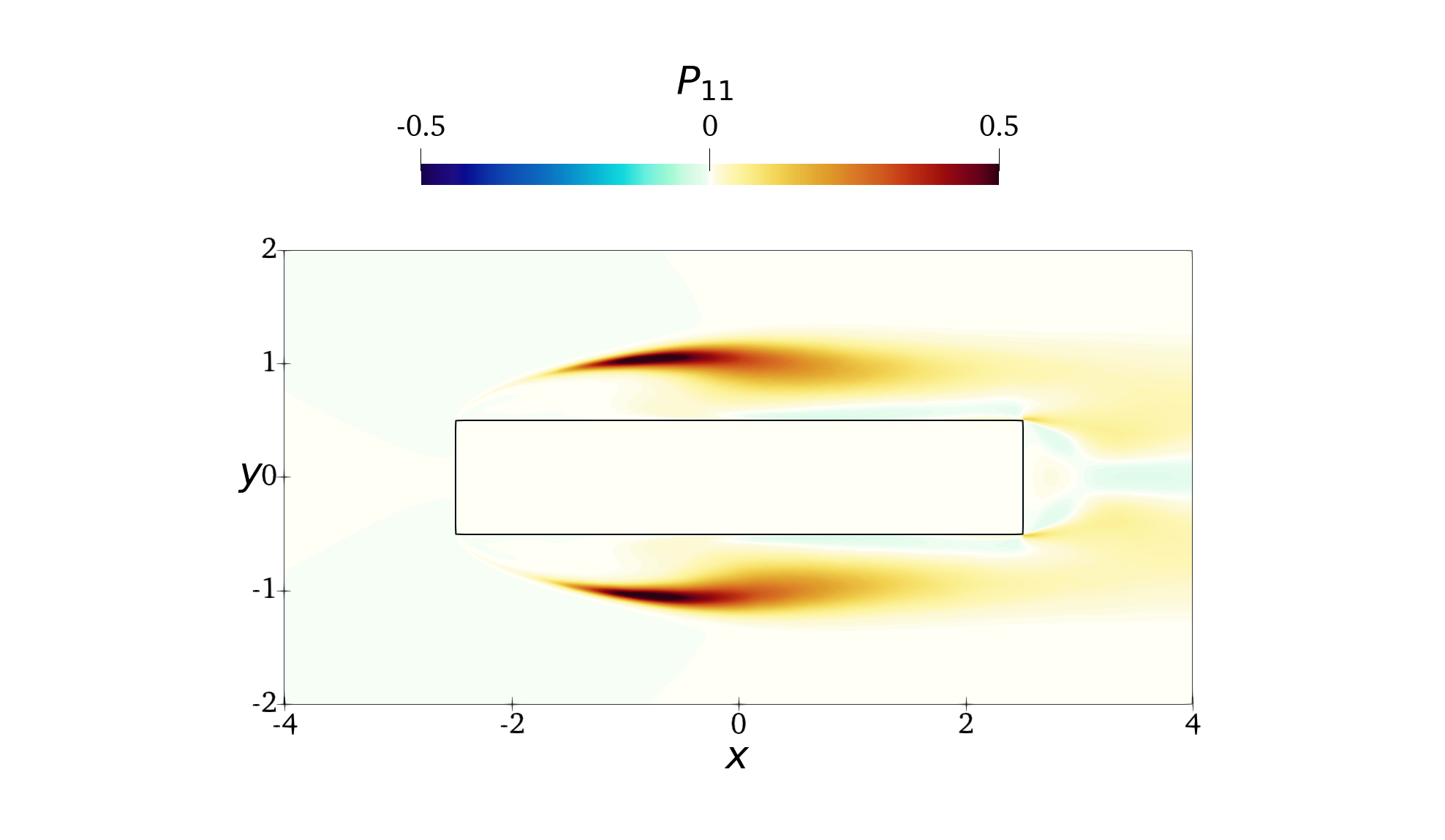}
\includegraphics[trim=290 100 290 80,clip,width=0.49\textwidth]{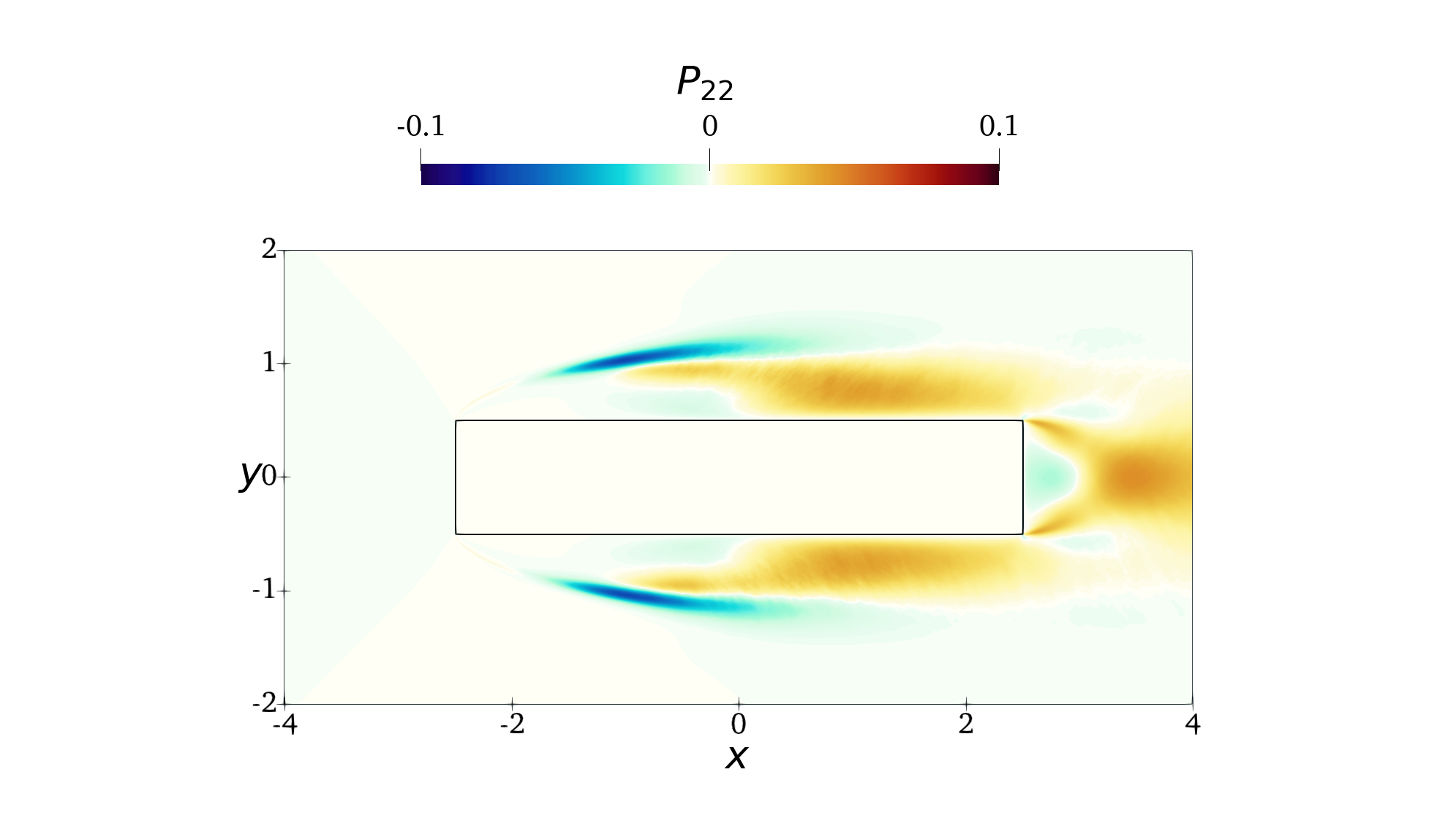}
\includegraphics[trim=290 100 290 80,clip,width=0.49\textwidth]{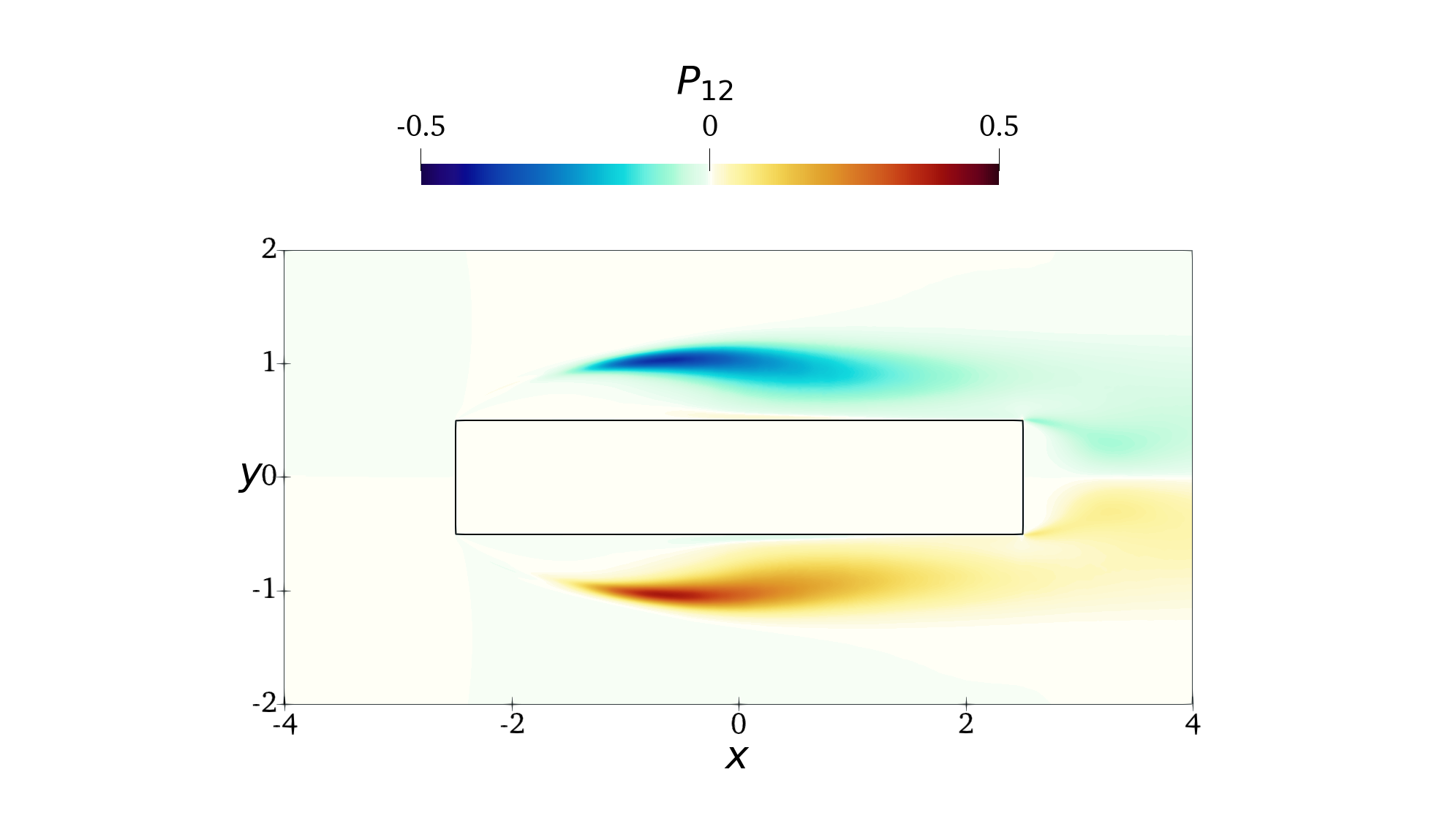}
\caption{Contours of the production terms $P_{11}$, $P_{22}$, and $P_{12}$.}
\label{fig:Prod}
\end{figure}
Figure \ref{fig:Prod} shows $P_{11}$, $P_{22}$ and $P_{12}$, i.e. the production terms for $\aver{uu}$, $\aver{vv}$ and $\aver{uv}$. We recall that $P_{33}=0$ because $\partial W / \partial x = \partial W / \partial y =0$, and that the sum of the diagonal terms is half the production of turbulent kinetic energy. In the shear layer for $x \ge -2$ $P_{11}$ is positive and $P_{22}$ is negative (albeit smaller). This implies that, once the shear layer instability takes place, energy is drained from the mean flow to feed streamwise fluctuations, whereas the opposite occurs for the cross-stream component $v$. Further downstream, in the core of the primary vortex $P_{11}$ and $P_{22}$ are both positive, and the mean flow feeds both $\aver{uu}$ and $\aver{vv}$. Furthermore, close to the wall in the downstream portion of the body, i.e. for $x \ge 0$, $P_{11}$ becomes mildly negative, hence a sink for $\aver{uu}$. In this region flow reversal takes place, where $\partial U / \partial{y} < 0 $, and the streamwise fluctuations are weaker. Similar considerations can be put forward for the wake. Indeed, $P_{11}$ is mildly positive in the recirculating region behind the trailing edge, but becomes negative more downstream along the centerline $y=0$, where indeed low values of $\aver{uu}$ take place. On the contrary, along the centerline line and outside the recirculation, $P_{22}$ is positive with relatively large values, denoting production of $v$ fluctuations as shown also in figure \ref{fig:Rss}. $P_{12}$ is shown in the bottom panel of figure \ref{fig:Prod}. It is worth noting \cite{gatti-etal-2020} that, since $\aver{uv}$ is not positive-definite, interpreting $P_{12}$, $\Pi_{12}$ and $\epsilon_{12}$ in terms of production or dissipation requires to account for the sign of $\aver{uv}$. Negative values of $P_{12}$ are observed everywhere -- except in a flat region very close to the cylinder side where it is slightly positive -- with a global minimum within the primary vortex at $(-0.59,1.03)$ and a local minimum in correspondence of the shear layer separating from the trailing edge at $(2.65,0.5)$. Overall, since in the region with large negative $P_{12}$ $\aver{uv}$ is negative too, a production of $\aver{-uv}$ takes place. With analogous reasoning, the negative $P_{12}$ in the first part of the shear layer separating from the leading edge corners indicates a sink for the positive $\aver{uv}$.

Let us now focus on the energy production, by examining the contributions to $P_{11}$ and $P_{22}$. Each production term can be split in two, to isolate contributions related to the streamwise variation of the mean flow from those related to its cross-stream variation:

\begin{equation}
  P_{11}=\underbrace{-2\aver{uu} \frac{\partial U}{\partial x}}_{P_{11,a}}  \underbrace{-2\aver{uv} \frac{\partial U}{\partial y}}_{P_{11,b}} \ \text{and} \ 
  P_{22}=\underbrace{-2\aver{uv} \frac{\partial V}{\partial x}}_{P_{22,a}} \underbrace{-2\aver{vv} \frac{\partial V}{\partial y}}_{P_{22,b}}.
  \label{eq:Prod1-Prod2}
\end{equation}
 
\begin{figure}
\centering
\includegraphics[trim=290 100 290 80,clip,width=0.49\textwidth]{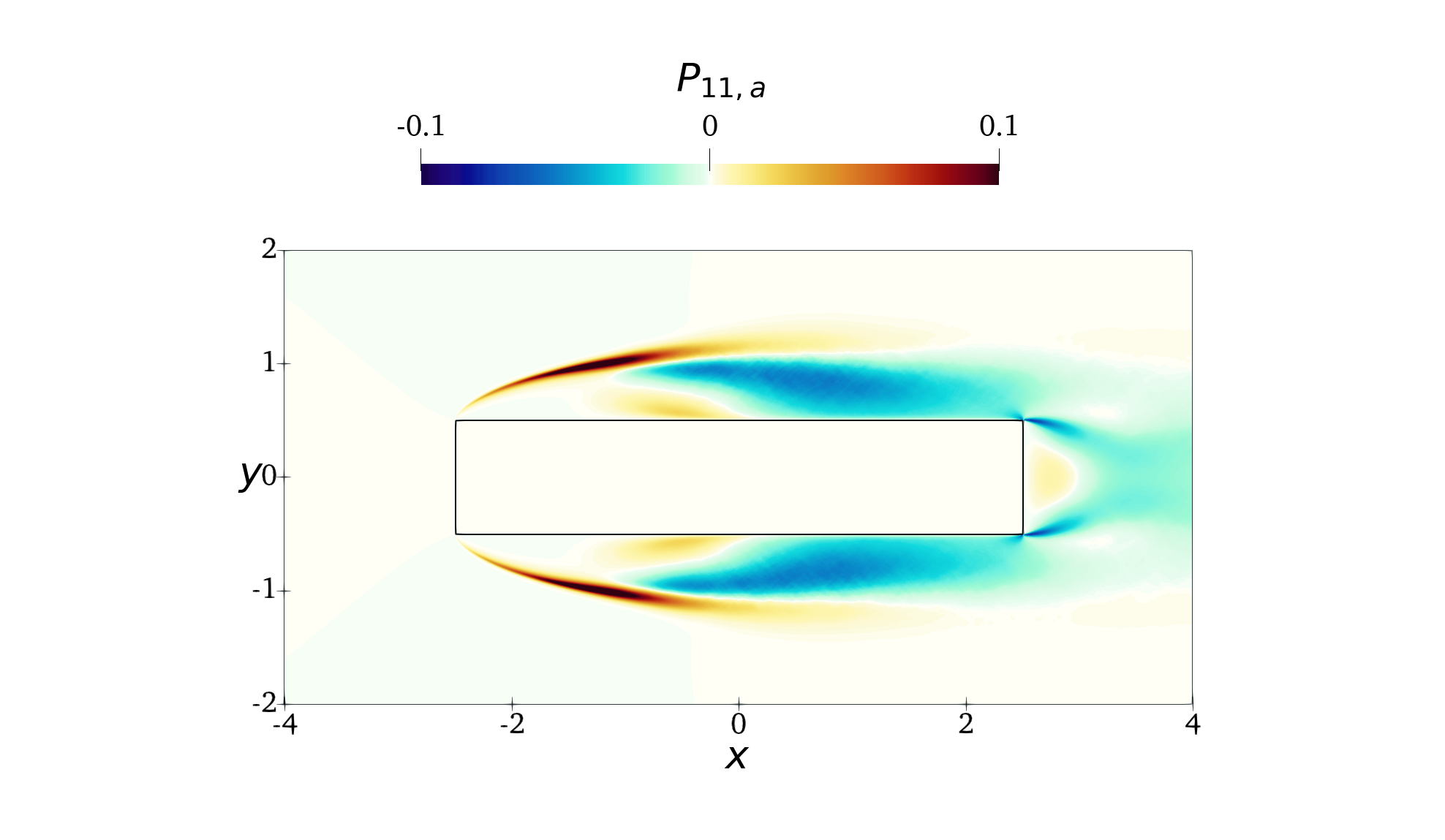}
\includegraphics[trim=290 100 290 80,clip,width=0.49\textwidth]{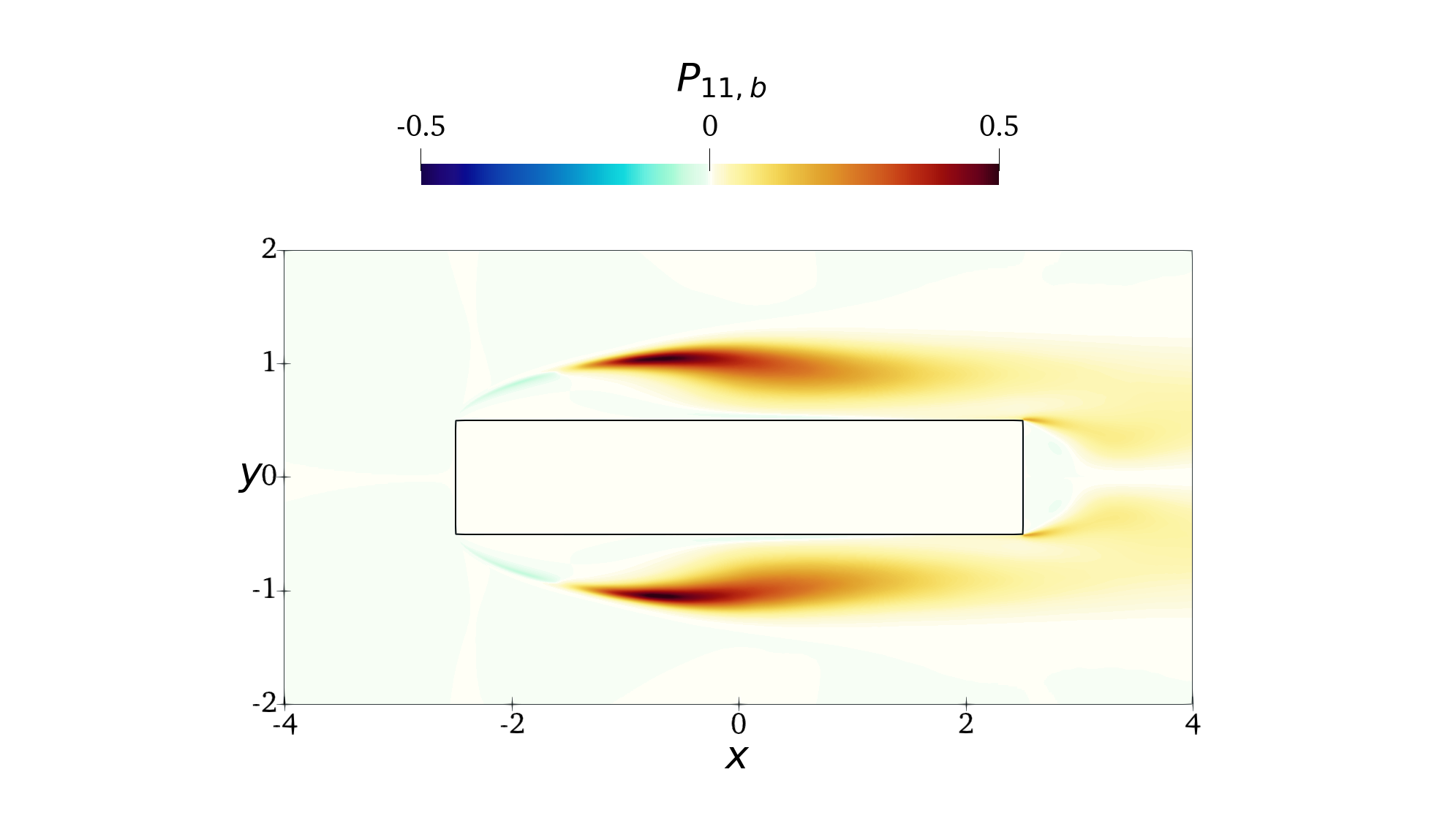}
\includegraphics[trim=290 100 290 80,clip,width=0.49\textwidth]{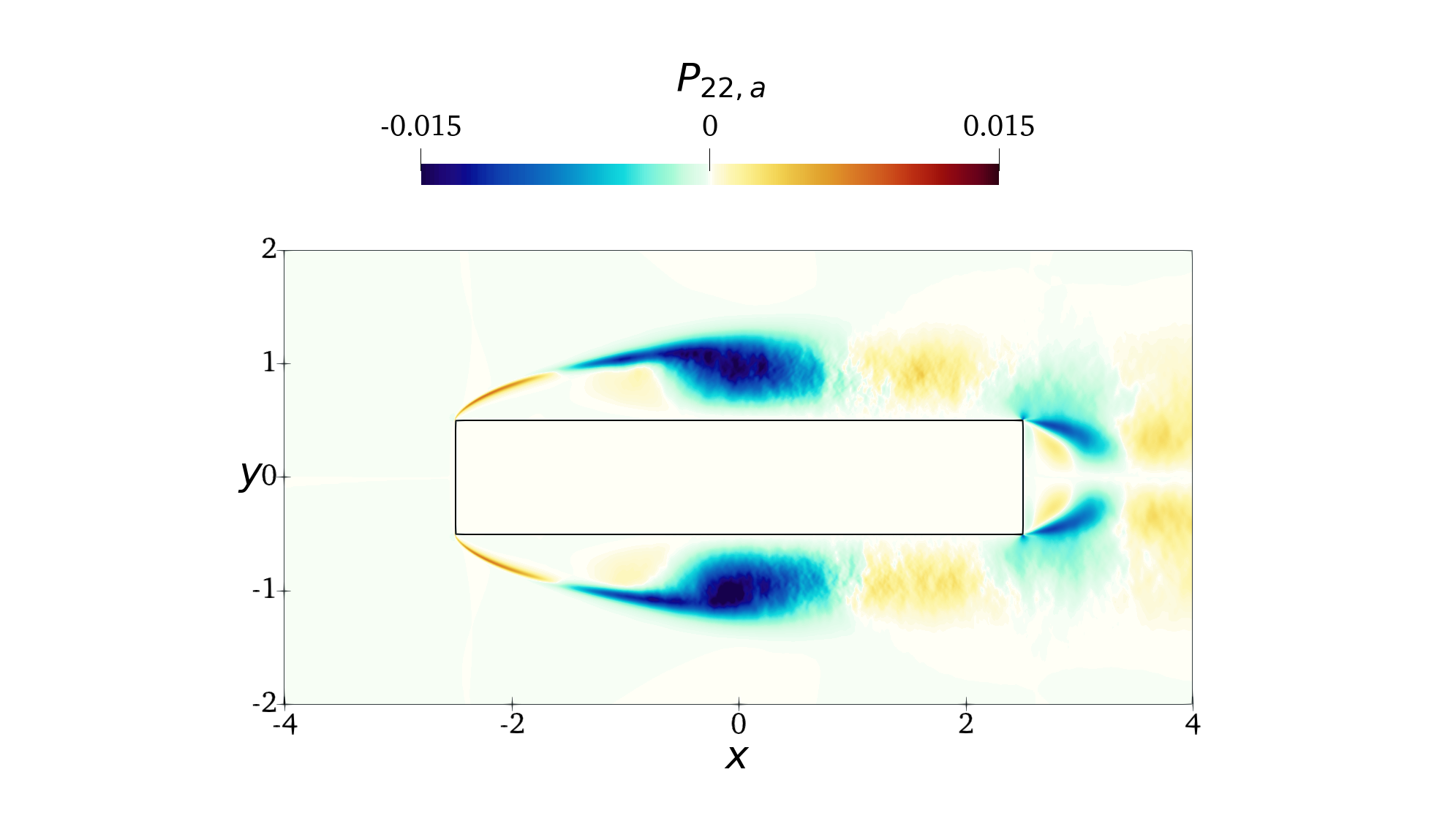}
\includegraphics[trim=290 100 290 80,clip,width=0.49\textwidth]{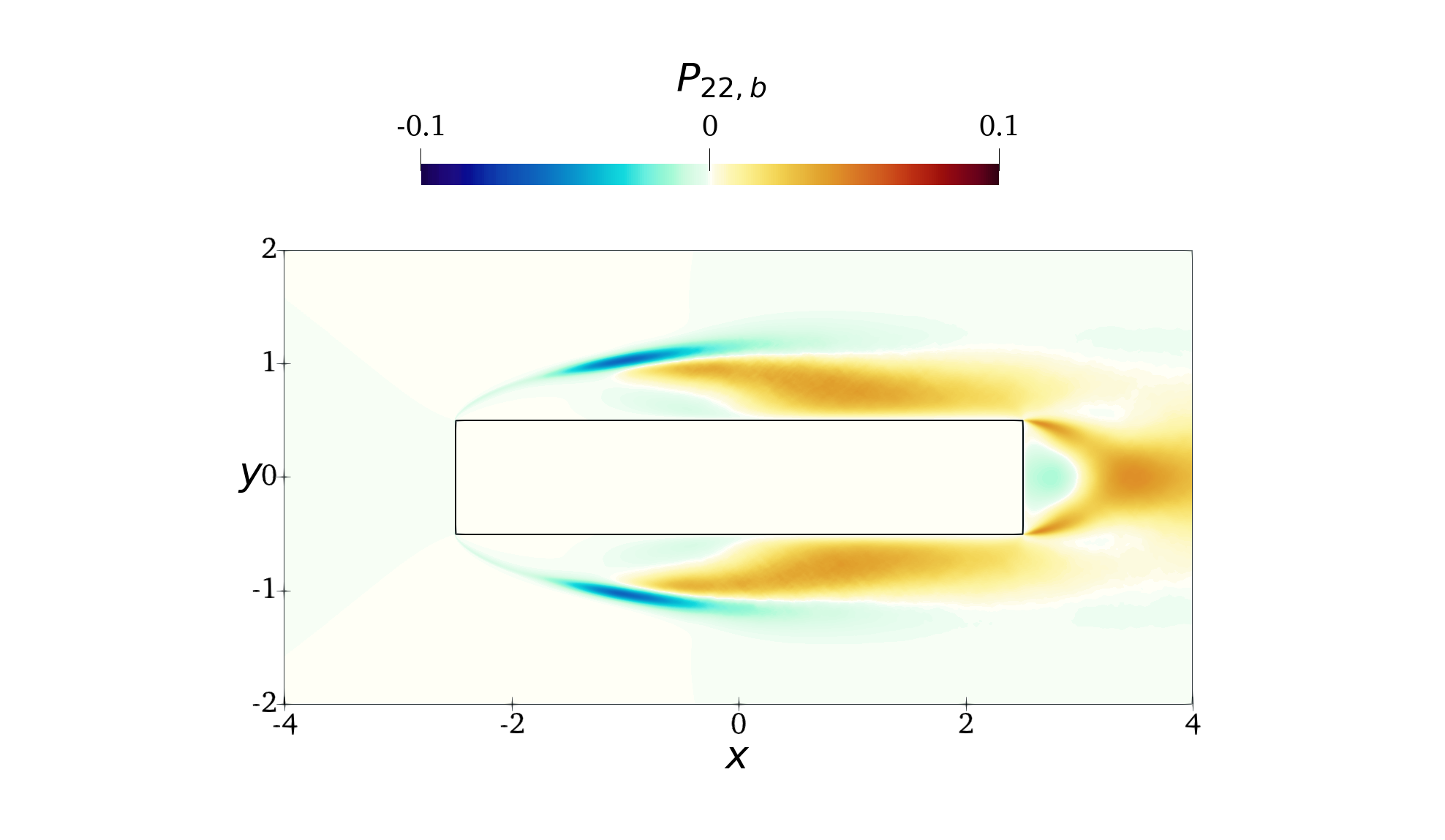}
\caption{Production of $\aver{uu}$ (top) and $\aver{vv}$ (bottom), with separated contributions related to streamwise variation (left) and cross-stream variation of the mean flow. The contributions are defined in equation \eqref{eq:Prod1-Prod2}.}
\label{fig:Prod1-Prod2}
\end{figure}

Figure \ref{fig:Prod1-Prod2} separately plots each of these four terms. $P_{11,a}$ and $P_{22,b}$ have opposite sign and similar appearance in the domain, owing to the incompressibility constraint $\partial U / \partial x = - \partial V \partial y$. The figure shows that the main contribution to $P_{11}$ changes depending on the region considered. On the shear layer, for $x \le -1.5$ positive production comes from $P_{11,a}$ and is due to the negative values of $\partial U / \partial x $ associated with the shear layer. Here $P_{11,b}$ is negative, since $\aver{uv}>0$ and $\partial U / \partial y $ is positive everywhere at this station. Further downstream, both $P_{11,a}$ and $P_{11,b}$ become positive, since $\aver{uv}<0$ (see figure \ref{fig:Rss}). Production of $\aver{uu}$ in the core of the main recirculating region is instead given by $P_{11,b}$, and is determined by the positive values of $\partial U / \partial y$: here $\partial U / \partial x <0$, leading to negative $P_{11,a}$. Near the body $P_{11}$ is dominated by $P_{11,a}$, as $P_{11,b} \rightarrow 0$ faster for $y \rightarrow 0$. 

Unlike $P_{11}$, the main contribution to $P_{22}$ comes from $P_{22,b}$, as the companion component $P_{22,a}$ is significantly smaller similarly to what found experimentally by Ref. \cite{moore-etal-2019} in the first part of the shear layer at $Re=13400$. Therefore, the main driver here is $\partial V / \partial y$. The negative production in the shear layer and in the reverse boundary layer for $x \le 0 $ is due to $ \partial V / \partial y > 0 $, whereas the positive values of $P_{22}$ in the core of the main recirculating region to $\partial V / \partial y < 0$. 

\begin{figure}
\centering
\includegraphics[trim=290 100 290 80,clip,width=0.49\textwidth]{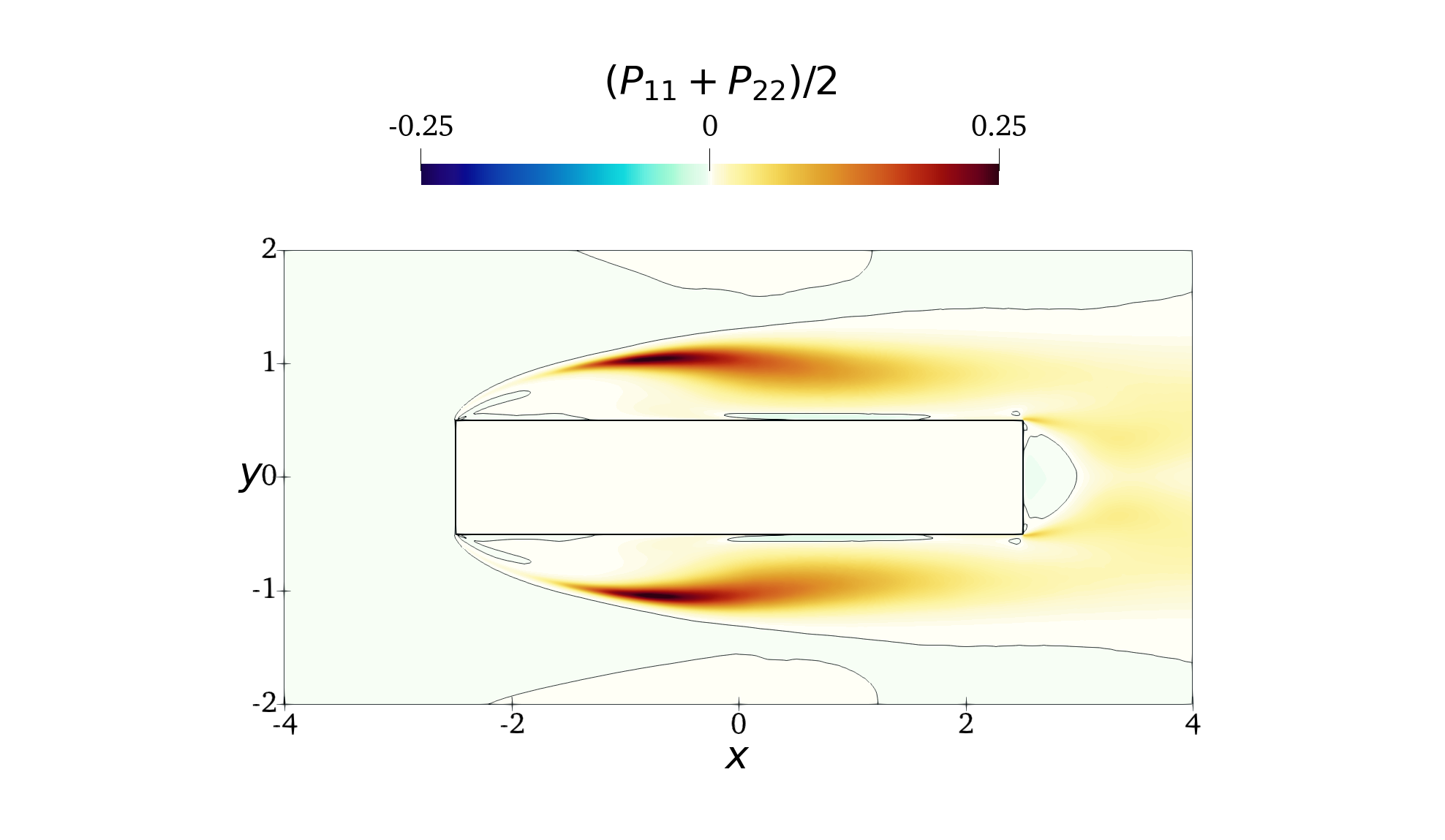}
\includegraphics[trim=290 100 290 80,clip,width=0.49\textwidth]{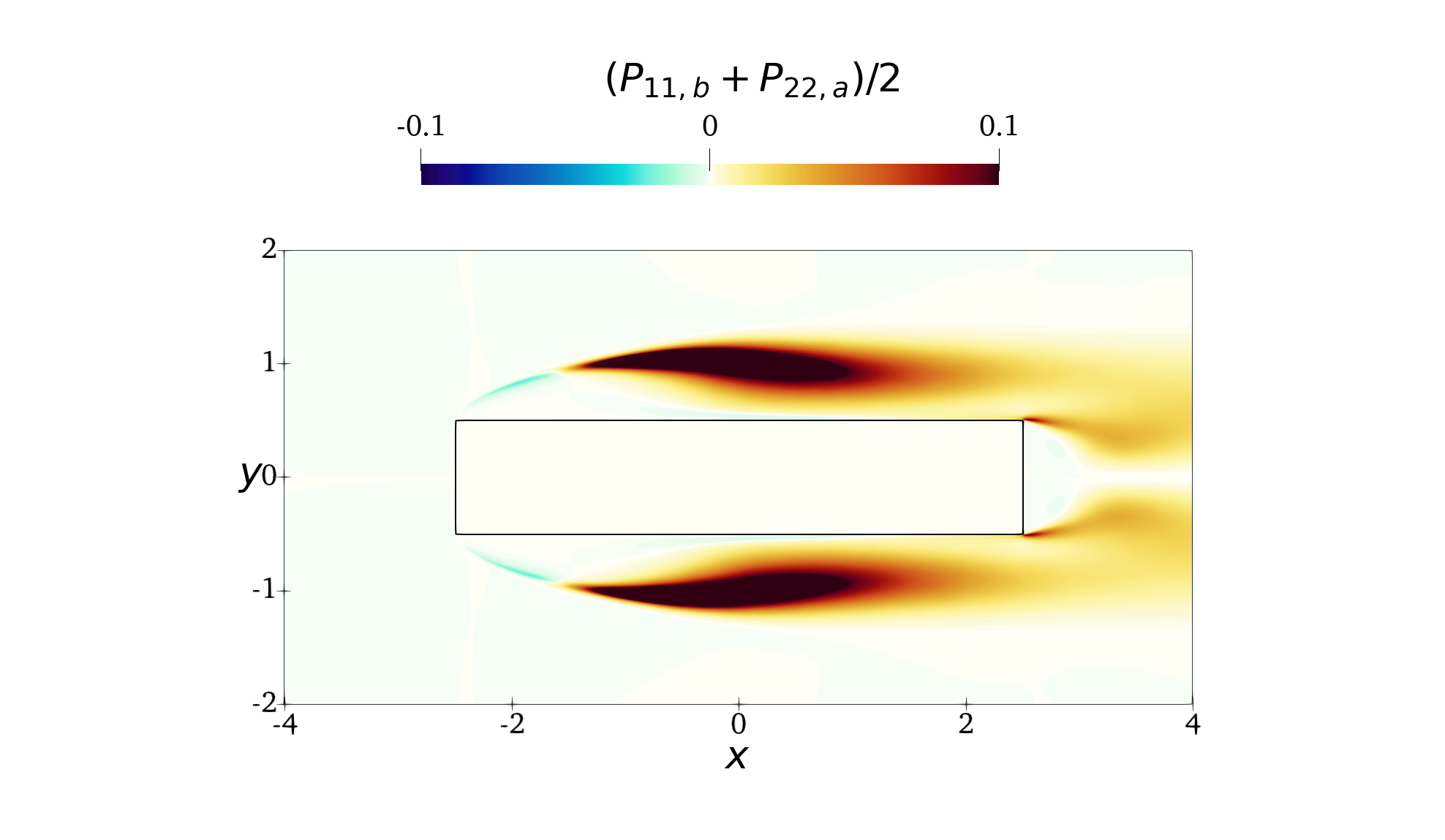}
\caption{Left: contour of the production term for the turbulent kinetic energy $P_k=(P_{11}+P_{22})/2$; the black line denotes the $0$ value. Right: contribution of to the production term for the turbulent kinetic energy from the interaction of $\aver{uv}$ and the mean shear $\partial U / \partial y + \partial V / \partial x$, i.e. $(P_{11,b}+P_{22,a})/2$.}
\label{fig:Prod-k}
\end{figure}
Cimarelli et al. in \cite{cimarelli-etal-2019-negative} discuss the presence of a thin region with significant negative production rate of turbulent kinetic energy, localised in the shear layer close to the leading-edge corners (see figure 2 of their paper). Such a negative production region is not observed here. In fact, as shown in the left panel of figure \ref{fig:Prod-k} where the total production $P_k=(P_{11}+P_{22})/2$ is plotted, mildly negative values are indeed found, but they are close to the leading-edge corners below the shear layer, close to the cylinder wall for $x>0$ and behind the trailing edge in the wake region. Overall, $P_{11}$ contributes to $P_k$ more than $P_{22}$ almost everywhere in the domain, as seen in figure \ref{fig:Prod}; this is true also for $x=-2.5$, unlike what found at larger $Re$ in Ref. \cite{rocchio-etal-2020}, where a larger contribution from $P_{22,b}$ is reported. In \cite{cimarelli-etal-2019-negative} negative production is linked to the interaction between $\aver{uv}$ and the mean streamwise and vertical shears $\partial U / \partial y$ and $\partial V / \partial x$. This is confirmed in the right panel of figure \ref{fig:Prod-k} where the sum $(P_{11,b}+P_{22,a})/2$ is plotted: a region with negative $(P_{11,b}+P_{22,a})/2<0$ is indeed observed in the first portion of the shear layer (more precisely, this region is due to $P_{11,b}$ alone, since $P_{22,a}$ is negligible). However, in this region of the domain $P_{11,a}$ is dominant, eventually resulting in a positive production term.

\begin{figure}
\centering
\includegraphics[trim=290 100 290 80,clip,width=0.49\textwidth]{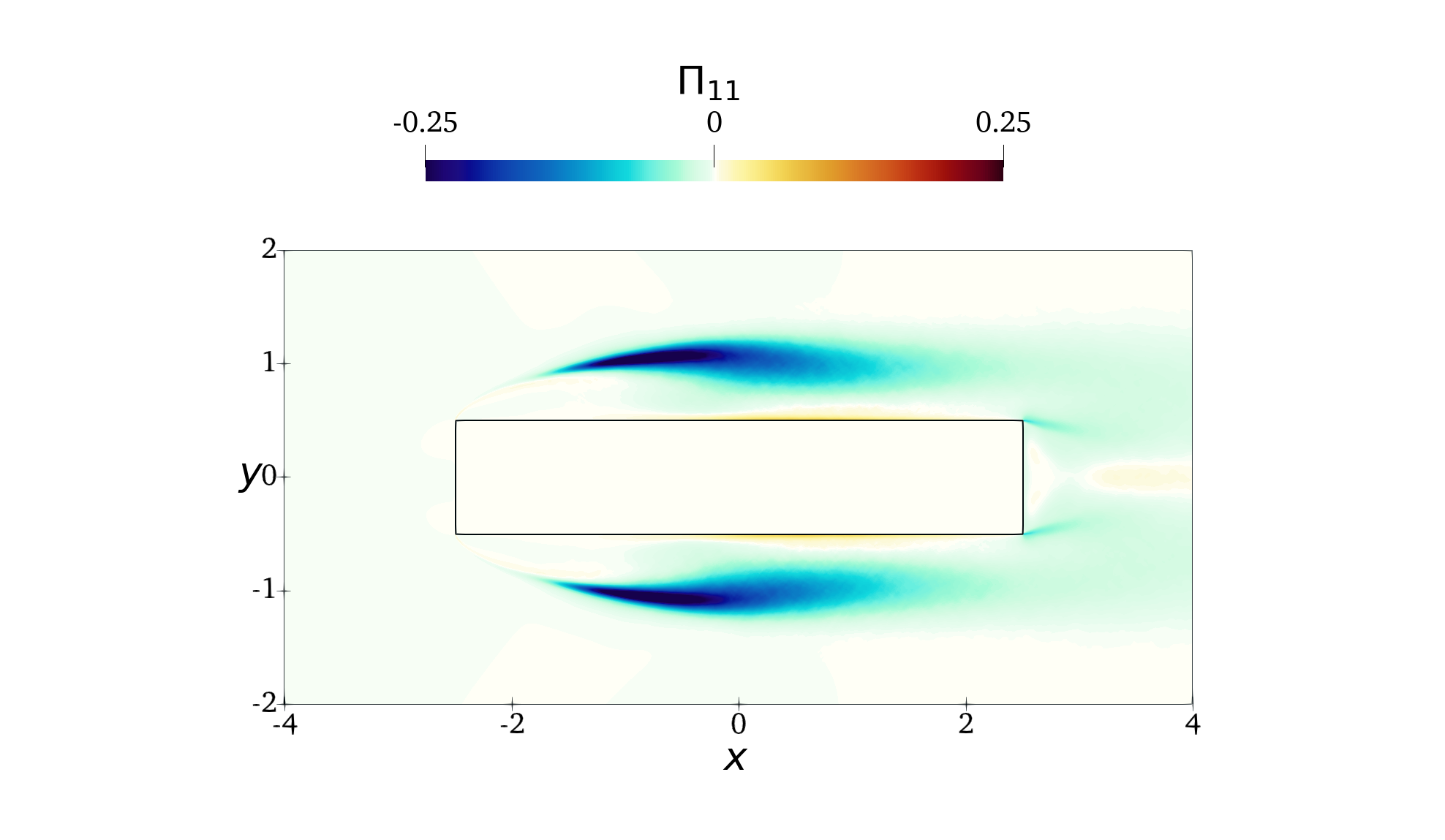}
\includegraphics[trim=290 100 290 80,clip,width=0.49\textwidth]{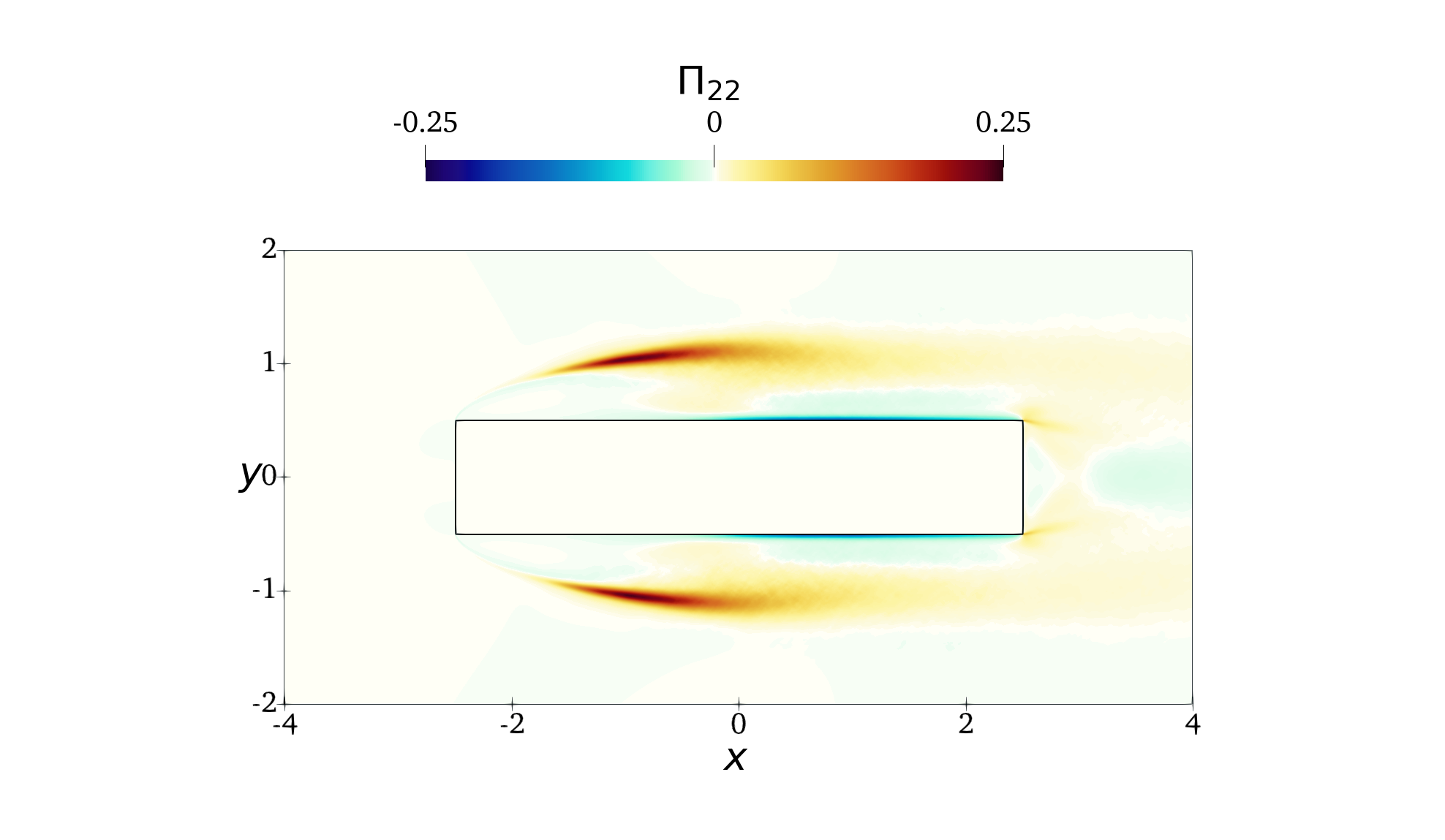}
\includegraphics[trim=290 100 290 80,clip,width=0.49\textwidth]{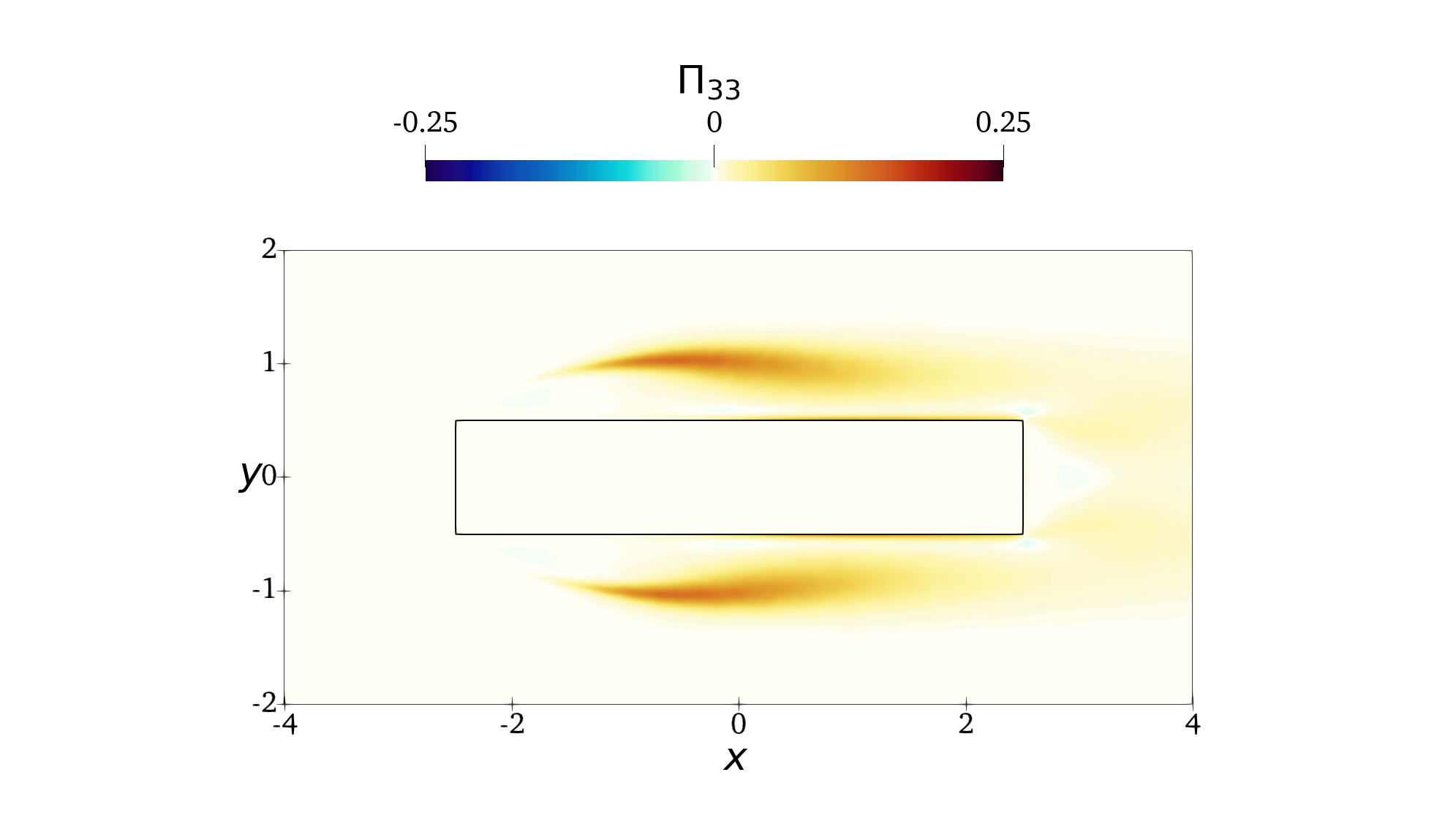}
\includegraphics[trim=290 100 290 80,clip,width=0.49\textwidth]{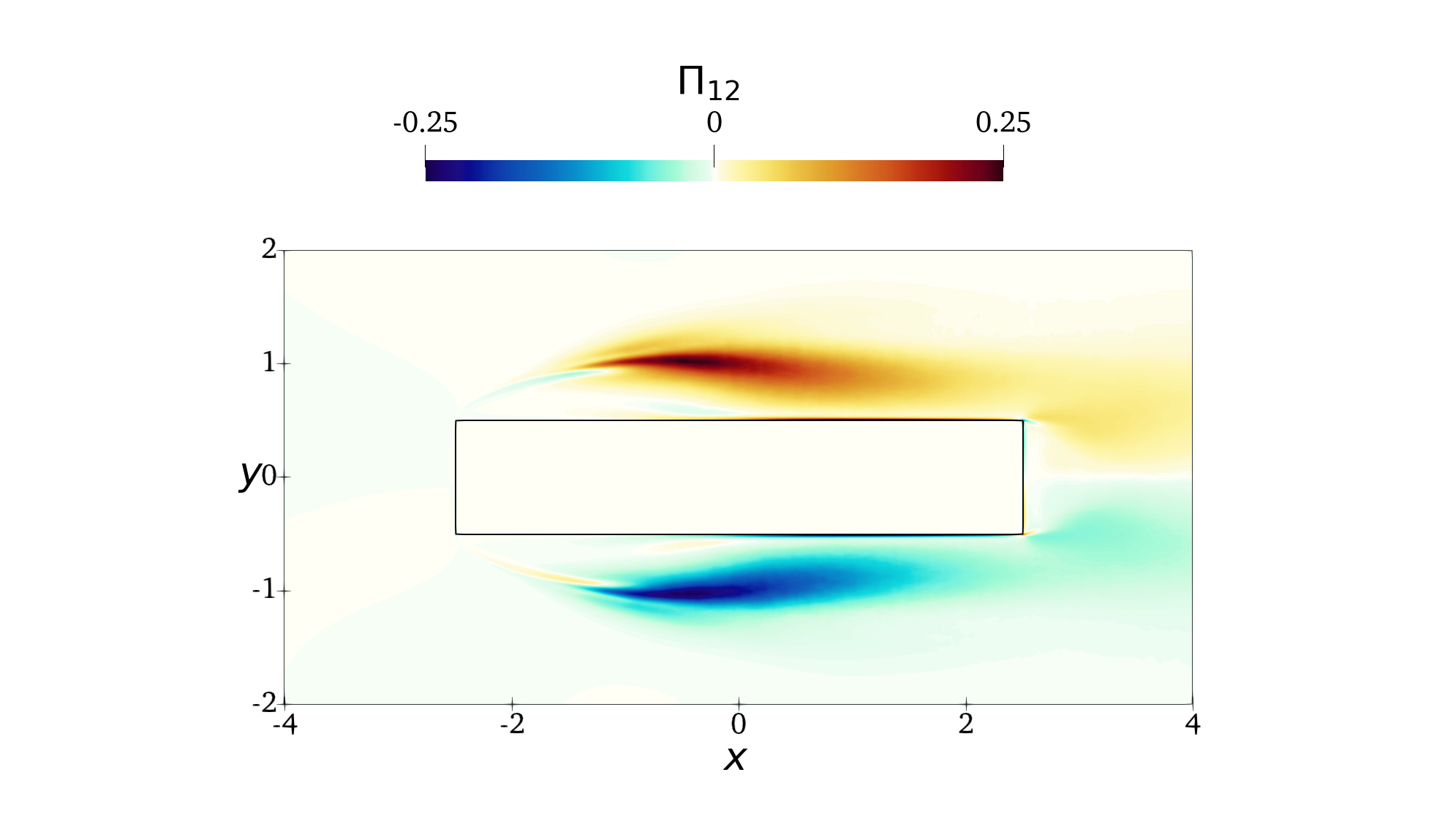}
\includegraphics[trim=290 100 290 200,clip,width=0.49\textwidth]{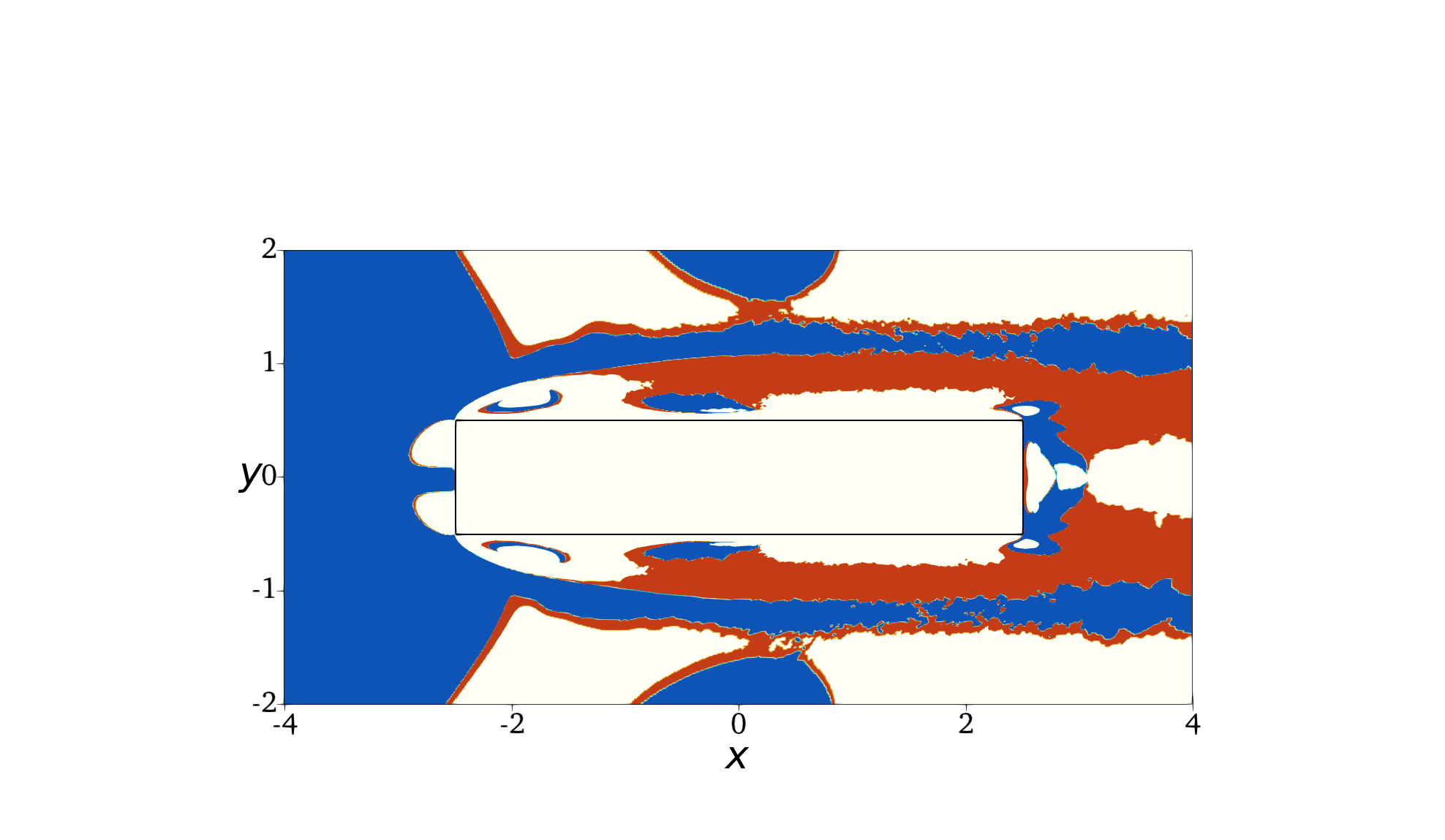}
\includegraphics[trim=290 100 290 200,clip,width=0.49\textwidth]{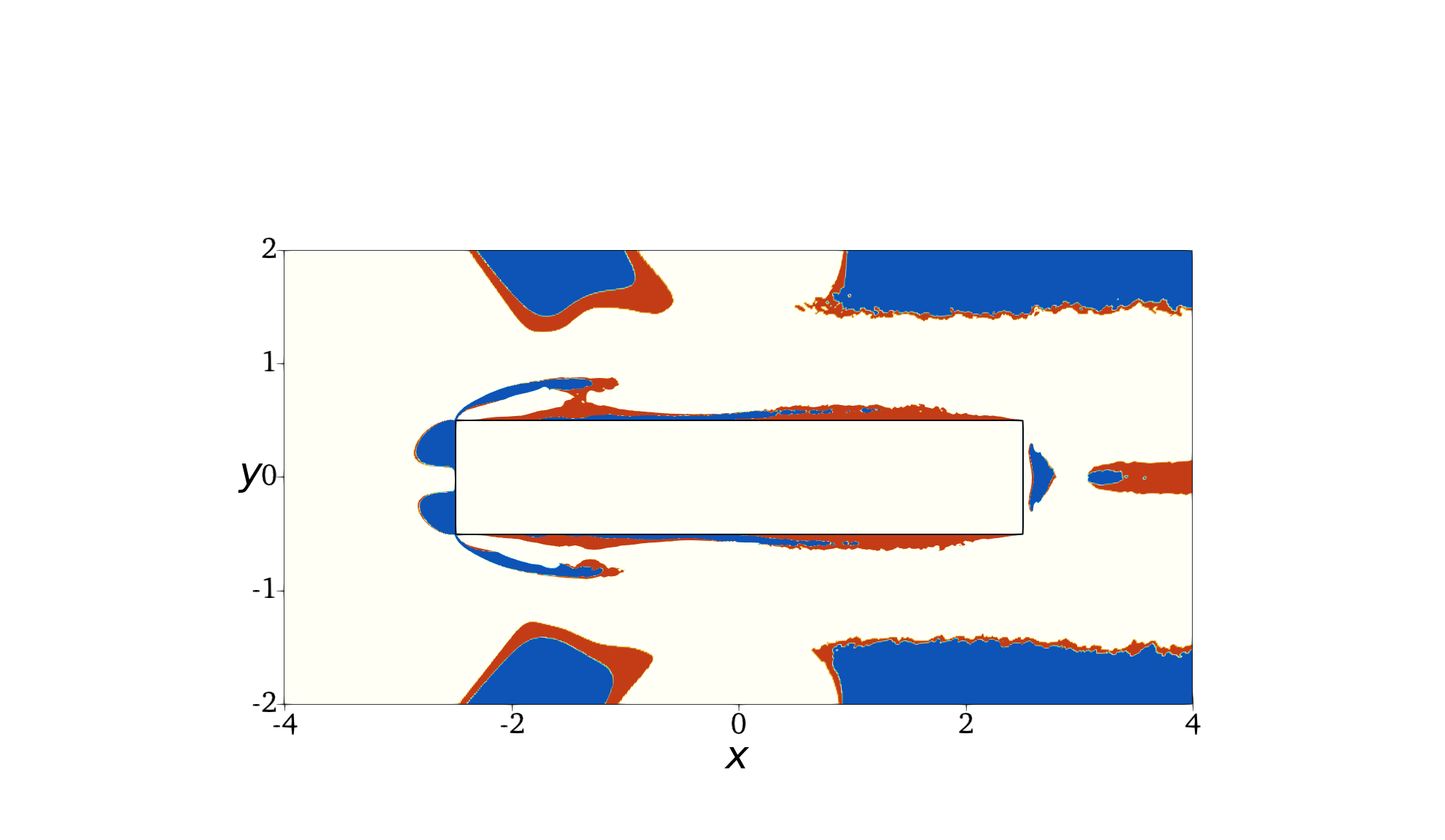}
\caption{Colour map of the pressure-strain tensor $\Pi_{ij}$. In the bottom left panel, 
blue indicates where $\Pi_{33}/|\Pi_{11}|<0.5$ when $\Pi_{11}<0$, $\Pi_{33}>0$ and $\Pi_{22}>0$, i.e $\Pi_{22}>\Pi_{33}$; 
red  indicates where $\Pi_{33}/|\Pi_{11}|>0.5$ when $\Pi_{11}<0$, $\Pi_{33}>0$ and $\Pi_{22}>0$, i.e. $\Pi_{33}>\Pi_{22}$. 
In the bottom right panel, 
blue indicates where $\Pi_{33}/|\Pi_{22}|<0.5$ when $\Pi_{22}<0$, $\Pi_{11}>0$ and $\Pi_{33}>0$, i.e $\Pi_{11}>\Pi_{33}$;
red  indicates where $\Pi_{33}/|\Pi_{22}|>0.5$ when $\Pi_{22}<0$, $\Pi_{33}>0$ and $\Pi_{11}>0$, i.e. $\Pi_{33}>\Pi_{11}$.}
\label{fig:Pstr}
\end{figure}

Figure \ref{fig:Pstr} shows the pressure-strain term, to elucidate how the pressure-mediated redistribution of energy affects the various components of the tensor of turbulent stresses. $\Pi_{11}$ is negative almost everywhere, implying that $\aver{uu}$ is redistributed to the other components. This takes place mainly in the shear layer and in the core of the primary vortex, where both $\Pi_{22}$ and $\Pi_{33}$ are positive, signaling that energy is received from $\aver{uu}$ by both $\aver{vv}$ and $\aver{ww}$. Along the shear layer $\Pi_{22} > \Pi_{33}$, so that $\aver{uu}$ preferentially provides energy to $\aver{vv}$; the opposite occurs in the core of the recirculating region, where $\aver{ww}$ is the largest receiver. Following \cite{gatti-etal-2020}, the bottom left panel visualises this concept, by plotting in red (or blue) the areas where energy is preferentially transferred from $\aver{uu}$ to $\aver{ww}$ (or $\aver{vv}$). In fact, the incompressibility constraint mandates that:
\begin{equation*}
\frac{\Pi_{22}}{\Pi_{11}} + \frac{\Pi_{33}}{\Pi_{11}} = -1.
\end{equation*}

Hence, this panel plots the quantity $\Pi_{33}/|\Pi_{11}|$ under the condition that $\Pi_{11}<0$, $\Pi_{22}>0$ and $\Pi_{33}>0$, and the colour scale is chosen to draw in red where the transfer from $\aver{uu}$ towards $\aver{ww}$ prevails over that towards $\aver{vv}$, and in blue the opposite case. Near the cylinder wall, predominantly in the downstream part, $\Pi_{22}$ is negative, whereas both $\Pi_{11}$ and $\Pi_{33}$ are positive. This essentially visualises the same splatting phenomenon observed for example by \cite{mansour-kim-moin-1988} in the plane turbulent channel flow: owing to the non-penetration boundary condition $v$ is redistributed among $u$ and $w$. Interestingly, near the BARC wall $\Pi_{33} > \Pi_{11}$, so that here the redistribution mainly occurs towards the spanwise velocity component. This is visualised in the bottom right panel of figure \ref{fig:Pstr} which plots the quantity $\Pi_{33}/|\Pi_{22}|$ under the condition that $\Pi_{22}<0$, $\Pi_{11}>0$ and $\Pi_{33}>0$; the colour scale is chosen to highlight in red (blue) the transfer of $\aver{vv}$ towards $\aver{ww}$ ($\aver{uu}$). This is in agreement with the observed prevalence of turbulent structures aligned with the $x$ direction and with larger intensity of $\aver{ww}$ compared to $\aver{uu}$ and $\aver{vv}$ in the near-wall region (recall also that $P_{11}<0$ is a sink for $\aver{uu}$ in this region of the flow).

The middle right panel of figure \ref{fig:Pstr} shows $\Pi_{12}$, the pressure-strain term for $\aver{uv}$. $\Pi_{12}$ is positive almost everywhere -- except for the first portion of the shear layer separating from the leading edge -- with the largest values at the cylinder wall for $x > 0$ and local maxima at $(x,y) \approx (-0.5,1.03)$ within the primary vortex and at $(x,y) \approx (2.7,0.53)$ in the shear layer separating from the trailing edge. Therefore, $\Pi_{12}$ is a sink for $\aver{uv}$ everywhere. Indeed, it dissipates positive $\aver{uv}$ in the shear layer for $x < -1$ and negative $\aver{uv}$ elsewhere with a peak of activity at the cylinder side.

\begin{figure}
\centering
\includegraphics[trim=290 100 290 80,clip,width=0.49\textwidth]{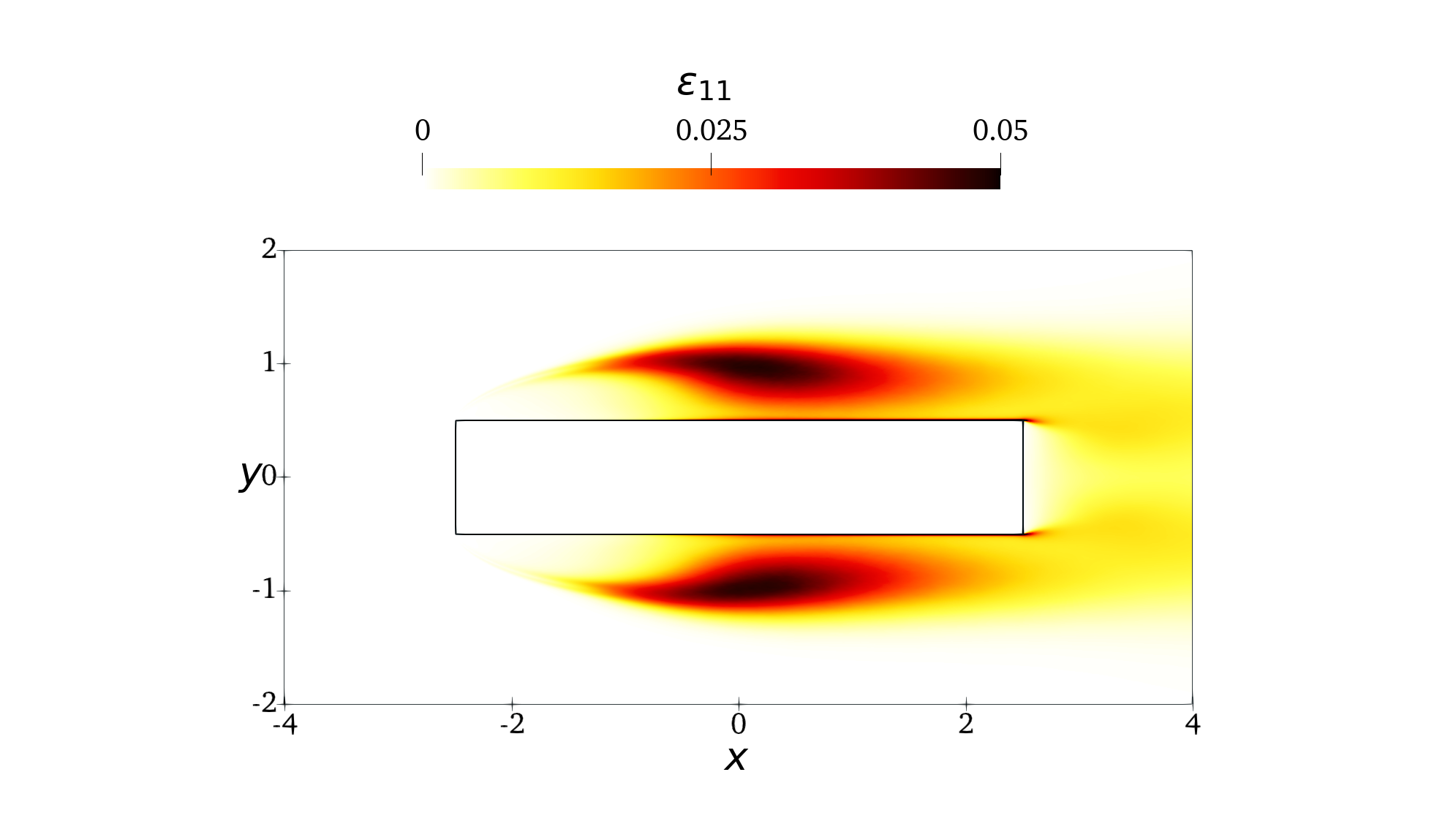}
\includegraphics[trim=290 100 290 80,clip,width=0.49\textwidth]{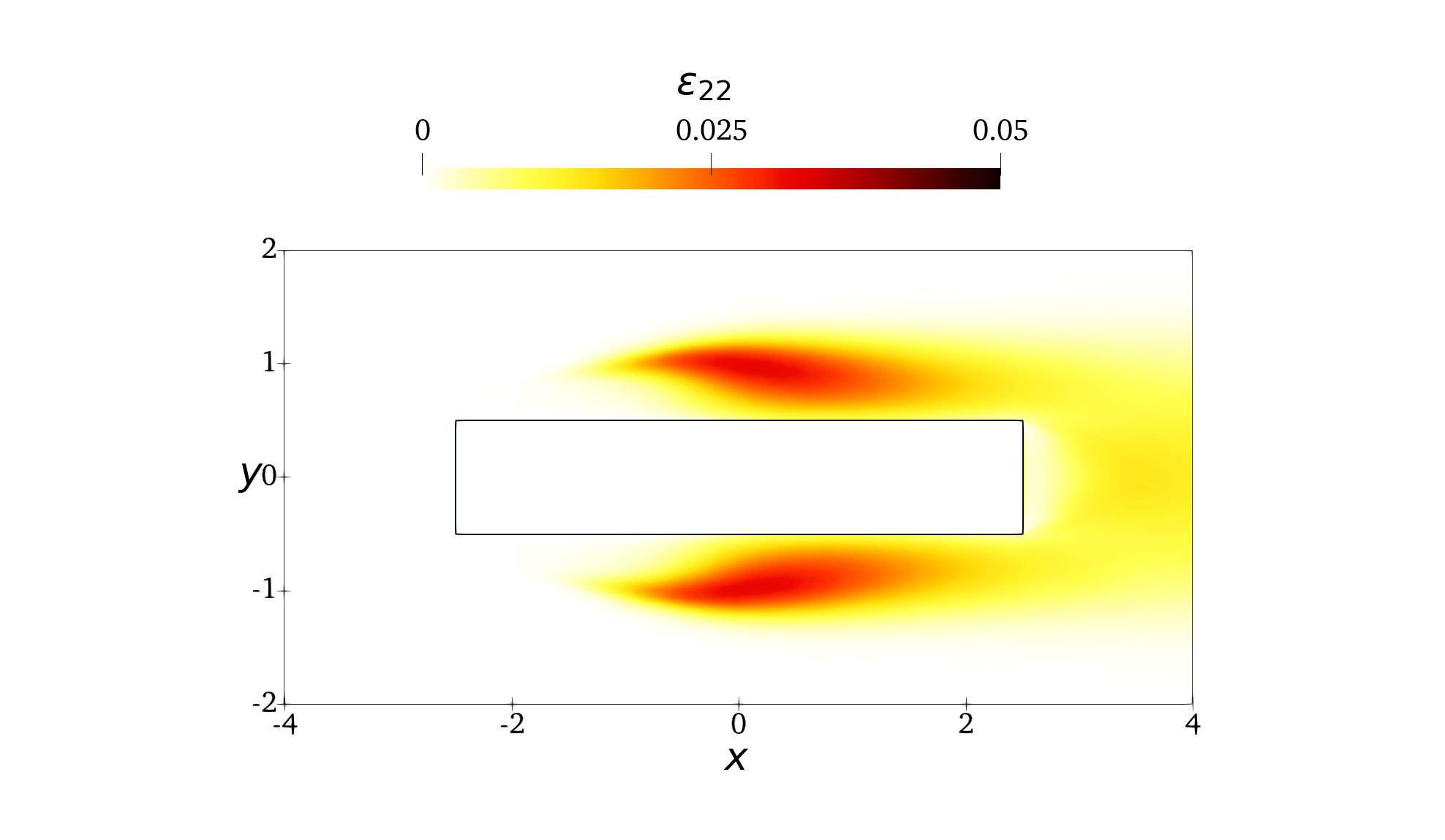}
\includegraphics[trim=290 100 290 80,clip,width=0.49\textwidth]{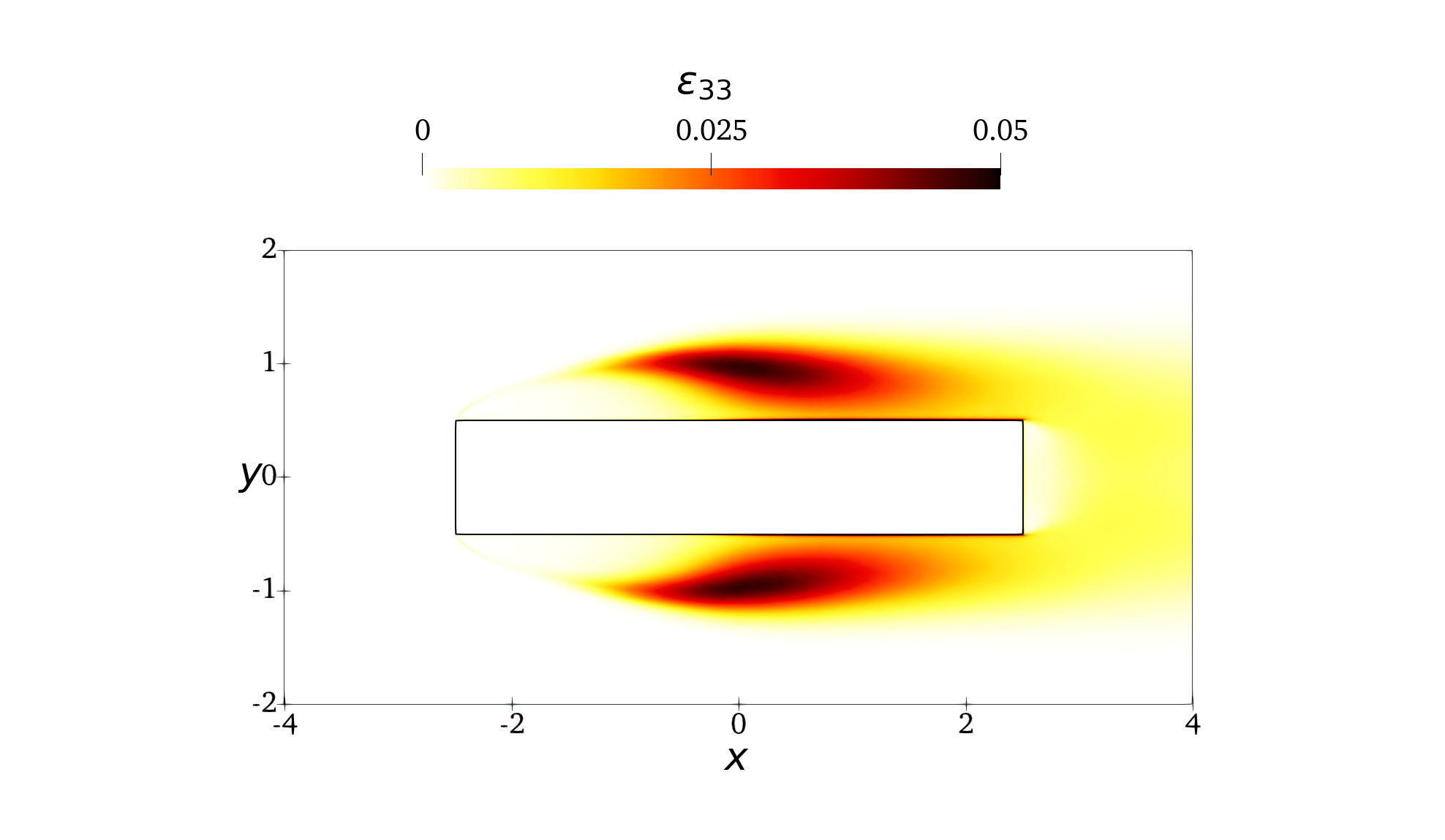}
\includegraphics[trim=290 100 290 80,clip,width=0.49\textwidth]{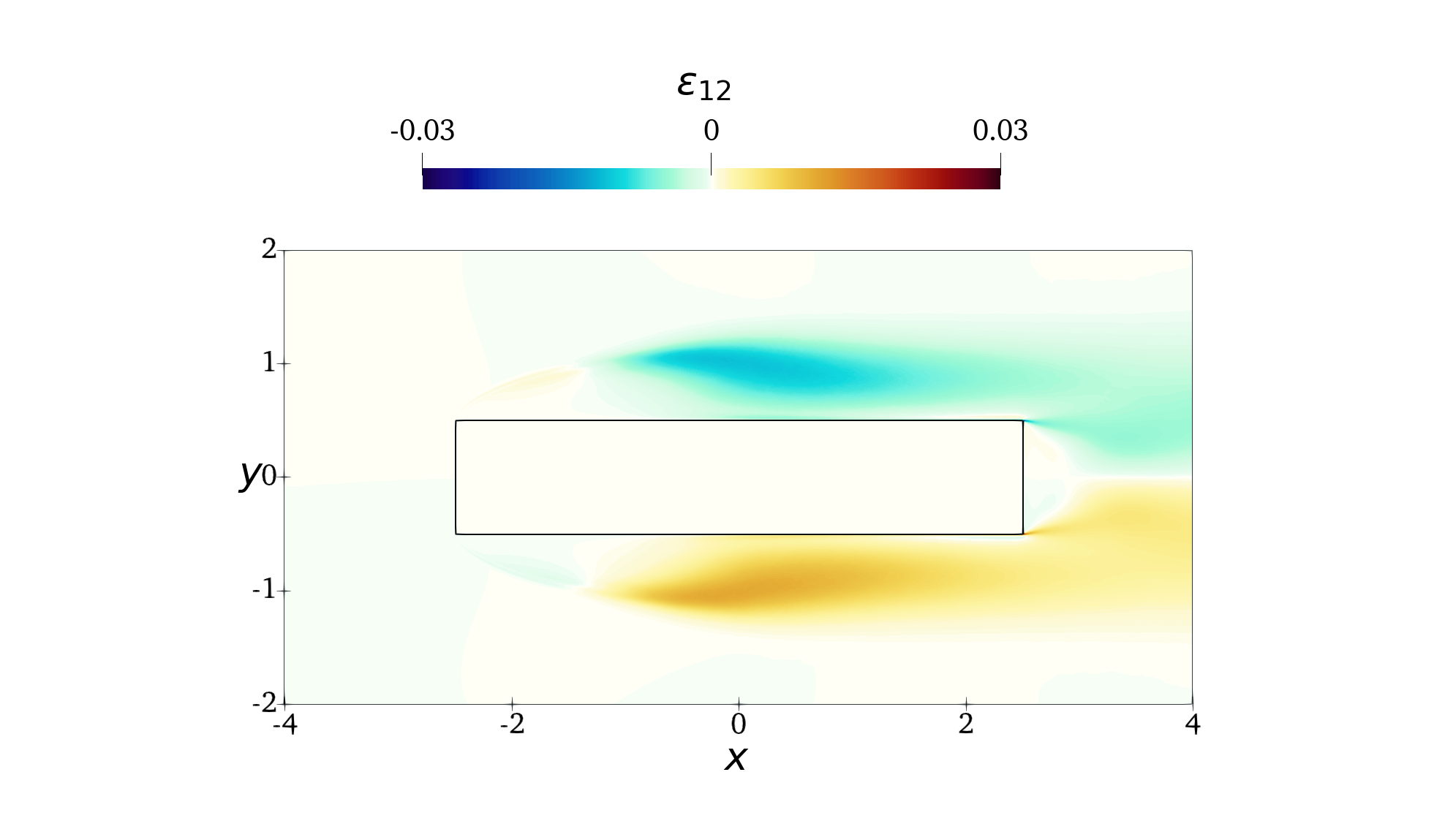}
\caption{Colour map of the dissipation tensor $\epsilon_{ij}$.}
\label{fig:Diss}
\end{figure}
The last term at the r.h.s. of equation \eqref{eq:budget} is the dissipation tensor. The relevant components are drawn in figure \ref{fig:Diss}. For the diagonal components, large values of the dissipation occur in the shear layer for $x \ge -1.5$ and in the core of the primary vortex. Interestingly, the largest values are observed for $\aver{uu}$ and $\aver{ww}$. Moreover, for both the streamwise and spanwise components, unlike for $\aver{vv}$, large dissipation is also seen close to the cylinder, where viscous effects are dominant. This qualitatively resembles the observations (see e.g. \cite{mansour-kim-moin-1988}) put forward for the channel flow, and confirms the larger degree of universality for the dissipative phenomena near a wall. The dissipation term of $\aver{uv}$, $\epsilon_{12}$, is shown in the bottom right panel of figure \ref{fig:Diss}. $\epsilon_{12}$ is negative almost everywhere, except in the first portion of the shear layer separating form the leading edge where it is slightly positive. The global minimum occurs at the trailing-edge corner where the viscous phenomena are dominant, whereas a local minimum, of almost a lower order of magnitude, is seen in the primary vortex at $(x,y) \approx (-0.1,1)$; similarly to $\epsilon_{22}$, $\epsilon_{12}$ close to the cylinder side $\epsilon_{12}$ is low. Overall, like for channel flow $\epsilon_{12}$ is of a lower order of magnitude compared to $P_{12}$ and $\Pi_{12}$ almost everywhere, and therefore its contribution to the production/dissipation of $\aver{uv}$ is negligible.

\begin{figure}
\centering
\includegraphics[trim=290 100 290 80,clip,width=0.49\textwidth]{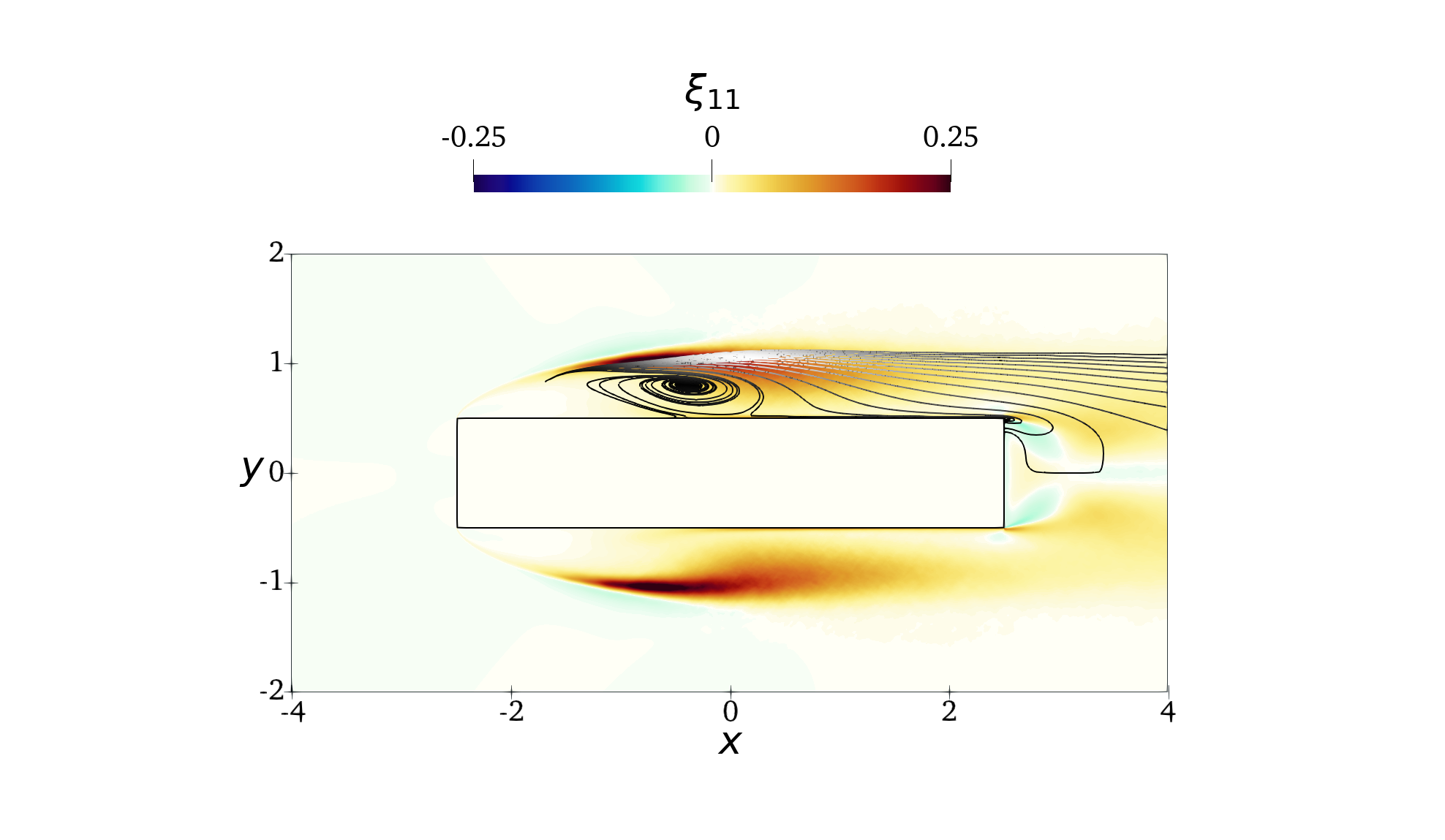}
\includegraphics[trim=290 100 290 80,clip,width=0.49\textwidth]{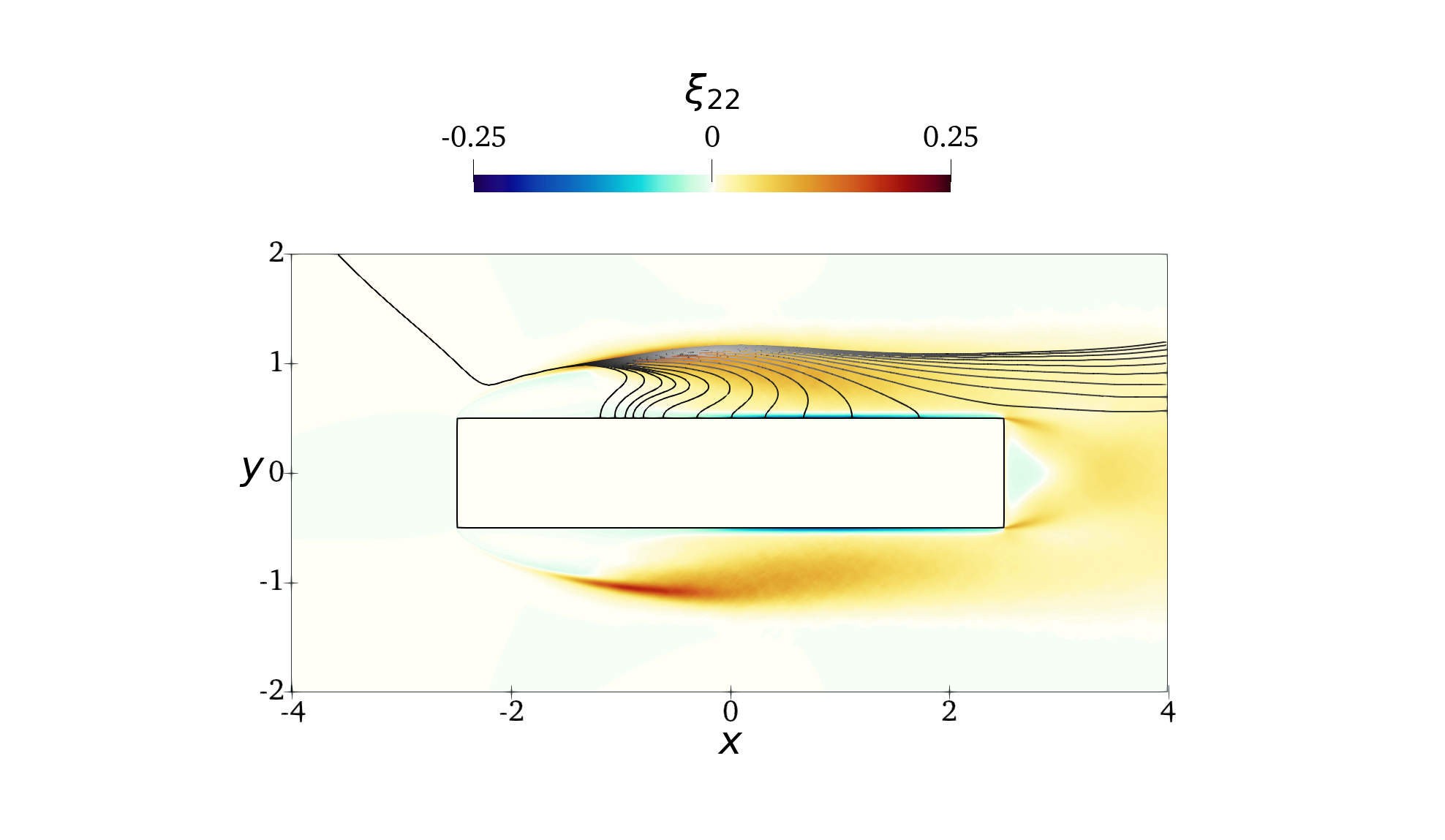}
\includegraphics[trim=290 100 290 80,clip,width=0.49\textwidth]{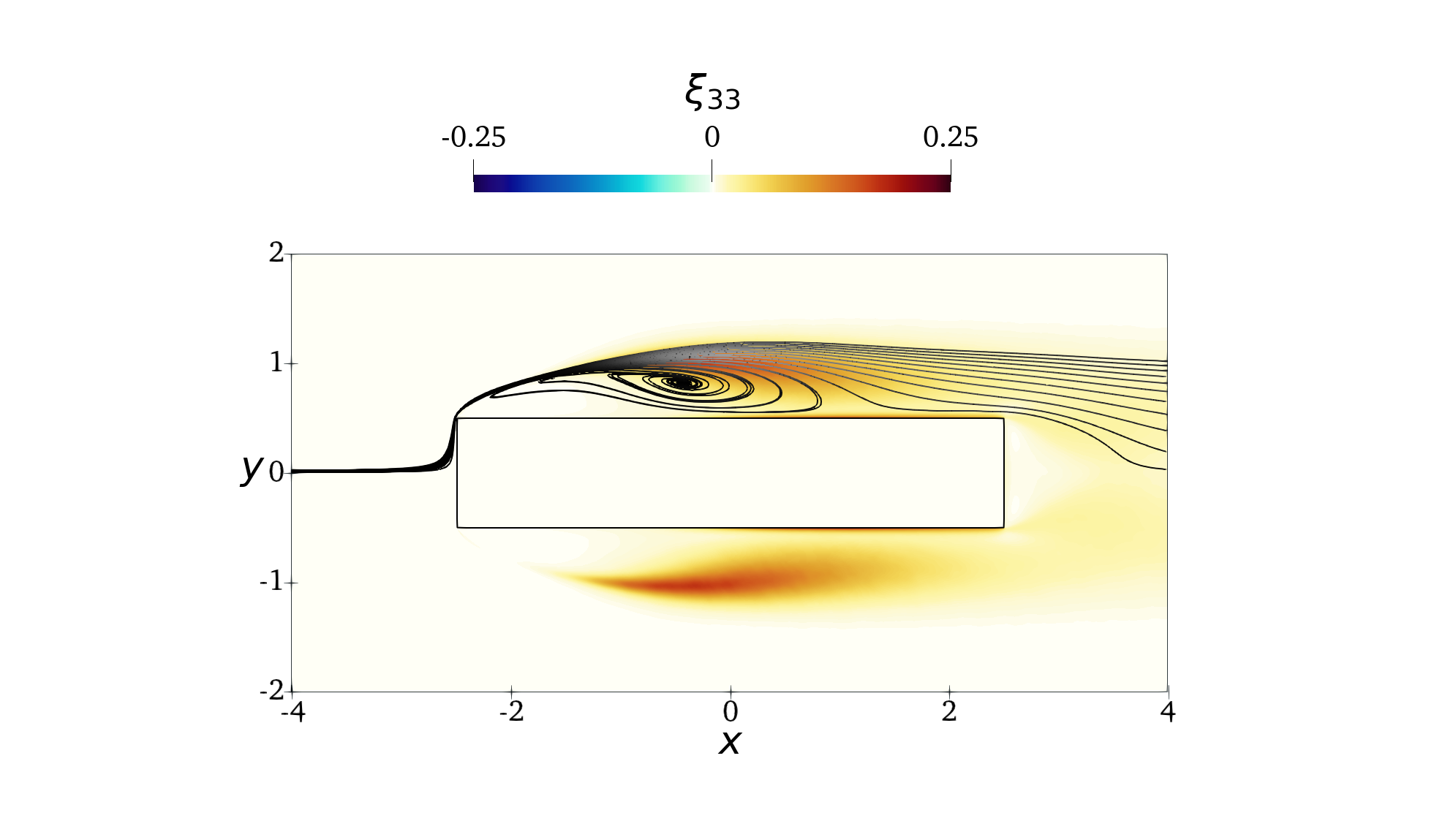}
\includegraphics[trim=290 100 290 80,clip,width=0.49\textwidth]{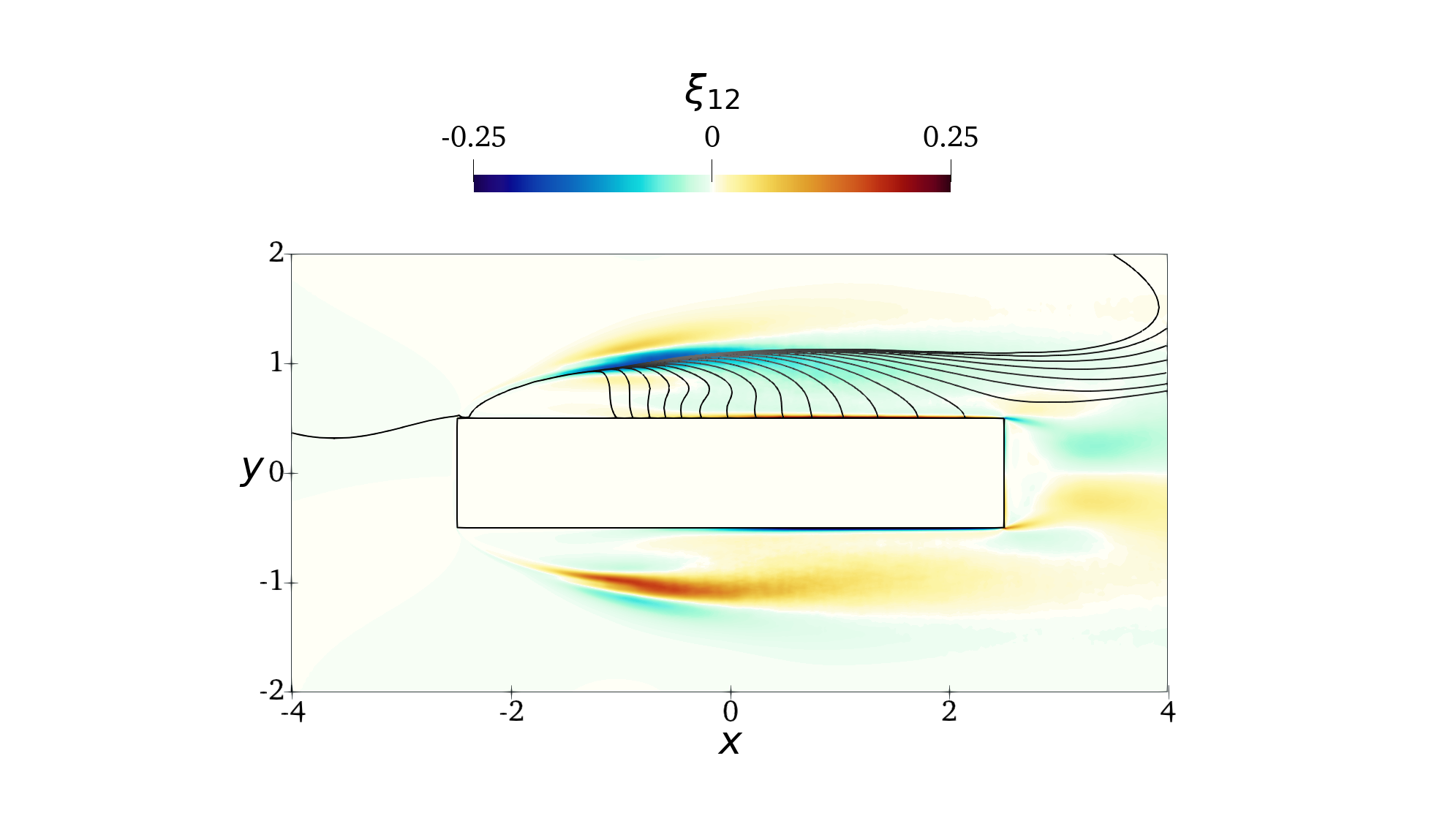}
\caption{Field lines of the fluxes superimposed to a colormap of the complete source term $\xi_{ij}=P_{ij}+\Pi_{ij}-\epsilon_{ij}$.}
\label{fig:phi}
\end{figure}
The production, pressure-strain and dissipation tensors can be put together as a single source term at the r.h.s. of equation \eqref{eq:budget}. Figure \ref{fig:phi} plots contour of the source term $\xi_{ij}=P_{ij} + \Pi_{ij} - \epsilon_{ij}$, together with the field lines of the flux vector with components $(\psi_{x,ij},\psi_{y,ij})$. These lines show how the excess of $\aver{u_i u_j}$ gets redistributed in the flow. For $\aver{uu}$, a positive $\xi_{11}$ shows an excess of production of $\aver{uu}$ which is redistributed along the field lines originating from a singularity point placed in the shear layer at $(x,y) \approx (-1.7,0.82)$, close to the maximum of $\xi_{11}$ (corresponding to the maximum of $P_{11}$). Following the field lines, some of them are observed to carry $\aver{uu}$ downstream towards the wake, whereas others enter the large recirculating region of the primary vortex. Much like the mean flow, these  lines show a spiraling motion and vanish because of dissipation close to the rotation centre of the vortex, at $(x,y) \approx (-0.4,0.8)$. Here, figure \ref{fig:Diss} confirms the importance of viscous effects. The field lines of $\aver{vv}$ are different. They originate in the shear layer at $(x,y) \approx (-2.12,0.82)$, close to the maximum of $\xi_{22}$ (where, as seen in figures \ref{fig:Prod} and \ref{fig:Pstr}, $\Pi_{22} > | P_{22} | $). Some of them carry part of the excess of $\aver{vv}$ downstream towards the wake region, whereas others are attracted by the solid wall. In fact, near the wall $\xi_{22}<0$ because of the splatting effect, so that the wall acts as a sink. The field lines of $\aver{ww}$ qualitatively resemble those for $\aver{uu}$. Indeed, $\xi_{33}$ is maximum in the shear layer, where the pressure strain produces spanwise fluctuations. As for $\aver{uu}$, some lines advect part of the excess of $\aver{ww}$ downstream towards the wake, whereas others enter and remain in the recirculating bubble, spiralling inwards towards the centre of rotation of the primary vortex, and ending at $(x,y) \approx (-0.44,0.83)$ due to viscous dissipation, i.e. slightly upstream the vanishing point of the lines of $\aver{uu}$. Interestingly, these lines reach the most upstream part of the primary vortex, and highlight the zone where the reverse flow separates because of the adverse pressure gradient. The field lines of $\aver{uv}$ have a distinct shape. Some originate at the solid wall where $\xi_{12}$ is maximum, whereas others originate in the wake and carry $\aver{uv}$ upstream. The line set is attracted by the large sink region with $\xi_{12}<0$ placed in the shear layer, and approach the leading edge corner. Again, when discussing these field lines, fluxes of $\aver{uv}$ should not be interpreted in terms of energy transfer, as $\aver{uv}$ is not a positive-definite quantity \cite{gatti-etal-2020}.  

\section{Conclusions}

We have studied with a Direct Numerical Simulation the BARC benchmark, i.e. a 5:1 rectangular cylinder immersed in a uniform flow, in the turbulent regime. Despite the simple and somewhat idealised shape of the cylinder, the BARC flow contains complex and fascinating features typical of separating and reattaching flows over complex geometries. The large scales associated with the flow instabilities coexist and interact with the smaller scales associated with turbulent motions to create a rich and intricate scenario. A fully reliable statistical characterisation of the BARC flow is still lacking: large scatter of data is observed already for first-order statistics \cite{bruno-salvetti-ricciardelli-2014}, as the flow is very sensitive to geometry details and various types of external disturbances which are unavoidable in  experiments, and to various modelling and discretisation choices in numerical simulations. Similarly, the agreement between experimental and numerical information is not entirely satisfying yet.

This study has replicated the first DNS of the BARC flow, recently carried out by Cimarelli et al. \cite{cimarelli-leonforte-angeli-2018b} at a value of the Reynolds number (based on the body thickness and the uniform incoming velocity) of $Re=3000$. The numerical toolbox employed here is quite different though, as we have used a in-house finite-differences solver instead of the finite-volumes OpenFOAM package. Moreover, we have used a finer spatial discretisation -- the total number of point is more than one order of magnitude larger -- and a longer averaging time. The goal of the study is to contribute to a solid and reliable DNS-based benchmark, that could be used as reference for RANS- or LES-based simulations of the BARC flow.

We have described and discussed the main features of the flow in terms of first- and second-order statistics, and the main differences with the available DNS results \cite{cimarelli-leonforte-angeli-2018b} have been reported. Our results present a near-perfect symmetry along the $y=0$ centerline, which demonstrates the adequacy of the statistical sample. The longitudinal extent of the largest primary vortex has been found to be $10\%$ larger, and the secondary vortex $10\%$ smaller than the reference study. The position of the recirculating regions are different too. Also, we do not confirm the presence of a spot in the leading-edge shear layer where the production rate of turbulent kinetic energy becomes negative \cite{cimarelli-etal-2019-negative}.

Moreover, the statistical understanding of the BARC flow has been furthered, and for the first time the whole set of terms involved in the budget equation for the tensor of the Reynolds stresses has been presented and discussed in detail. The analysis highlights the strongly anisotropic and inhomogeneous nature of the flow. The shear layer separating from the leading-edge corner is initially laminar, but its instability soon leads to the fluctuating field draining energy from the mean flow and feeding $\aver{u'u'}$, which is the dominant contributor to the turbulent kinetic energy. The other components, $\aver{v'v'}$ and $\aver{w'w'}$, are instead produced by a redistribution of $\aver{u'u'}$ driven by the pressure-strain term, and have a lower intensity. The cross-stream component of the mean flow in the shear layer is fed from the fluctuating field. The excess of turbulent kinetic energy is partially advected towards the wake, and partially transported within the large primary vortex. Here, $\aver{u'u'}$ and $\aver{v'v'}$ are fed by energy drained from the mean flow, and redistribution moves energy away from $\aver{u'u'}$ towards $\aver{v'v'}$ and $\aver{w'w'}$. In that portion of the primary vortex where reverse flow takes place (close to the cylinder wall), a splatting effect is observed, with $\aver{v'v'}$ being redistributed to $\aver{u'u'}$ and (mainly) $\aver{w'w'}$. In this region the production term for $\aver{u'u'}$ is negative, indicating that energy is drained from $\aver{u'u'}$ to feed the mean flow. Hence, in the reverse boundary layer the turbulent kinetic energy is mainly organised in spanwise fluctuations. Energy dissipation takes place mainly in the core of the primary vortex, where it is comparable for all the normal stresses, and in the vicinity of the cylinder wall, where the largest dissipation is observed for the wall-parallel components. 

While the present state of affairs calls for further reduction of the uncertainty in the statistical description of this flow, the present dataset -- which is made available to the research community at the following DOI: {\tt https://doi.org/10.5281/zenodo.4472682} -- provides an useful addition to the existing knowledge on the BARC benchmark, an interesting flow for turbulence modelling.

\begin{acknowledgements}
Computing time has been provided by the Italian supercomputing center CINECA under the ISCRA C projects TAWBF and AGKEbump. 
\end{acknowledgements}

\section*{Conflict of interest}
The authors declare that they have no conflict of interest.

\bibliographystyle{spmpsci}

\end{document}